\documentclass[envcountsame]{llncs}
\usepackage{amsfonts}

\newcommand{\ra}{\rightarrow}
\newcommand{\Ra}{\Rightarrow}
\newcommand{\La}{\Leftarrow}
\newcommand{\la}{\leftarrow}
\newcommand{\lra}{\leftrightarrow}
\newcommand{\LRa}{\Leftrightarrow}

\newcommand{\lc}{{\lceil}}
\newcommand{\rc}{{\rceil}}
\newcommand{\lf}{{\lfloor}}
\newcommand{\rf}{{\rfloor}}
\newcommand{\PTM}{{\rm PTM}}
\newcommand{\TM}{{\rm TM}}
\newcommand{\Prv}{{\rm Pr}}
\newcommand{\N}{\mathbb{N}}

\newcommand{\CA}{{\rm CA}}
\newcommand{\CR}{{\rm CR}}

\newcommand{\CD}{{\rm CD}}
\newcommand{\Acc}{{\rm Acc}}

\newcommand{\Acpt}{{\rm Acpt}}
\newcommand{\NP}{{\rm NP}}
\newcommand{\bm}{\boldmath}
\newcommand{\ov}{\overline}
\newcommand{\U}{{\rm U}}

\newcommand{\Size}{{\rm Size}}
\newcommand{\TSAT}{{\rm 3SAT}}

\newcommand{\ISET}{{\rm ISET}}
\newcommand{\TRANS}{{\rm TRANS}}
\newcommand{\EVAL}{{\rm EVAL}}
\newcommand{\dec}{\triangleright}

\newcommand{\Bit}{{\rm Bit}}

\newcommand{\NNN}{{\mathfrak N}}
\newcommand{\Se}{\mathbf S}
\newcommand{\0}{\mathbf 0}
\newcommand{\1}{\mathbf 1}
\newcommand{\2}{\mathbf 2}
\newcommand{\3}{\mathbf 3}
\newcommand{\4}{\mathbf 4}

\newcommand{\aaa}{\mathbf a}
\newcommand{\bbb}{\mathbf b}
\newcommand{\ccc}{\mathbf c}
\newcommand{\ddd}{\mathbf d}
\newcommand{\eee}{\mathbf e}

\newcommand{\iii}{\mathbf i}
\newcommand{\jjj}{\mathbf j}
\newcommand{\kkk}{\mathbf k}
\newcommand{\lll}{\mathbf l}
\newcommand{\mmm}{\mathbf m}
\newcommand{\nnn}{\mathbf n}

\newcommand{\sss}{\mathbf s}
\newcommand{\ttt}{\mathbf t}
\newcommand{\uuu}{\mathbf u}
\newcommand{\vvv}{\mathbf v}
\newcommand{\www}{\mathbf w}
\newcommand{\xxx}{\mathbf x}
\newcommand{\yyy}{\mathbf y}
\newcommand{\zzz}{\mathbf z}

\newcommand{\hik}{\dot{-}}

\newcommand{\PA}{{\rm PA}}

\newcommand{\Zn}{\mathbb{Z}/n\mathbb{Z}}
\newcommand{\Znc}{\mathbb{Z}/n^c\mathbb{Z}}

\renewcommand{\phi}{\ensuremath{\varphi}}

\newcommand{\SAT}{\mbox{\rm SAT}}
\newcommand{\CSAT}{{\cal SAT}}
\newcommand{\cCSAT}{co{\mbox{\rm -}}{\cal SAT}}

\newcommand{\DecSAT}{\mbox{\rm DecSAT}}

\newtheorem{rem}{Remark}

\begin{document}
\title{Resource Bounded Unprovability of \\
Computational Lower Bounds \\
(Part 1)
}
\author{Tatsuaki Okamoto$^{*}$ \ \ \ \ \  Ryo Kashima$^{**}$ }
\institute{* NTT Laboratories, Nippon Telegraph and Telephone Corporation \\
1-1 Hikarino-oka, Yokosuka-shi, Kanagawa, 239-0847 Japan \\
 {**} Dept. of Mathematical and Computing Sciences,
Tokyo Institute of Technology \\
1-12-1 O-okayama Meguro-ku, Tokyo, 152-8552 Japan \\
 \ \\
March 30, 2005
}

\maketitle

\begin{abstract}
This paper introduces new notions of asymptotic 
proofs, PT(polynomial-time)-extensions,
PTM(polynomial-time Turing machine)-$\omega$-consistency, 
etc. 
on formal theories of arithmetic
including PA (Peano Arithmetic).
An asymptotic proof is a set of infinitely many
formal proofs, which is introduced to 
define and characterize a property, PTM-$\omega$-consistency,
of a formal theory. 
Informally speaking, 
PTM-$\omega$-consistency is
a {\it polynomial-time bounded}
version (in asymptotic proofs) of $\omega$-consistency, and 
characterized in two manners:
(1) (in the light of the {\it extension of PTM to TM})
the resource {\it unbounded} version of
PTM-$\omega$-consistency is equivalent to $\omega$-consistency,
and (2) (in the light of {\it asymptotic proofs by PTM})
a PTM-$\omega$-{\it inconsistent} theory
includes an axiom
that only a super-polynomial-time Turing machine
can prove asymptotically over PA, under some assumptions.
This paper shows that 
{\it P$\not=$NP (more generally, any super-polynomial-time lower bound 
in PSPACE) is unprovable in a PTM-$\omega$-consistent theory $T$}, 
where $T$ is a consistent PT-extension of PA (although 
this paper does not show that P$\not=$NP is unprovable in PA,
since PA has not been proven to be
\PTM-$\omega$-consistent). 
This result implies that 
to prove P$\not=$NP
by any technique
requires a PTM-$\omega$-{\it inconsistent} theory,
which should include an axiom that only a super-polynomial-time machine
can prove asymptotically over PA
(or implies a super-polynomial-time computational upper bound)
under some assumptions.
This result is a kind of generalization 
of the result of ``Natural Proofs'' by Razborov and Rudich \cite{RazRud94},
who showed that to prove ``P$\not=$NP'' by a class of 
techniques called ``Natural Proofs'' implies a 
super-polynomial-time (e.g., sub-exponential-time) algorithm 
that can break a typical cryptographic primitive, a pseudo-random generator. 
Our result also implies that
any relativizable proof of P$\not=$NP
requires the {\it resource unbounded version}
of \PTM-$\omega$-{\it inconsistent} theory,
$\omega$-{\it inconsistent} theory,
which suggests another negative result 
by Baker, Gill and Solovay \cite{BakGilSol75} that
no relativizable proof can prove ``P$\not=$NP''
in PA, which is a $\omega$-consistent theory. 
Therefore, our result gives a unified view to  
the existing two major negative results on proving P$\not=$NP,
Natural Proofs and relativizable proofs,
through the two manners of characterization of PTM-$\omega$-consistency. 
We also show that the PTM-$\omega$-consistency of $T$
cannot be proven in any PTM-$\omega$-consistent theory $S$,
where $S$ is a consistent PT-extension of $T$.
That is, to prove the independence of P vs NP
from $T$ by proving the PTM-$\omega$-consistency of $T$
requires a PTM-$\omega$-{\it inconsistent} theory,
or implies a super-polynomial-time computational upper bound
under some assumptions.
This seems to be related to 
the results of 
Ben-David and Halevi \cite{BenHal92}
and Kurz, O'Donnell and Royer \cite{KurODnRoy87},
who showed that to prove the independence of P vs NP from PA
using any currently known mathematical paradigm implies 
an extremely-close-to-polynomial-time (but still super-polynomial-time) 
algorithm that can solve NP-complete problems.  
Based on this result, we show that
{\it the security of any computational
cryptographic scheme is unprovable}
in the setting 
where adversaries and provers are modeled as
polynomial-time Turing machines and 
only a PTM-$\omega$-consistent theory is allowed to
prove the security.

\end{abstract}



\tableofcontents

\section{Introduction}\label{sec:intro}

\subsection{Background}\label{sec:back}
It looks very mysterious that proving computational
lower bounds is extremely difficult, although 
many people believe that there exist various natural intractable
problems that have no efficient algorithms that can solve them.
A classical technique, diagonalization,
can separate some computational classes like 
P $\not=$ EXP, but it fails to separate 
computational classes between P and PSPACE,
which covers almost all practically interesting computational problems. 
Actually, we have very few results on the lower bounds 
of computational natural problems between P and
PSPACE. 
The best known result of computational lower bounds
(in standard computation models such as Turing machines and Boolean circuits)
of a computational natural problem is about $5n$ in 
circuit complexity \cite{IwaMor02}, where $n$ is problem size.
Therefore, surprisingly, it is still very hard for us
to prove even the $6n$ lower bound of TQBF, a PSPACE complete
problem, which is considered to be much more intractable than
NP complete problems.

Considering this situation, it seems natural to think 
that there is some substantial reason  
why proving computational lower bounds is so difficult.
The ultimate answer to this question 
would be to show that
such computational lower bounds are impossible to prove,
e.g., showing its independence from 
a formal proof system like Peano Arithmetic 
(a formal system for number theory) 
and ZFC (a formal system for set theory).   

This paper gives a new type of impossibility result,
{\it resource bounded} impossibility, in the proof of
computational lower bounds.

\subsection{Our Results}\label{sec:our}
Let theory $T$, on which we are assumed to try to prove P$\not=$NP,
be a consistent PT-extension of PA,
throughout this paper (and hereafter in this section),
where theory $T$ is called PT-extension if there exists a
polynomial-time algorithm that, given $n \in \N$, decides 
whether $n$ is the G{\"o}del number of an axiom of $T$
(Section \ref{pt-ext}).

This paper shows the following results.

\subsubsection{New Notions} 
We introduce notions of asymptotic proofs, polynomial-time 
proofs, polynomial-time decisions, PT-extensions,
PTM-$\omega$-consistency etc. 
on formal theories of arithmetic
including PA (Peano Arithmetic).

\begin{itemize}
\item Asymptotic Proofs (Section \ref{asymp-proof}):
$Q_1 \xxx_1 \ \cdots Q_k \ \xxx_k  \  \phi(\xxx_1, \ldots, \xxx_k)$
has an asymptotic proof over $T$
if
$$ 
Q_1 x_1 \in \N \ \cdots \ Q_k x_k \in \N \ \ \ 
T \vdash \phi(\xxx_1, \ldots, \xxx_k),$$
where a boldfaced symbol (e.g, $\xxx$) denotes
a variable in theory $T$ or numeral 
(e.g., $\xxx$ is the numeral of $x \in \N$),
and $Q_i$ ($i \in \{1,\ldots,k\}$) denotes an unbounded quantifier.   

\item
Polynomial-time proofs (Section \ref{PT-proofs}):
$$\PTM_{e}(x) \vdash_T \phi(\xxx)$$
denotes that 
a PTM (polynomial-time Turing machine)
coded by $e \in \N$, given $x \in \N$ and the G{\"o}del number
of the expression of $\{ \phi(\aaa) \mid a \in \N  \}$ (constant in $|x|$),  
produces a proof (tree) of formula $\phi(\xxx)$ in theory $T$.

\item
Let theory $S$ be a PT-extension of theory $T$.
PTM-$\omega$-consistency (Definition \ref{PTM-omega-con}):
Theory $S$ is PTM-$\omega$-consistent for $\Delta_1$-formula 
$\phi(\eee^*,\xxx)$ over theory $T$,
if the following condition holds. 
\begin{eqnarray*}  
& & 
\forall e \in \N \ \ \exists e^* \in \N 
\ \ \exists \ell  \in \N \ \ \forall n \geq \ell 
\ \ \forall c \in \N \ \ \
\PTM_{e}(n) \ 
\not\vdash_T \ 
\exists \xxx \ (\nnn \leq \xxx < \nnn+|\nnn|^{\ccc}) \ \ 
\phi(\eee^*,\xxx)
\\
\Ra \ \ \ 
& & 
\forall e \in \N \ \ \exists e^* \in \N \ \ 
\exists \ell  \in \N \ \ \forall n \geq \ell \ \ \ 
\PTM_{e}(n) \ 
\not\vdash_S \ 
\exists \xxx \geq \nnn \ \ \phi(\eee^*,\xxx),
\end{eqnarray*} 
where 
$|\nnn|$ denotes the numeral of $|n|$ 
(see Section \ref{sec:notation}). 

Theory $T$ is PTM-$\omega$-consistent for $\phi(\eee^*,\xxx)$,
if $T$ is PTM-$\omega$-consistent for $\phi(\eee^*,\xxx)$
over $T$.


\end{itemize}

\subsubsection{Formalization of P$\not=$NP} 
 
We formalize P$\not=$NP as follows
(Definition \ref{vartheta}): 
\begin{equation}
{\ov{\mbox{P$\not=$NP}}} \ \ \equiv \ \
\forall \eee \ \ \forall \nnn \ \ \exists  \xxx \geq \nnn \ \ \
\neg\DecSAT(\eee,\xxx),
\end{equation}
where 
$\DecSAT(\eee,\xxx)$
is a formula in PA which informally means that a PTM coded by $e$
correctly decides the satisfiability or unsatisfiability
of a 3CNF coded by $x$.


\subsubsection{Unprovability of ${\ov{\mbox{P$\not=$NP}}}$
in a PTM-$\omega$-consistent theory}  

$\ov{\mbox{P$\not=$NP}}$ cannot be proven in $T$
that is PTM-$\omega$-consistent
for any $\Delta_2^P$ formula (Theorem \ref{mainthm5-dec}):
$$
T \not\vdash \
{\ov{\mbox{\rm P}\not=\mbox{\NP}}}.
$$

\subsubsection{Unprovability of PTM-$\omega$-consistency
in a PTM-$\omega$-consistent theory}  

Let theory $S$ be a consistent PT-extension of theory $T$,
and $S$ be PTM-$\omega$-consistent for any 
$\Delta_2^P$-formula. 
Then, 
PTM-$\omega$-consistency of $T$ 
for a $\Delta_2^P$-formula cannot be proven in $S$.
(Theorem \ref{thm:ptm-omega-con-formal})

Thus, the independence 
of P vs NP from $T$ by proving PTM-$\omega$-consistency
of $T$ for a $\Delta_2^P$-formula (i.e., through Theorem \ref{mainthm5-dec})
cannot be proven in $S$. 

In fact, the existence of PTM-$\omega$-consistent theory $T$ for a 
$\Delta^P_2$-formula has not been proven, 
and the independence of P vs NP from PA has not been
proven. 

\subsubsection{Unprovability of the Security of Computational
Cryptography}  

The one-wayness of any function family 
is {\it unprovable}
in the setting where an adversary and a prover are modeled to be 
polynomial-time Turing machines, and the security proof should be made 
in a PTM-$\omega$-consistent theory $T$ 
for $\Delta^P_2$
(Theorem \ref{imp-crypto}).
In other words,
the security of any computational
cryptographic scheme is unprovable under this setting.

\subsection{An Implication of Our Results}\label{sec:our-result-implication}
To interpret our results,  
let assume the following hypotheses:
\begin{itemize}
\item \
(Hypothesis 1) \ 
$ \NNN \models \ {\ov{\mbox{P$\not=$NP}}}$,
where $\NNN$ is the standard model of natural numbers
(i.e., P$\not=$NP is true.)
\item \
(Hypothesis 2) \
PA is PTM-$\omega$-consistent for $\Delta^P_2$.
\end{itemize}

We then have the following consequence from our results.
\begin{itemize}
\item
{\it P vs NP is independent from PA}.

This is because
${\ov{\mbox{P$\not=$NP}}}$ is consistent with PA
from Hypothesis 1, and $\neg{\ov{\mbox{P$\not=$NP}}}$ 
is consistent with PA, since  
$\PA \not\vdash {\ov{\mbox{P$\not=$NP}}}$
(from Theorem \ref{mainthm5-dec} and Hypothesis 2).

\item
{\it Hypothesis 2 cannot be proven 
in a PTM-$\omega$-consistent theory $T$}, 
where $T$ is a consistent PT-extension of PA. 

That is, even if P vs NP is independent from PA,
the independence (by proving Hypothesis 2)  
cannot be proven in a PTM-$\omega$-consistent theory $T$. 
\end{itemize}

\subsection{Characterization of PTM-$\omega$-consistency}\label{sec:implication}

Informally speaking, 
PTM-$\omega$-consistency is 
a {\it polynomial-time bounded}
version (in asymptotic proofs) of $\omega$-consistency, and 
characterized in two manners:
\begin{enumerate}
\item
(Characterization in the light of the {\it extension of PTM to TM})
The resource {\it unbounded} version of
PTM-$\omega$-consistency is equivalent to $\omega$-consistency.
\item
(Characterization in the light of {\it asymptotic proofs by PTM})
A PTM-$\omega$-{\it inconsistent} theory
includes an axiom
that only a super-polynomial-time Turing machine
can prove asymptotically over PA, under some assumptions.
\end{enumerate}

First, 
PTM-$\omega$-consistency can be
extended to a resource {\it unbounded} TM (Turing machine) version of 
PTM-$\omega$-consistency, 
TM-$\omega$-consistency, which is equivalent to $\omega$-consistency
(Remark \ref{rem:5} of Definition \ref{PTM-omega-con} 
in Section \ref{sec:ptm-omega-con}).

Second,
PTM-$\omega$-consistent theory $T$ 
is a formal theory, but is characterized by asymptotic proofs of 
PTM provers over $T$. 
A proof in a formal theory itself is a finite
length proof and has no asymptotic property
as well as no implication of prover's computational capability.
However, 
PTM-$\omega$-consistency is defined through asymptotic proofs of 
PTM provers, and 
an axiom in a PTM-$\omega$-consistent theory may be
characterized by asymptotic proofs of
a PTM prover.

For example, 
a PTM-$\omega$-{\it inconsistent} theory $T$,
which is a consistent PT-extension of PA,
should include an axiom outside PA
that only a super-polynomial-time Turing machine
can prove asymptotically over PA, 
assuming that PA is PTM-$\omega$-consistent and  
deduction in $T$ can be made asymptotically by PTM
(Remark \ref{rem:7} of Definition \ref{PTM-omega-con}
in Section \ref{sec:ptm-omega-con}).
Let $T$ be a theory in which an axiom, $X$, 
outside PA is added to PA.
Although $X$ cannot be proven in PA,
it can be {\it asymptotically} proven over PA
if it is true, 
since any true $\Delta_1$-sentence can be proven
in PA.
Therefore, a resource {\it unbounded} Turing machine
can always produce an asymptotic proof of $X$ over PA,
but a resource {\it bounded} (e.g., 
polynomial-time) Turing machine may produce
no asymptotic proof of $X$ over PA.
Hence, axiom $X$ (and theory $T$) can be characterized
by the computational complexity
of a prover for producing an asymptotic proof of $X$.
If $T$ is 
PTM-$\omega$-{\it inconsistent},
the computational complexity
of a prover for producing an asymptotic proof of $X$
should be super-polynomial-time,
under the above-mentioned assumption
(Remark \ref{rem:7} of Definition \ref{PTM-omega-con}). 
Thus, PTM-$\omega$-consistency
bridges a formal proof and
prover's (asymptotic) computational capability
through asymptotic proofs.


In accordance with the two manners of characterization 
of PTM-$\omega$-consistency,
our main result that    
$\ov{\mbox{P$\not=$NP}}$ cannot be proven in a PTM-$\omega$-consistent
theory (Theorem \ref{mainthm5-dec}) 
suggests two avenues towards negative results:
\begin{itemize}
\item
To prove $\ov{\mbox{P$\not=$NP}}$
requires a PTM-$\omega$-{\it inconsistent} theory,
which should include an axiom that only a super-polynomial-time 
machine can prove asymptotically over PA
(or implies a super-polynomial-time computational upper bound),
under the assumption.
This is a kind of generalization 
of the result of ``Natural Proofs'' by Razborov and Rudich 
\cite{RazRud94}.
See Section \ref{sec:related}.
\item
To prove $\ov{\mbox{P$\not=$NP}}$ by a relativizable proof,
i.e., 
to prove $\ov{\mbox{P$^A\not=$NP$^A$}}$ with oracle $A$ 
requires a PTM$^A$-$\omega$-{\it inconsistent} theory
(Proposition \ref{prop:super-C}). 
Therefore, if there exists a relativizable proof of 
$\ov{\mbox{P$\not=$NP}}$, which implies  
a proof of $\ov{\mbox{P$^A\not=$NP$^A$}}$ for any oracle $A$,
it will require an $\omega$-{\it inconsistent} theory, 
since a PTM$^A$-$\omega$-{\it inconsistent} theory
with any oracle $A$ is equivalent to a 
$\omega$-{\it inconsistent} theory.   
This suggests the result that
no relativizable proof can prove ``P$\not=$NP''
in PA (or any $\omega$-consistent theory),
which was shown by Baker, Gill and Solovay \cite{BakGilSol75}.
See the remark of Theorem \ref{mainthm5-dec}.
\end{itemize} 

Therefore, 
our result, Theorem \ref{mainthm5-dec}
(and its generalization, Proposition \ref{prop:super-C}),  
gives a unified view to the existing two 
major negative results on proving P$\not=$NP,
Natural Proofs and relativizable proofs,
through the above-mentioned two manners of characterization  
of PTM-$\omega$-consistency.

PTM-$\omega$-consistency has also the following properties:
\begin{itemize}
\item
\PTM-$\omega$-consistency 
and $\omega$-consistency do not imply each other.
(Remark \ref{rem:4} of Definition \ref{PTM-omega-con})
\item
Although the PTM-$\omega$-consistency of PA seems to  
be as natural as the $\omega$-consistency of PA,
no PTM-$\omega$-consistent theory $T$, which is a consistent 
PT-extension of PA, 
can prove the PTM-$\omega$-consistency of PA.
(Theorem \ref{thm:ptm-omega-con-formal}
and Remark \ref{rem:6} of Definition \ref{PTM-omega-con}) 
\end{itemize}

%

\subsection{Related Works}\label{sec:related}


\subsubsection{Self-defeating results}

Our result is considered to be a kind of generalization of
or a close relation to 
the previously known self-defeating results as follows:
\begin{itemize}
\item
Our result that \PTM-$\omega$-consistent theory cannot prove 
$\ov{\mbox{P$\not=$NP}}$ 
(Theorem \ref{mainthm5-dec}) 
implies a {\it self-defeating} property such that
to prove a super-polynomial-time lower bound like P$\not=$NP
requires a \PTM-$\omega$-{\it inconsistent} theory,
which should include an axiom that only a super-polynomial-time machine
can prove asymptotically over PA
(or implies a super-polynomial-time computational upper bound)
under the assumption described in the previous section.

``Natural Proofs'' 
by Razborov and Rudich \cite{RazRud94}
showed that to prove a computational lower bound
(e.g., a super-polynomial-time lower bound like P$\not=$NP) 
by a class of techniques called ``Natural Proofs'' implies a comparable level of 
computational upper bound (e.g., a super-polynomial-time algorithm 
to break a typical cryptographic primitive, 
a pseudo-random generator).
In other words, to prove P$\not=$NP by a ``Natural Proof''
requires an additional axiom $X$ that implies a
super-polynomial-time (e.g., sub-exponential-time) algorithm 
to break a pseudo-random generator and that can be
proven asymptotically only by a super-polynomial-time machine, 
since no polynomial-time machine
is considered to be able to {\it asymptotically} prove  
an upper bound property of a super-polynomial-time machine.
Therefore, to prove P$\not=$NP by a specific type of proof called ``Natural Proof''
requires a specific type of PTM-$\omega$-{\it inconsistent} theory,
which is PA + $X$. That is, the negative result regarding
``Natural Proofs'' is considered to be a special case of our result, 
Theorem \ref{mainthm5-dec}.

\item
Our results 
imply another {\it self-defeating} property such that
\PTM-$\omega$-consistent theory $S$ over $T$ cannot prove 
the independence 
of P vs NP from $T$ by proving PTM-$\omega$-consistency
of $T$ for a $\Delta_2^P$-formula 
(Theorem \ref{thm:ptm-omega-con-formal}). 
In other words,
to prove the independence of P vs NP from $T$ 
through Theorem \ref{mainthm5-dec}
(i.e., to prove $T \not\vdash \ov{\mbox{P$\not=$NP}}$ 
by proving PTM-$\omega$-consistency of $T$
and to prove $T \not\vdash \ \neg\ov{\mbox{P$\not=$NP}}$ by some way)
requires \PTM-$\omega$-{\it inconsistent} theory over $T$, 
or implies a super-polynomial-time upper bound
under the above-mentioned assumption.

Ben-David and Halevi \cite{BenHal92}
and Kurz, O'Donnell and Royer \cite{KurODnRoy87}
showed that to prove the independence of P vs NP from PA
using any currently known mathematical paradigm implies 
a comparable level of computational upper bound,  
an extremely-close-to-polynomial time algorithm to solve NP-complete 
problems.  
In other words, to prove the independence of P vs NP from PA
using any currently known mathematical paradigm
requires an additional axiom $Y$ that implies 
an extremely-close-to-polynomial time (but still super-polynomial-time)
algorithm to solve NP-complete problems and that can be
proven asymptotically only by a super-polynomial-time machine. 
Therefore, to prove the independence of P vs NP from PA 
by a specific type of proof using currently known mathematical paradigms
requires a specific type of PTM-$\omega$-{\it inconsistent} theory,
which is PA + $Y$. That is, the negative result by Ben-David et.al. 
is considered to be a special case of our result, 
Theorem \ref{thm:ptm-omega-con-formal}, provided that
Hypothesis 1 in Section \ref{sec:our-result-implication} is
true and $\PA \not\vdash \ov{\mbox{P$\not=$NP}}$
implies Hypothesis 2.

\end{itemize}

\subsubsection{Relativizable proofs}

Our result that \PTM-$\omega$-consistent theory cannot prove 
$\ov{\mbox{P$\not=$NP}}$ 
(Theorem \ref{mainthm5-dec}) 
suggests the result by 
Baker, Gill and Solovay \cite{BakGilSol75},
who showed that there is no relativizable proof
of ``P$\not=$NP'', and
the result by Hartmanis and Hopcroft \cite{HarHop76,Hartmanis78},
who showed that
for any reasonable theory $T$ we can effectively construct 
a TM $M$ such that 
relative to oracle $L(M)$, ``P$\not=$NP'' cannot be 
proven in $T$. 
(See the remark of Theorem \ref{mainthm5-dec}.)
 
Our result might be related to the result by
da Costa and Doria \cite{CosDor03}, but the relationship between their result
and ours is unclear for us.

\subsubsection{Mathematical logic approaches}\label{sec:logic-app}

The results of this paper are constructed on the theory and
techniques of mathematical logic, especially proof theory.
Several mathematical logic approaches to solve the P vs NP problem 
have been investigated such as  
bounded arithmetic \cite{Buss86,Krajicek95}, 
propositional proof length \cite{BeaPit98,Krajicek95,Razborov02} and
descriptive complexity \cite{Fagin74}.

Bounded arithmetic characterizes an analogous notion 
of PH (polynomial hierarchy of computational complexity),
which is a hierarchy of weak arithmetic theories, so-called 
bounded arithmetic classes,
wherein only bounded quantifiers are allowed.
The target of the bounded arithmetic approach 
is to separate one class from another in bounded arithmetic,
which may imply a separation of one class from another in PH
(i.e., typically P$\not=$NP).

The proof length of propositional logic can characterize 
the NP vs co-NP problem, since TAUT, the set of propositional
tautologies, is co-NP complete.
Therefore, the main target of this approach is to prove NP$\not=$co-NP 
by showing a super-polynomial length lower bound of 
a formal propositional proof of TAUT. 
In this approach,
the lower bounds of the proof lengths and limitation of provability 
of some specific propositional proof systems (e.g., resolution, Frege system and extended
Frege system)
have been investigated.
  
The descriptive complexity characterizes NP by 
a class of problems definable by 
existential second order formulas and P by a class of problems definable
in first order logic with an operator. 
The target of this approach is
to separate P and NP using these logical characterizations.

This paper characterizes the concepts of P and P$\not=$NP etc.,
by formulas in Peano Arithmetic (PA).
A novel viewpoint of our approach is to 
introduce the concept of
an asymptotic proof produced by a polynomial-time Turing machine
as a prover,
to characterize a property of a formal theory, 
PTM-$\omega$-consistency, by using this concept, 
and to show that no PTM-$\omega$-consistent theory
can prove a super-polynomial-time computational lower bound
such as P$\not=$NP.
 
To the best of our knowledge, no existing approach has studied 
computational lower bounds from such a viewpoint. 
\footnote{
A prover is modeled as a Turing machine
in the interactive proof system theory, and 
the computational complexity of a prover has been 
investigated \cite{Goldreich99,Goldreich01}.
However, no proof system with a polynomial-time Turing machine 
prover that produces an asymptotic proof of  
a computational lower bound has been studied.}


\subsubsection{Proof systems}

In order to define the PTM-$\omega$-consistency,
this paper introduces a new concept of proof systems,
{\it asymptotic proofs} and  
{\it polynomial-time proofs} 
where the computational complexity of (prover's) proving 
a set of statements asymptotically is bounded by polynomial-time.
In the conventional proof theory,
the properties and capability of a proof system 
(e.g., consistency, completeness, incompleteness etc.)
are of prime interest, but
the required properties and 
capability of the prover are not considered
(i.e., no explicit restriction nor condition is placed on the prover).

Note that the bounded arithmetic approach 
seems to follow this conventional paradigm and bounds 
the capability of the proof system (axioms and rule of inferences)
to meet the capability of resource bounded computational classes.
That is, the prover is still thought to exceed the scope of the approach. 
  
In this paper, the computational complexity of a prover
is investigated through the concept of  
an asymptotic proof system.
An asymptotic proof is a set of an infinite number of
formal proofs, and a resource bounded 
(e.g., polynomial-time bounded or exponential-time bounded etc.) 
prover asymptotically produces an asymptotic proof
of a set of infinitely many formal statements.   

This paper then introduces a new concept,
PTM-$\omega$-consistency,
which is a property of a conventional proof system, 
but is defined and characterized by the concept of asymptotic proofs with
a polynomial-time bounded prover.
PTM-$\omega$-consistency plays a key role in our results 
(for example, see Section \ref{sec:implication}).

\subsubsection{Undecidability}

Although the computational complexity theory is a resource bounded 
version of the recursion theory,
to the best of our knowledge,
little research has been made on 
resource bounded undecidability of formal statements.

This paper introduces a resource bounded 
(asymptotic) decision system, which corresponds to a resource
bounded (asymptotic) proof system, 
and presents the incompleteness theorems
(Sections \ref{sec:pt-dec} and \ref{sec:rbicthm-dec}).
Using the incompleteness theorem of
resource bounded 
(asymptotic) decision systems yields the 
resource bounded unprovability
of $\ov{\mbox{P$\not=$NP}}$
(Section \ref{sec:unprove-dec}).

\subsection{Key Ideas of Our Results}

In order to obtain our main result
(Theorem \ref{mainthm5-dec}:
$\ov{\mbox{P$\not=$NP}}$ cannot be
proven in a \PTM-$\omega$-consistent theory),
this paper introduces the concept of {\it polynomial-time 
decision systems} (Section \ref{sec:pt-dec}).
In a proof system, we usually consider only one side,
a proof of a true statement. In a decision system, however,
we have to consider two sides, CA (correctly accept: 
accept of a true statement) and CR (correctly reject:
reject of a false statement).
CD (correctly decide) means CA or CR.

The key idea to prove Theorem \ref{mainthm5-dec} 
is a {\it polynomial-time decision
version of incompleteness theorems}.
Informally speaking, we introduce a special sentence, $\rho_{e}^A(\xxx)$,
(an analogue of the so-called G{\"o}del sentence) 
like ``this statement, $\rho_{e}^A(\xxx)$, cannot be correctly accepted 
by a polynomial-time Turing machine (PTM) encoded by $e$.''
(Hereafter, ``a PTM encoded by $e$'' is called ``PTM $e$'')  
If $\rho_{e}^A(\xxx)$ can be correctly accepted by PTM $e$, 
it contradicts the definition of $\rho_{e}^A(\xxx)$.
It follows that $\rho_{e}^A(\xxx)$ cannot be correctly accepted by PTM 
$e$.
We also define another sentence, $\rho_{e}^R(\xxx)$,
which cannot be correctly rejected by PTM $e$.
(First incompleteness theorems of polynomial-time decisions:
Theorems \ref{rbic-1-thm-dec-ca} and \ref{rbic-1-thm-dec-cr}).
Based on these theorems, we show that, for any formula set
$\{\psi(\xxx) \mid x \in \N\}$
(e.g., formula set on the satisfiability of 3CNF),
for any PTM $e$, there exists another PTM $e^*$
such that PTM $e$, on input $x \in \N$, cannot asymptotically prove
that PTM $e^*$ cannot correctly decide $\psi(\xxx)$
(Second incompleteness theorem of polynomial-time decisions: 
Theorem \ref{rbic-2nd-dec}).
By using Theorem \ref{rbic-2nd-dec}, we show that 
no PTM can prove $\ov{\mbox{P$\not=$NP}}$ 
asymptotically (Lemma \ref{mainthm1-dec}).

This paper then introduces the {\it \PTM-$\omega$-consistency} of $T$,
which is a PTM version of $\omega$-consistency
and plays a key role in our result
(for its semantics and rationale, see Section \ref{sec:implication}
and the remarks of 
Definition \ref{PTM-omega-con}).
Combining Lemma \ref{lemma-sigma}
and 
\PTM-$\omega$-consistency of $T$,
we can show that $\ov{\mbox{P$\not=$NP}}$ cannot be
proven in \PTM-$\omega$-consistent theory $T$
(Theorem \ref{mainthm5-dec}).

This paper also introduces 
the notion of {\it polynomial-time proof systems}, and 
obtains a {\it polynomial-time proof
version of incompleteness theorems}
(Sections \ref{sec:formalization} and \ref{sec:rbicthm}).
Informally speaking,
we introduce a special sentence, $\rho_{e,T}$,
like ``this statement, $\rho_{e,T}$, cannot be proven 
by a polynomial-time Turing machine (PTM) $e$ in theory $T$.''
If $\rho_{e,T}$ can be proven by PTM $e$ in $T$, it contradicts the
definition of $\rho_{e,T}$, assuming that $T$ is consistent.
It follows that $\rho_{e,T}$ cannot be proven by PTM $e$ in $T$,
although another PTM can prove it 
(First incompleteness theorem of polynomial-time proofs: 
Theorem \ref{rbic-1}).
Based on this theorem, we show that, for any formula set
$\{\psi(\xxx) \mid x \in \N\}$
for any PTM $e$, there exists another PTM $e^*$
such that PTM $e$, on input $x \in \N$, cannot asymptotically prove
that PTM $e^*$ cannot prove $\psi(\xxx)$
(Second incompleteness theorem of polynomial-time proofs: 
Theorem \ref{rbic-2nd}).

By using Theorem \ref{rbic-2nd}
and \PTM-$\omega$-consistency, we show that 
the \PTM-$\omega$-consistency of $T$
cannot be proven in a PTM-$\omega$-consistent theory $S$, where $S$ is 
a consistent PT-extension of $T$ (Theorem \ref{thm:ptm-omega-con-formal}).
(In fact, we have not shown the existence of 
a consistent and PTM-$\omega$-consistent 
PT-extension of PA; therefore, we have not shown 
the unprovability of $\ov{\mbox{P$\not=$NP}}$ in PA.)

Finally, based on Theorem \ref{mainthm5-dec},
the unprovability of the security of
the computational cryptography 
is obtained (Theorems \ref{imp-crypto} and \ref{imp-trapdoor})
in a setting that provers as well as adversaries are modeled as PTMs 
and only PTM-$\omega$-consistent theory is allowed to
prove the security.





\section{Polynomial-Time Proofs}\label{sec:formalization}

This paper follows the standard notions and definitions 
of computational complexity theory (e.g., definitions of P and NP)
and mathematical logic (e.g., definition of a formal proof 
in Peano Arithmetic).  
See \cite{Sipser97} for such standard notions and definitions of computational 
complexity theory and see \cite{Barwise77,Enderton01,Shoenfield67} 
for the standard notions and definitions of 
mathematical logic. 

The central interest of this paper is the difficulty of proving 
the lower bound of computational problems by resource bounded Turing machines.
For this purpose, first, we need to formalize the notion of a formal proof 
produced by a resource bounded Turing machine.
This section introduces our formalization of a proof 
produced by a polynomial-time Turing machine (polynomial-time proof: PTP)
in a theory that is an extension of Peano Arithmetic 
(hereafter Peano Arithmetic is abbreviated to PA).


\vspace{10pt}
\noindent
{\bf Remark:} \\
This paper is based on the standard notion of 
formal proofs in first order logic 
\cite{Barwise77,Enderton01,Shoenfield67}.
There are, however, many possible ways of 
formalizing such formal proofs, especially 
with regard to the style of formalizing the deduction system; 
alternatives to the selection of 
logical axioms and rules of inference.
There are two typical styles:
one is the Hilbert-style, which has several logical axioms 
and a few rules of inference, and the other is
the Gentzen-style, which has just one logical axiom and
several rules of inference. 
However, the results in this paper are
not affected by the way of formalizing 
the deduction system,
and almost all descriptions in this paper 
are independent of the style of formal deduction system
adopted.
When we need to make an explicit description on a specific deduction system, 
this paper adopts the Hilbert style, which has
two rules of inference; Modus Ponens and 
Generalization rules. 

 

\subsection{Notations}\label{sec:notation}
Let $\N$ be the set of natural numbers including 0. 

When $w$ is a bit string, $|w|$ denotes the bit length of $w$.

When $w \in \N$, $[w]$ denotes the binary
representation of $w$, i.e., bit string ${w_{k-1} w_{k-2} \cdots w_0}$
with $w = w_{k-1}2^{k-1}+w_{k-2}2^{k-2}+ \cdots w_0$,
$k = \lf\log_2{w}\rf+1$ ($w>0$), and $w_i \in \{0,1\}$ for 
$i=0,1,2,\ldots,k-1$. 
When $w=0$, $[w], i.e., [0]$, denotes the binary representation, 0.
When $w \in \N$, 
$|w|$ denotes the bit length of $[w]$. 

PA has a constant symbol, $\0$, intended
to denote the number 0, and 
has three function symbols, $\Se$, $+, \cdot$, where
$\Se$ is 
a one-place function symbol intended to denote 
the successor function $S: \N \ra \N,$ i.e., 
the function for which $S(n) = n+1$, and 
symbols $+$ and $\cdot$ are two-place function symbols of
addition and multiplication, respectively.
PA also has symbols of predicate logic 
such as logical symbols ($\neg,$ $\land,$ $\lor,$ $\ra$, 
$\forall,$ $\exists$, etc.),
relation symbols ($=$, $<$, etc.), and
variable symbols ($\xxx,\yyy, \zzz$, etc.).
 
The numerals of PA are denoted by boldfaced number symbols such as 
$\1$, $\2$, $\3$, $\ldots$, for $\Se\0$, $\Se\Se\0$, $\Se\Se\Se\0$, $\ldots$ .
Boldfaced alphabet symbols such as $\xxx$, $\yyy$, $\xxx_1$, $\xxx_i$, etc., 
are also used for variables in theory $T$.

Throughout this paper, we assume that
the numeral, \(\overbrace{\Se\Se\cdots \cdot\Se}^{\mbox{$n$ times}}
\mbox{\bm $0$}\), of natural number $n$ 
is expressed by the following binary form
in a theory including PA:
$$\nnn_0 + \nnn_1\cdot\Se\Se\mbox{\bm $0$} + \cdots +
\nnn_{k-1}\cdot 
\overbrace{\Se\Se\mbox{\bm $0$}\cdot\Se\Se\mbox{\bm $0$} 
\cdots \cdot\Se\Se\mbox{\bm $0$}}^{\mbox{$k-1$ times}},$$
where $n = n_0 + n_1\cdot 2 + \cdots n_{k-1}\cdot 2^{k-1}$,
$n_i \in \{0,1\}$, 
$\nnn_i = \mbox{\bm $0$}$ if $n_i = 0$ and 
$\nnn_i = \1 (= \Se\mbox{\bm $0$}$) if $n_i = 1$  ($i=0,1,\ldots,k-1$).
Here we denote this expression of the numeral of natural number $n$
by $\Se^n\mbox{\bm $0$}$ or just $\nnn$. 
Similarly, if alphabet $a$ denotes a natural number,
$\aaa$ denotes $\Se^a\mbox{\bm $0$}$.

We will now introduce two additional function symbols in PA.
(A function symbol, $f$, of a primitive recursive function 
is considered to be implicitly included in PA, i.e., 
$f$ can be identified with a formula, $\rho_f$, in PA, since
$f$ is representable by a $\Delta_1$formula, $\rho_f$,
in PA and 
$ \PA \vdash \forall \xxx_1 \cdots \forall \xxx_k \exists !\yyy \ 
\rho_f(\xxx_1,\ldots,\xxx_k,\yyy)$.
See Subsection \ref{convention}.)

Here it is worth noting that
these function symbols, which correspond to primitive recursive
functions, are introduced for improving 
the readability of formulas, not 
for increasing proving ability.
Therefore, in this paper we assume that no G{\"o}del number
for a function symbol of a primitive recursive
function (except $\Se$, $+$ and $\cdot$ ) 
is provided (see the next section for
G{\"o}del numbers). The G{\"o}del number
of a formula including such a function symbol
is calculated on the formula without using the function 
symbol, i.e., the formula in which only function symbols in \PA \
are employed.
This assumption is applied for any theory $T$
which is a PT-extension of \PA \  throughout this paper.
Hence the G{\"o}del number of a formula in $T$
is uniquely defined even if 
some function symbols of primitive recursive functions
are employed in the formula.

If $\xxx$, $\yyy$ and $\zzz$ are numerals in PA,
$\xxx \dot{-} \yyy$ denotes a two-place function: 
$(\xxx,\yyy) \mapsto \zzz$, such that
$\zzz = \Se^{\max\{x - y, 0\}}\0$, $\xxx=\Se^x\0$, $\yyy=\Se^y\0$, 
$x \in \N$ and $y \in \N$.

If $\xxx$, $\yyy$ and $\zzz$ are numerals in PA,
$\xxx^{\yyy}$ denotes a two-place function: 
$(\xxx,\yyy) \mapsto \zzz$, such that
$\zzz = \Se^{x^y}\0$, $\xxx=\Se^x\0$, $\yyy=\Se^y\0$, 
$x \in \N$ and $y \in \N$.

By using these function symbols (notations),
the notation of a numeral, $\nnn$, is defined by
$$\nnn_0 + \nnn_1\cdot\2^{\1} + \cdots +
\nnn_{k-1}\cdot\2^{\kkk \dot{-} \1} 
.$$

When $\nnn$ is a numeral,
$|\nnn|$ denotes the numeral of 
$|n|$. The function symbol, $| \cdot |$,
is justified by the first claim in the proof of
Theorem \ref{representability}.

Some other notations are:
\begin{itemize}
\item
$\psi \lra \phi$ denotes
$$(\psi \ra \phi) \land (\psi \la \phi),$$ 
\item
$\exists ! \yyy \ \phi(\yyy)$ denotes
$$\exists \yyy \ \phi(\yyy) \ \land \ \forall \yyy_1 \forall \yyy_2
(\phi(\yyy_1) \land \phi(\yyy_2) \ra \yyy_1=\yyy_2),$$
which means $\yyy$ {\it uniquely} exists to satisfy $\phi(\yyy)$. 
\item
$\forall \xxx \geq \nnn \ \phi(\xxx)$ denotes
$$ \forall \xxx \ (\xxx \geq \nnn \ \ra \ \phi(\xxx)).$$
\item
$\exists \xxx \geq \nnn \ \phi(\xxx)$ denotes
$$ \exists \xxx \ (\xxx \geq \nnn \ \land \ \phi(\xxx)).$$
\item
Some basic notations in proof theory \cite{Barwise77}:
$$ T \vdash \phi,$$
which informally denotes 
``the truth of formula $\phi$ is provable in theory $T$''.
$$ \Prv_T(\lc \phi \rc),$$
which denotes a formula in $T$, which informally means 
``there exists a proof for the truth of formula $\phi$ in theory $T$''.
Here $\lc \phi \rc$ denotes $\Se^{{\#}\phi}\0$.
\item
$T$ is {\it inconsistent} if 
there exists a formula $\phi$ in $T$ such that
$T \vdash \phi$ and $T \vdash \neg\phi$,
which is also denoted by $T \vdash \bot$.
$T$ is {\it consistent} if 
there exists no such formula $\phi$ in $T$. 
Here, $\bot \ \equiv \ \neg\forall \xxx (\xxx=\xxx)$. 
\item
$T$ is {\it $\omega$-inconsistent} if 
there exists a formula $\phi(\xxx)$ in $T$ such that
$$T \vdash \exists \xxx \ \phi(\xxx), \ \ \  \mbox{and}$$ 
$$\forall a \in \N \ \ T \vdash \neg\phi(\aaa).$$
$T$ is {\it $\omega$-consistent} if 
there exists no such formula $\phi(\xxx)$ in $T$. 
(If $T$ is $\omega$-consistent, $T$ is also consistent.
The reverse is not always true.)
\item
$\NNN$ is the standard model of natural numbers.
When $\phi$ is a formula in PA,
$$\NNN \models \phi$$ 
denotes that $\phi$ is true in $\NNN$.
\end{itemize}

\subsection{G{\"o}del numbers}\label{G-numbers}

There are many ways of defining
the G{\"o}del numbers, and the way introduced in this section 
differs from those described in G{\"o}del's original paper
and textbooks (e.g., \cite{Enderton01}), since
in this paper we require a polynomial time algorithm
to make unique encoding and decoding.
We basically follow the approach introduced by \cite{Buss86}.
(We can also adopt a coding method employed in actual current
computer systems.)

Let ${\#}\phi$ be a G{\"o}del number of $\phi$. 
First, we define G{\"o}del numbers of basic symbols
in $L$ as follows: (for example) 
${\#}\forall$ is 0, 
${\#}($  is 1,
${\#}\mbox{\bm $0$}$ is 2, 
${\#})$ is 3, 
${\#}\Se$ is 4,
${\#}\neg$ is 5,
${\#}<$ is 6,
${\#}\ra$ is 7,
${\#}+$ is 8,
${\#}=$ is 9,
${\#}\cdot$ is 10,
${\#},$ is 11,
${\#}\aaa_1$ is 20,
${\#}\xxx_1$ is 22,
${\#}\aaa_2$ is 24,
${\#}\xxx_2$ is 26,
etc.
   
We then use the following method to obtain the 
G{\"o}del number of a sequence of natural numbers,
$a_1, a_2, \ldots, a_k$ \cite{Buss86}:
\begin{enumerate}
\item Represent $a_i$ by the binary representation
with the least significant bit on the right, as is traditional.
Then, $a_1, a_2, \ldots, a_k$ can be
represented by the sequence of three symbols `0', `1' and `,'.
\item Reverse the order of the sequence of `0', `1' and `,',
and replace `0' by `10', `1' by `11' and `,' by `01'.
We then obtain a sequence of `0' and `1'.
\item The natural number whose binary representation is 
this sequence is the G{\"o}del number of the number sequence,
$a_1, a_2, \ldots, a_k$.
It is denoted by $\langle a_1, a_2, \ldots, a_k \rangle$.
\end{enumerate}

For example,
$\langle 3, 4, 5 \rangle$ is a natural number, 
whose binary representation is 11101101101011011111, 
because
$3,4,5$ is binary-represented along with commas by
$11,100,101$ and is encoded to a binary sequence,
11101101101011011111.

When $\phi$ is an expression in language $L$, 
it is a sequence of symbols, $s_0 s_1 \cdots s_k$, of $L$.
We then define the G{\"o}del number, ${\#}\phi$, of $\phi$ as follows:
$${\#}\phi \equiv
\langle {\#}s_0, {\#}s_1, \ldots, {\#}s_k \rangle.$$
For example, when 
$\phi$ is $\neg ( \forall \xxx_1 (\xxx_1 < \Se\mbox{\bm $0$}))$,

$${\#}\phi 
\equiv 
\langle {\#}\neg, {\#}(, {\#}\forall, {\#}\xxx_1, {\#}(, {\#}\xxx_1, {\#}<,
{\#}\Se, {\#}\mbox{\bm $0$}, {\#}),{\#}) \rangle $$
$$ \equiv \langle 5, 1, 0, 13, 1, 13, 6, 4, 2, 3, 3 \rangle.$$

Remember here that numeral $\nnn$ ($\equiv \Se^n\0$) denotes 
the binary form, i.e.,
$$\nnn_0 + \nnn_1\cdot\Se\Se\mbox{\bm $0$} + \cdots +
\nnn_{k-1}\cdot 
\overbrace{\Se\Se\mbox{\bm $0$}\cdot\Se\Se\mbox{\bm $0$} 
\cdots \cdot\Se\Se\mbox{\bm $0$}}^{\mbox{$k-1$ times}}.$$
Hence,
$|{\#}\nnn|$ (i.e., $|{\#}\Se^n\mbox{\bm $0$}|$) is
of the order of $\log_2^2{n}$.

Here also remember that we provide no G{\"o}del numbers of  
additionally introduced function symbols of primitive
recursive functions 
such as ${\2}^{\nnn}$.
That is, the G{\"o}del number
of a formula including such a function symbol
is calculated on the formula with only function symbols in \PA.
For example, 
$${\#}{\2}^{\nnn} \equiv 
{\#}\overbrace{\Se\Se\mbox{\bm $0$}\cdot\Se\Se\mbox{\bm $0$} 
\cdots \cdot\Se\Se\mbox{\bm $0$}}^{\mbox{$n$ times}}
\equiv \langle {\#}\Se, {\#}\Se, {\#}\0, {\#}\cdot,  
\ldots, {\#}\0 \rangle.
$$
Therefore, $|{\#}{\2}^{\nnn}| = O(n)$.\footnote{ 
If we have the G{\"o}del number of the 
function symbol EXP such as EXP$(\xxx, \yyy)={\xxx}^{\yyy}$,
then $|{\#}{\2}^{\nnn}| = $
$|{\#}{\rm EXP}(\2,\nnn) |$  
$ = |{\#}{\rm EXP}(\Se\Se\0, 
\nnn_0 + \nnn_1\cdot\Se\Se\mbox{\bm $0$} + \cdots +
\nnn_{k-1}\cdot\Se\Se\mbox{\bm $0$}\cdots\Se\Se\mbox{\bm $0$})|$
$= | \langle {\#}{\rm EXP}, {\#}(, {\#}\Se, \ldots, {\#}\0 \rangle|$ 
$= O(\log_2{n})$, since $k = O(\log_2{n})$.}

We then introduce a concatenation operation $||$ of
two G{\"o}del numbers, ${\#}\phi$ and ${\#}\psi$,
where 
${\#}\phi \equiv \langle {\#}s_0, {\#}s_1, \ldots, {\#}s_k \rangle$ 
and 
${\#}\psi \equiv \langle {\#}t_0, {\#}t_1, \ldots, {\#}t_l \rangle$.
${\#}\phi || {\#}\psi$ is defined by
$$ \langle {\#}s_0, {\#}s_1, \ldots, {\#}s_k, 
{\#}t_0, {\#}t_1, \ldots, {\#}t_l \rangle.$$




\subsection{Polynomial-Time Extension of PA}\label{pt-ext}

Let formula {\bf Axiom}$_{T}(\nnn)$ be true if and only if 
$n$ is the G{\"o}del number of an axiom of $T$.
If the truth of {\bf Axiom}$_{T}(\nnn)$ can be correctly decided 
by a polynomial-time algorithm in $|n|$, 
on input $n$, 
we say that $T$ is polynomial-time axiomizable.    
If $T$ is an extension of $T_0$ and polynomial-time axiomizable,
then we say that $T$ is a polynomial-time (PT) extension of
$T_0$.

Using the notations introduced in Section \ref{pt-dec},
a polynomial-time axiomizable theory, $T$, is defined as follows:
Let ${\cal AX} \equiv \{ {\bf Axiom}_{T}(\nnn) \mid n \in \N  \}$
and $\Size_{\cal AX}(n) = |n|$.
There exists $e \in \N$ such that for all $n \in \N$ 
$$
\PTM_{e}^{\cal AX}(n) \dec \ {\bf Axiom}_{T}(\nnn)
\ \ \ \lor \ \ \ 
\PTM_{e}^{\cal AX}(n) \dec \ \neg{\bf Axiom}_{T}(\nnn).
$$



\subsection{Representability Theorem in Mathematical Logic}\label{convention}

This section introduces the representability theorem
in the conventional mathematical logic 
\cite{Barwise77,Enderton01,Shoenfield67}.
This theorem plays an important role in
many situations as well as in
constructing the polynomial-time version of the 
representability theorem (Theorem \ref{representability}), 
which is essential to formalize 
the execution of PTM in PA. 

In this paper, we use the standard notions and notations of mathematical
logic, such as $T \vdash \phi$ (informally, a sentence $\phi$ is 
provable in theory $T$), with no introduction
(see \cite{Barwise77,Enderton01,Shoenfield67}).

\begin{definition}
\begin{enumerate}
\item
Let $R$ be a $k$-ary relation on $\N$; i.e., $R \subseteq \N^k$.
A formula $\rho_R(x_1,\ldots,x_k)$ (in which only $x_1,\ldots,x_k$ occur free)
will be said to {\it represent} a relation $R$ in theory $T$

if and only if 
for every $a_1,\ldots,a_k$ in $\N^k$
\begin{eqnarray*}
& & 
(a_1,\ldots,a_k) \in R \ \ \Ra \ \ 
T \vdash \rho_R(\aaa_1,\ldots,\aaa_k),
\\
& & 
(a_1,\ldots,a_k) \not\in R \ \ \Ra \ \ 
T \vdash \neg\rho_R(\aaa_1,\ldots,\aaa_k).
\end{eqnarray*}

A relation $R$ is said to be {\it representable} 
in $T$ if and only if 
there exists some formula $\rho_R$ 
that represents $R$ in $T$. 

\item
Let $f$ be a $k$-place function on the natural numbers.
A formula $\rho_f(x_1,\ldots,x_k,y)$ (in which only $x_1,\ldots,x_k,y$ occur free)
will be said to {\it functionally represent} $f$ in theory $T$
if and only if
for every $a_1,\ldots,a_k$ in $\N^k$
$$T \vdash \forall y ( \rho_f(\aaa_1,\ldots,\aaa_k,\yyy)
\lra \yyy = \Se^{f(a_1,\ldots,a_k)}\0 ).$$

A function $f$ is said to be {\it functionally representable} 
in $T$ if and only if 
there exists some formula $\rho$ that 
functionally represents $f$ in $T$. 
\end{enumerate}
\end{definition}

\begin{proposition}\label{c-representability}
(Representability Theorem) 
For any {\it primitive recursive} relation on $\N^k$, $R$, and 
any {\it primitive recursive} function on $\N^k$, $f$,
there exist formulas, 
$\rho_R(\xxx_1,\ldots,\xxx_k)$
and $\rho_f(\xxx_1,\ldots,\xxx_k,\yyy)$,
such that:
\begin{itemize}
\item
$\rho_R(\xxx_1,\ldots,\xxx_k)$
represents $R$, 
and $\rho_f(\xxx_1,\ldots,\xxx_k,\yyy)$ 
functionally represents $f$ in PA.
\item
$\rho_R(\xxx_1,\ldots,\xxx_k)$
and $\rho_f(\xxx_1,\ldots,\xxx_k,\yyy)$
are $\Delta_1$ in PA.
\item
$$ \PA \vdash \forall \xxx_1 \cdots \forall \xxx_k \exists !\yyy \ 
\rho_f(\xxx_1,\ldots,\xxx_k,\yyy).$$
\end{itemize}

\end{proposition}


\begin{proposition}\label{extension}
Let only $\xxx_1,\ldots,\xxx_k,\yyy $ occur free
in formula $\phi(\xxx_1,\ldots,\xxx_k,\yyy)$. \\
If 
$ T \vdash \forall \xxx_1 \cdots \forall \xxx_k \exists !\yyy \ 
\phi(\xxx_1,\ldots,\xxx_k,\yyy),$
then
theory $T'$
$$\equiv \  T \cup 
\{ \forall \xxx_1 \cdots \forall \xxx_k \forall \yyy \ 
(
\phi(\xxx_1,\ldots,\xxx_k,\yyy)
\lra f(\xxx_1 \cdots \xxx_k)=\yyy 
)
\}
$$
is a conservative extension of $T$.
\end{proposition}

 From Proposition \ref{extension},
we can identify theory $T'$, which has function symbol $f$,
with theory $T$, in the light of provability
and representability.
In other words, we can consider that function symbol $f$ 
(
and the corresponding axiom, 
$\forall \xxx_1 \cdots \forall \xxx_k \forall \yyy$ 
$( \ \phi(\xxx_1,\ldots,\xxx_k,\yyy)$
$\lra f(\xxx_1 \cdots \xxx_k)=\yyy )$ 
)
is implicitly included in theory $T$.  
Therefore, from Propositions \ref{c-representability} and
\ref{extension},
we can consider that a function symbol of any primitive recursive function 
is implicitly included in PA.  
Later in this paper, 
we will introduce several primitive recursive function symbols
in theory $T$ which is a PT-extension of PA. 

\begin{proposition}\label{composition}
Let $g$ be an $n$-place function, let $h_1,\ldots,h_n$ be
$m$-place functions, and let $f$ be defined by
$$v = f(x_1,\ldots,x_m) \equiv g(h_1(x_1,\ldots,x_m),\ldots,(x_1,\ldots,x_m)).$$
Let formulas, $\psi$ and $\theta_1,\ldots,\theta_n$, 
functionally represent $g$ and $h_1,\ldots,h_n$, and
formula $\rho_f$ be defined as follows: 
\begin{eqnarray*}
\rho_f((\xxx_1,\ldots,\xxx_m,\vvv) &\equiv& 
\\   
\exists \yyy_1 \ldots \exists \yyy_k \
&(& \theta_1(\xxx_1,\ldots,\xxx_m,\yyy_1) \land \ldots \land 
\theta_k(\xxx_1,\ldots,\xxx_m,\yyy_k)
\land \psi(\yyy_1,\ldots,\yyy_k,\vvv) \ ).
\end{eqnarray*}
Then, $\rho_f$ functionally represents $f$.
\end{proposition}

\subsection{Turing Machines}\label{TM}

A Turing machine (TM) is represented by 
$(Q, \Sigma, \Gamma, \delta, q_0, q_{accept}, q_{reject})$,
where $Q$ is a set of states, $\Sigma = \{0,1\}$
is the input alphabet, $\Gamma$ is the tape alphabet with
blank symbol $\sqcup$ and $\{0,1\}$,
$\delta: Q \times \Gamma \ra Q\times \Gamma \times$ $\{{\rm L},{\rm R}\}$  
is transition function,
$q_0$ is the start state, $q_{accept}$ is the accept state, 
and $q_{reject}$ is the reject state
\cite{Sipser97}.

The computation process of a Turing machine 
can be represented by the sequence of
{\it configurations}, $C_0, C_1, \ldots, C_k$.
Each configuration $C_i$ consists of three items,
the current state, $q_i \in Q$, the current tape contents,
and the current tape head location.
It is convenient to represent a configuration
by triple $(u, q, v)$, where 
the current state is $q$, the current tape
contents is $uv$ 
and the current head location is
the leftmost bit of $v$, where $uv$ denotes 
the concatenation of bit strings $u$ and $v$. 
When a configuration $C_i$ is $(ua, q, bv)$ ($a, b \in \{0,1\}$),
transition function $\delta$ yields 
configuration $C_{i+1}$ such that
$$ C_{i+1} = (u, q', acv) \ \ \mbox{if} \ \delta(q,b)=(q',c,{\rm L}),$$
$$ C_{i+1} = (uac, q', v) \ \ \mbox{if} \ \delta(q,b)=(q',c,{\rm R}).$$

We can also define a Turing machine whose output is
not just accept/reject, but a finite sequence of 
$\Sigma$. Here, $q_{halt}$ is used in place of
$q_{accept}$ and $q_{reject}$.
The output value is the tape contents 
in state $q_{halt}$.

Let $e_M$ be a natural number whose binary representation, $[e_M]$, 
is a part of the input to a universal Turing machine $\U$, and denotes 
the description of Turing machine $M$.
In other words, $\U$ can simulate $M$ by reading 
$[e_M]$.
Let $w$ be a natural number and $[w]$ be the input to $M$.
We then use $\U(e_M, w)$ ($ = M(w)$) 
to denote a natural number whose binary representation, $[\U(e_M, w)]$, 
is the output of $\U$
with input $[e_M]$ and $[w]$.
So, we abuse notation $\U$ for a function over natural numbers,
which is defined by universal Turing machine $\U$.




\subsection{Polynomial-Time Turing Machines}\label{PTM}

Let $M$ be a {\it polynomial-time Turing machine} (PTM). 
W.l.o.g., we assume $[e_M]$ consists of a pair of 
bit strings, $(t,[c])$: 
$t$ is a description of a Turing machine
that allows a universal Turing machine to simulate $M$,
and $c$ is a constant natural number such that 
$M$'s running time is bounded by $\Size(w)^c$.
Here $w \in \N$,  $[w] \in W$ ($W$: set of input strings to $M$)
is an input to $M$. 

$\Size(\cdot)$ is a function, 
$$\Size: \N \ra \N, \ \ \ \  \Size: w \mapsto \Size(w),$$ 
which determines the size (bit length) of input $[w] \in W$
such that, 
for positive constants $c_1$ and $c_2$, for all $[w] \in W$,
$|[w]|^{c_1} \leq \Size(w)| \leq |[w]|^{c_2}$, 
and $\Size(\cdot)$ is a polynomial-time (in $|w|$)
computable function.
The size function, $\Size(\cdot)$,
is uniquely determined by each class of problems
such as 3SAT and Hamiltonean circuit.
If $\Size(\cdot)$ is not explicitly defined,
$\Size(a) \equiv |a|$.
For example, if the underlying class of problems is 3SAT,
$[w]$ is a binary-code description of a 3CNF formula, 
and $\Size(w)$ is the number of variables of the 3CNF formula.
Then, we may use $\Size_{\TSAT}(w)$ 
to explicitly represent this function for a specific
problem, 3SAT.
If $[w]$ is not a (syntactically) valid value 
that describes a 3CNF formula,
the function value of $\Size_{\TSAT}(w)$ is defined to be $|w|$,
and a PTM specific to 3SAT, which reads such an 
invalid input value, immediately moves to
the reject state (or outputs ``invalid input'' etc).
(In Section \ref{formaldef-pnp}, function $\Size_{\CSAT}$
is more simply defined by $\Size_{\CSAT}(w) \equiv |w|$ for all $w\in \N$.)

It is easy to convert any Turing machine described by $t$
into a Turing machine described by $(t,[c])$,
by just adding a 
running step number counter (with a specific tape for counting).
Note that the counter does not count the running steps for counting.
$M$ accepts $[w]$ if it accepts $[w]$ within $\Size_W(w)^c$ steps,
and it rejects $[w]$ otherwise.
Given $(t,[c])$, universal Turing machine $\U$ simulates 
PTM $M$
by description $t$ and counts the running step number
of the simulated machine up to $\Size_W(w)^c$ and halts the machine
when the number exceeds $\Size_W(w)^c$.
Such a special universal Turing machine for PTMs, which
only accepts the form of $(t,[c])$ as input $[e_M]$,
is denoted by $\U_{\PTM}$ in this paper.
We assume a single (fixed) $\U_{\PTM}$ for PTMs. 
Then, natural number $e_M$ implies a unique PTM $M$.
It is clear that any PTM $M$ can be 
simulated by $\U_{\PTM}$ with input $[e_M]$ with form of $(t,[c])$. 
That is, $M(w)$ is exactly simulated by 
$\U_{\PTM}(e_M, w)$.

Here, w.l.o.g., we assume 
$\U_{\PTM}$ can syntactically check 
whether bit string $t$ is a syntactically correct description 
of a Turing machine for $\U_{\PTM}$.
Such syntactic rules of describing a Turing machine for 
$\U_{\PTM}$ can be clearly specified. $\U_{\PTM}$ can effectively 
check whether $t$ is a syntactically correct description or not,
in a manner similar to that used by computer 
language compilers.     
$\U_{\PTM}$ can also effectively check whether the format of
$[e_M] = (t,[c])$ is syntactically valid or not.
If $\U_{\PTM}$ recognizes $[e_M]$ to be syntactically incorrect
(e.g., the part of $[c]$ is not syntactically recognized),
$\U_{\PTM}$ outputs a special string denoting 
``syntactically invalid code''.
Here it is essential that
$\U_{\PTM}$ be able to correctly simulate a PTM if 
$(t,[c])$ is valid, and such a valid string 
$[e_M]$, which is syntactically recognized valid by
$\U_{\PTM}$, always exists for any PTM $M$.
Note: it is not essential how well $\U_{\PTM}$
can find an invalid string. If $\U_{\PTM}$ incorrectly
recognizes an invalid string as a valid one, and executes the input,
then $\U_{\PTM}$ may run abnormally (e.g., runs in an infinite loop or 
immediately halts). 
If it immediately halts (i.e., in a halting state), it is the output
of the execution.
If it runs in an infinite loop, the step counter of 
$\U_{\PTM}$ executes independently and halts when the number of
steps exceeds $\Size_W(w)^c$.


We then use $\U_{\PTM}(e_M, w)$ to denote 
a natural number whose binary representation, 
$[\U_{\PTM}(e_M, w)]$, 
is the output of $\U_{\PTM}$
with input $[e_M]$ and $[w]$.
Therefore, similarly to $\U$,
we also abuse the notation of $\U_{\PTM}$ for a function over 
natural numbers: 
$(e_M, w) \mapsto \U_{\PTM}(e_M, w)$.
Clearly, it is a totally recursive (for input $(e_M, w)$)
and polynomial-time (in $\Size_W(w)$) function.

When the input to PTM $M$ is a tuple of natural 
numbers, $(w_1,w_2,\ldots,w_k) \in W$,
we denote $\U_{\PTM}(e_M, (w_1,w_2,\ldots,w_k))$  
as its output natural number.
Here, we can consider $\U_{\PTM}$ as a totally recursive function
over $(k+1)$-tuple natural numbers $(e_M, w_1,w_2,\ldots,w_k)$
and a polynomial-time (in $\Size_W(w_1,w_2,\ldots,w_k)$) function.

We then introduce a classical result on the relationship 
between the time complexity of a Turing machine 
and Boolean circuit complexity (Theorem 9.25 in \cite{Sipser97}).

\begin{proposition}\label{circuit}
Let $t: \N \ra \N$ be a function, where $t(n) \geq n$,
and $X = \bigcup_{n \in \N} X_n$, where $X_n \equiv \{x_n \mid n \in \N \}$ 
is a set of problems $x_n$ with $\Size_{X_n}(x_n)=n$. 
If all problems in $X_n$ can be computed/decided 
by a Turing machine within time $t(n)$,
then they can be computed/decided by a Boolean circuit with size 
$O(t^2(n))$.
\end{proposition}

This proposition implies that the functionality of a polynomial-time Turing 
machine can be
realized by a polynomial size (uniform) Boolean circuit.
This property is used in the proof of Theorem \ref{representability}.



\subsection{Polynomial-Time Proofs}\label{PT-proofs}

A formal proof, $\pi$, of a formula, $\phi$, 
is expressed in tree form, called a proof tree, as follows:
A proof tree consists of nodes and directed branches.
When node $a$ is connected with node $b$ 
through a branch directed from $b$ to $a$ (i.e., $b \ra a$),
$a$ is called a child of $b$ and $b$ is called a parent of $a$:
we denote the relation as $a[b]$. If $b$ and $c$ are parents 
of $a$, the relation is denoted by $a[b, c]$.
If $a[b [c, [d, ],] ]$,
then $a$ is called a descendent of $c$ and $d$, and $c$ and $d$
are called ancestors of $a$.
A node with no child node is called a root,  
and a node with no parent node is called a leaf.
A proof tree has only one root node.
(Thus, the image of a proof tree is similar to an actual tree: 
the root is located at the bottom
of a tree and the leaves are at upper on branches.)
Node $x$ has form 
$<x_0, x_1>$, where $x_0$ is a formula
and $x_1$ is a rule of inference of the predicate logic in theory $T$.
If $a,b,c$ are nodes of a proof tree of $\pi \equiv a[b, c]$, and
$a \equiv <a_0,a_1>$, $b \equiv <b_0,b_1>$ and $c \equiv <c_0,c_1>$,     
then $\pi$ means that formula $a_0$ is deduced from formulas 
$b_0$ and $c_0$
through a rule of inference, $a_1$. 
If no rule of inference is used for the deduction,
the part of $a_1$ is empty.
If node $a \equiv <a_0,a_1>$ is a leaf,
then $a_0$ is an axiom of the underlying theory $T$
of the proof tree, and $a_1$ is empty.
If node $r \equiv <r_0,r_1>$ is the root of a proof tree of formula $\phi$, 
$r_0 = \phi$. 

If theory $T$ is a polynomial-time extension of PA,
the G{\"o}del numbers of all axioms and rules of inference 
in $T$ 
are polynomial-time (in the size of axioms) decidable.
Hence it is clear that the validity of proof tree $\pi$
can be verified within polynomial-time in the size of axioms and  
the number of nodes of $\pi$.
At the end of this subsection, we will show 
a more precise description of the polynomial-time
algorithm to verify the validity of proof tree $\pi$.


Let $\Phi \equiv \{ \phi(\aaa) \mid a \in \N  \}$
be a set of an infinite number of formulas in $T$.
The size function, $\Size_{\Phi}(\cdot)$, over natural numbers 
$\{ a \in \N  \}$, 
is uniquely defined in each $\Phi$.
If $\Size_{\Phi}(\cdot)$ is not explicitly defined,
$\Size_{\Phi}(a) \equiv |a|$.
Let ${\#}\Phi$ be the G{\"o}del number 
of the (finite-size) expression of the symbol sequence, 
$\{ \phi(\aaa) \mid a \in \N  \}$
(or the G{\"o}del number of any description of $\Phi$).
Note that the size of ${\#}\Phi$ is finite i.e., 
a constant in $|a|$. 


If $\U_{\PTM}(e,(p,{\#}\Phi,a))={\#}\pi$ and
$\pi$ is a valid proof tree of $\phi(\aaa) \in \Phi$ 
in theory $T$,
we denote 
$$\PTM_{e}(a) \vdash_T \phi(\aaa).$$
Here $p$ denotes a natural number (e.g., 0), which
indicates that the output target of $\U_{\PTM}(e,\cdot)$ 
is a proof of the formula's truth. 

If a natural number (e.g., 1), $d$, is input to
$\U_{\PTM}(e,\cdot)$ in place of $p$, it 
indicates that the output target of $\U_{\PTM}(e,\cdot)$ 
is a decision (accept or reject) of the formula's truth.
That is, ``$\U_{\PTM}(e,(d,{\#}\Phi,a))$ accepts''
implies that $\U_{\PTM}(e,\cdot)$, given $(d,{\#}\Phi,a)$, 
decides that formula $\phi(\aaa)$ is true.
(See Section \ref{pt-dec}.)
 
In other words, 
\begin{eqnarray*}
& & \PTM_{e}(a) \vdash_T \phi(\aaa) 
\\
& & \mbox{ \ \ }  \  
\LRa   \ \ \ 
\U_{\PTM}(e,(p,{\#}\Phi,a))={\#}\pi  \  \land  \  
\U_{\PTM}(v_{T},({\#}\phi(\aaa), {\#}\pi)) \ \mbox{accepts},
\end{eqnarray*}
where  
$\U_{\PTM}(v_{T},({\#}\phi(\aaa),{\#}\pi)) \mbox{accepts}$,
if and only if
PTM $\U_{\PTM}(v_{T},\cdot)$ accepts 
its input $({\#}\phi(\aaa),{\#}\pi)$ as 
$\pi$ is a valid proof tree of $\phi(\aaa) \in \Phi$ in theory $T$.
Here, $|{\#}\pi|$ is clearly polynomially (in $\Size_{\Phi}(a)$)
bounded, since ${\#}\pi$ is the output of 
$\U_{\PTM}(e, (p,{\#}\Phi,a))$. In addition,
we use the notation 
$$\PTM_{e}(a) \not\vdash_T \phi(\aaa)$$
if and only if 
$ \neg( \ \PTM_{e}(a) \vdash_T \phi(\aaa) \ )$.

We now describe PTM $\U_{\PTM}(v_{T},\cdot)$
more precisely.

\begin{enumerate}
\item \  
(Input to $\U_{\PTM}(v_{T},\cdot)$) \ $({\#}\phi(\aaa),{\#}\pi)$,
where $\phi(\aaa)$ is a formula and $\pi$ is a proof tree.
\item
$({\#}\phi(\aaa),{\#}\pi)$ is interpreted as the G{\"o}del numbers
of $\phi(\aaa)$ and $\pi$ in the manner described in Subsection \ref{G-numbers}.
\item
Check the validity of the syntactic form of $\pi$. 
\item
Search all nodes of $\pi$, and, for each node,
decide whether the node is leaf, root or other (say ``middle nodes''). 
\item
Repeat the following procedure for all leaf nodes, $a^{(i)}$ ($i=1, \ldots, \ell)$:

Pick up $a^{(i)}$, and  
check whether $a^{(i)}_0$ is an axiom of theory $T$, 
where $a^{(i)} = <a^{(i)}_0, a^{(i)}_1>$ and $a^{(i)}_1$ is empty string.

\item

Repeat the following procedure for all ``middle nodes''
and the root note, $b^{(i)}$ ($i=1, \ldots, m)$:

Pick up $b^{(i)}$ along with its parent nodes
(say $c^{(i,j)}$ ($j=1, \ldots, p_i)$), and  
check whether $b^{(i)}_0$ is deduced from $c^{(i,j)}_0$ ($j=1, \ldots, p_i)$,
by using a rule of inference $b^{(i)}_1$
(or by using no rule of inference when $b^{(i)}_1$ is empty),
where $b^{(i)} = <b^{(i)}_0, b^{(i)}_1>$, and 
$c^{(i,j)} = <c^{(i,j)}_0, c^{(i,j)}_1>$.

\item
Let $r = <r_0,r_1>$ be the root node.
Check whether $r_0 = \phi(\aaa)$.

\item
If all of the above-mentioned checks are passed correctly,
the machine accepts the input.
Otherwise rejects. 

\end{enumerate}

A series of formal proofs produced by a PTM is 
called a series of ``polynomial-time proofs''.
Here, each polynomial-time proof is a formal proof, $\pi$, of
each formula $\phi(\aaa)$ in theory $T$
(i.e., ``a polynomial-time proof'' does not mean 
a set of formal proofs. We will introduce a notion of 
a set of formal proofs in the next section).

In addition,
we introduce the following notation:
\begin{eqnarray*}
& & \TM_{e}(a) \vdash_T \phi(\aaa) 
\\
& & \mbox{ \ \ }  \  
\LRa   \ \ \ 
\U(e,(p,{\#}\Phi,a))={\#}\pi  \  \land  \  
\U_{\PTM}(v_{T},({\#}\phi(\aaa), {\#}\pi)) \ \mbox{accepts}.
\end{eqnarray*}

\subsection{Asymptotic Proofs}
\label{asymp-proof}

This section introduces a notion called ``asymptotic proof''.

\begin{definition}\label{def:asymp-proof}
Let $T$ be a theory,
$\Phi \equiv \{ \phi(\aaa_1,\ldots, \aaa_k) 
\mid (a_1,\ldots, a_k) \in \N^k  \}$
be a set of (infinite number of) formulas,
$\phi(\aaa_1,\ldots, \aaa_k)$ in $T$,
and
$$ 
Q_1 x_1 \in \N \ \cdots \ Q_k x_k \in \N \ \ \ 
T \vdash \phi(\xxx_1, \ldots, \xxx_k),$$
where $Q_1, \ldots, Q_k$ are unbounded quantifiers
(including partially bounded ones like $\exists x \geq n$ ). 
 
Then, 
a set of an infinite number of formal proofs, $\Pi$,  
in $T$, is called an ``asymptotic proof'' of 
$Q_1 \xxx_1 \ \cdots Q_k \ \xxx_k  \  \phi(\xxx_1, \ldots, \xxx_k)$,
over $T$ if 
\begin{eqnarray*}
& & Q_1 a_1 \in \N \ \ \cdots Q_k a_k \in \N  
\\
& & \mbox{ \ \ \ } \ 
( \ \pi(\aaa_1,\ldots, \aaa_k) \in \Pi \ \land \ 
\U_{\PTM}(v_{T},
({\#}\phi(\aaa_1,\ldots, \aaa_k),{\#}\pi(\aaa_1,\ldots, \aaa_k)))
\mbox{accepts} \ ).
\end{eqnarray*}

\end{definition}

The descriptive size of an asymptotic proof,
$\Pi$, can be infinite.
Therefore, such an asymptotic proof, $\Pi$, cannot be formulated
as a conventional formal proof in $T$, which should be finite-length.



The following lemma demonstrates 
the difference in 
provability between formal proofs
and asymptotic proofs.

\begin{lemma}\label{dif-prov}
Let $T$ be a primitive recursive extension of
PA and consistent.

There exists an {\it asymptotic} proof of the consistency of $T$ 
over PA.

On the other hand, there exists no {\it formal} proof of 
the consistency of $T$ in $T$.
\end{lemma}

\begin{proof}

Let {\bf Prov}$_T$ be a relation 
over $(n,m) \in \N^2$ such that  
$(n,m) \in$ {\bf Prov}$_T$ 
if and only if 
$n$ is the G{\"o}del number
of a formula (say $\psi$) and $m$ is the G{\"o}del number
of the proof of $\psi$ in $T$.

Then, $T$ is consistent if and only if  
$$ 
\forall m \in \N \ \ \  
(n^*,m) \not\in \mbox{{\bf Prov}}_T,
$$

where $n^*$ is the G{\"o}del number of $\bot$
($\bot$ is $\phi \land \neg\phi$ for a formula $\phi$). 

Since $T$ is a primitive recursive extension of PA,
{\bf Prov}$_T$ is a primitive recursive relation.  
Then, from Proposition \ref{c-representability},
there exists a $\Delta_1$-formula, Prov$_T(\xxx,\yyy)$, 
that represents relation {\bf Prov}$_T$ in PA.

Therefore, there exists an  
{\it asymptotic} proof of the consistency of $T$ over PA
as follows:
\begin{equation}\label{eq:dif-prov}  
\forall x \in \N \ \ \ 
\PA \vdash \ \neg\mbox{Prov}_T(\lc \bot \rc,\xxx)
\end{equation}
if and only if 
$T$ is consistent.

On the other hand,
even if $T$ is consistent,
$$ 
T \not\vdash \ \forall \xxx \  \neg\mbox{Prov}_T(\lc \bot \rc,\xxx)
$$

by the second G{\"o}del incompleteness theorem.

\begin{flushright}
$\dashv$
\end{flushright}
\end{proof}

We now consider the computational complexity
of producing an asymptotic proof.
Section \ref{PT-proofs}
introduced the concept of a polynomial-time proof,
that is a proof produced by a PTM.
Then, we have a combined concept,
an {\it asymptotic proof produced by a PTM} as follows:

\begin{definition}
If an asymptotic proof of 
$Q_1 \xxx_1 \ \cdots Q_k \ \xxx_k  \  \phi(\xxx_1, \ldots, \xxx_k)$ 
is produced by a PTM, i.e., 
$$
\exists e \in \N \ Q_1 x_1 \in \N \ \cdots \ Q_k x_k \in \N \ \ \ 
\PTM_{e}(x_1,\ldots, x_k) \vdash_T 
\phi(\xxx_1, \ldots, \xxx_k),
$$
then we say 
``a PTM asymptotically produces a proof of 
$Q_1 \xxx_1 \ \cdots Q_k \ \xxx_k  \  \phi(\xxx_1, \ldots, \xxx_k)$ 
over $T$,'' or
``$Q_1 \xxx_1 \ \cdots Q_k \ \xxx_k  \  \phi(\xxx_1, \ldots, \xxx_k)$
has a polynomial-time proof (is polynomial-time provable) 
over $T$.'' 

Similarly,
if an asymptotic proof of 
$Q_1 \xxx_1 \ \cdots Q_k \ \xxx_k  \  \phi(\xxx_1, \ldots, \xxx_k)$ 
is produced by a machine in computational class ${\cal C}$, 
then 
we say 
``a machine in ${\cal C}$ asymptotically produces a proof of 
$Q_1 \xxx_1 \ \cdots Q_k$ \ 
$\xxx_k  \  \phi(\xxx_1, \ldots, \xxx_k)$ 
over $T$,'' or
``$Q_1 \xxx_1 \ \cdots Q_k \ \xxx_k  \  \phi(\xxx_1, \ldots, \xxx_k)$
has a ${\cal C}$ proof (is ${\cal C}$ provable) over $T$.'' 

\end{definition}

A variant of Lemma \ref{dif-prov} demonstrates 
an example of asymptotic proofs produced by a PTM.

If $T$ is a ``PT-extension'' of PA,
then $\mbox{Prov}_T(\lc \bot \rc,\xxx)$ in Lemma \ref{dif-prov}
can be equivalent to
$\mbox{\PTM-\Acpt}(\vvv_T, \lc \bot \rc, \xxx)$
(see Section \ref{formal-ptm}
for the notation of $\mbox{\PTM-\Acpt}(\vvv_T, \cdot, \cdot )$). 
Next, we obtain the following lemma by Theorem \ref{representability}.

\begin{lemma}\label{con-poly-proof}
Let $T$ be a consistent PT-extension of PA.

There exists an {\it polynomial-time} proof of the consistency 
of $T$ over PA.
That is,
\begin{equation}\label{eq:con-poly-proof}   
\exists e \in \N \ \ \forall x \in \N \ \ \ 
\PTM_{e}(x) \vdash_{\PA} \ 
\neg\mbox{\PTM-\Acpt}(\vvv_T, \lc \bot \rc, \xxx).
\end{equation}
\end{lemma}


\subsection{Representability Theorem of Polynomial-Time Proofs}
\label{ptm-rep}

\begin{definition}
\begin{enumerate}
\item
Let $R$ be a $k$-ary relation on $\N$, 
i.e., $R \subseteq \N^k$.
A formula $\rho_R(\xxx_1,\ldots,\xxx_\kkk)$ 
(in which only $\xxx_1,\ldots,\xxx_k$ occur free)
will be said to {\it polynomial-time represent} 
relation $R$ in theory $T$
if and only if 
there exists $e_R \in \N$ such that 
for every $a_1,\ldots,a_k$ in $\N^k$,
\begin{eqnarray*}
& & 
(a_1,\ldots,a_k) \in R \ \ \Ra \ \ 
\PTM_{e_R}(a_1,\ldots,a_k) 
\vdash_T \rho_R(\aaa_1,\ldots,\aaa_k),
\\
& & 
(a_1,\ldots,a_k) \not\in R \ \ \Ra \ \ 
\PTM_{e_R}(a_1,\ldots,a_k) 
\vdash_T \neg\rho_R(\aaa_1,\ldots,\aaa_k).
\end{eqnarray*}



\item
Let $f$ be a $k$-place function on natural numbers $a_1,\ldots,a_k$.
A formula $\rho_f(\xxx_1,\ldots,\xxx_k,\yyy)$ 
(in which only $\xxx_1,\ldots,\xxx_k,\yyy$ occur free)
will be said to {\it functionally polynomial-time represent} 
$f$ in theory $T$
if and only if
there exists $e_f$ such that 
for every $a_1,\ldots,a_k$ in $\N^k$
$$\PTM_{e_f}(a_1,\ldots,a_k) \vdash_T 
\forall \yyy ( \rho_f(\aaa_1,\ldots,\aaa_k,\yyy)
\lra \yyy = \Se^{f(a_1,\ldots,a_k)}\0).$$



\end{enumerate}
\end{definition}




\begin{theorem}\label{representability}
(Polynomial-Time Representability Theorem)
For any {\it polynomial-time computable} relation on $\N^k$, $R$, and 
any {\it polynomial-time computable} function on $\N^k$, $f$,
there exist formulas, 
$\rho_R(\xxx_1,\ldots,\xxx_k)$
and $\rho_f(\xxx_1,\ldots,\xxx_k,\yyy)$,
such that:
\begin{itemize}
\item
$\rho_R(\xxx_1,\ldots,\xxx_k)$
polynomial-time represents $R$, 
and $\rho_f(\xxx_1,\ldots,\xxx_k,\yyy)$ 
functionally polynomial-time represents $f$ in PA.
\item
$\rho_R(\xxx_1,\ldots,\xxx_k)$
and $\rho_f(\xxx_1,\ldots,\xxx_k,\yyy)$
are $\Delta_1$ in PA.
\item
$$ \PA \vdash \forall \xxx_1 \cdots \forall \xxx_k \exists !\yyy \
\rho_f(\xxx_1,\ldots,\xxx_k,\yyy).$$
\end{itemize}
\end{theorem}

\begin{proof}
For simplicity of description, we consider the case of relation $R$ 
with only one free variable $x$.
It is straightforward to extend this result to the cases 
with multiple free variables and functional representability.  

First, we will introduce two function symbols,
$| \cdot |$ and $\Bit(\cdot, \cdot)$,
which are intended to denote the length of the binary 
representation of a numeral and the $i$-th rightmost 
numeral ($\0$ or $\1$) of the binary 
representation of a numeral, respectively.

\begin{claim}\label{claim1}
$$ 
\PA \vdash \forall \xxx > \0 \ \exists ! \nnn \
{\2}^{\nnn \hik \1} \leq \xxx < {\2}^{\nnn}.
$$
\end{claim}

\begin{proof}
We will use the induction axiom in PA.
We can prove the following
by using the axioms of PA easily
(e.g., by proving $\1 + \1 = \2 = \2^{\1}$):
$$
\PA \vdash \
({\2}^{\0} \leq \1 < {\2}^{\1}).
$$
In addition,
we can also prove the following
by using the axioms of PA
(e.g., by using the axiom, 
$\forall \xxx \forall \yyy (\xxx+\Se \yyy) = \Se(\xxx+\yyy)$ etc.):
\begin{eqnarray*}  
& & \PA \vdash \
\forall \xxx \ 
( \  
\exists ! \nnn \
(  {\2}^{\nnn \hik \1} \leq \xxx < {\2}^{\nnn} \dot{-} \1 )  
\ra
\exists ! \nnn \
( {\2}^{\nnn \hik \1} \leq \xxx + \1  < {\2}^{\nnn} )
\  )
,
\\
& &
\PA \vdash \
\forall \xxx \ 
(
\exists ! \nnn \ (\xxx = {\2}^{\nnn} \dot{-} \1 )  
\ra
{\2}^{\nnn} = \xxx + \1 
\\
& & \mbox{ \ \ \ \ \ \  } \ \ \ 
\ra
{\2}^{\nnn} \leq \xxx + \1 < {\2}^{\nnn + \1)} 
)
\ra 
\exists ! \nnn' \ ({\2}^{\nnn' \hik \1} \leq \xxx + \1 < {\2}^{\nnn'}) 
). 
\end{eqnarray*}

Combining the above results, we obtain
$$
\PA \vdash \  \
({\2}^{\0} \leq \1 < {\2}^{\1}) \ \land \ 
\forall \xxx \ (
(\exists ! \nnn \
{\2}^{\nnn \hik \1} \leq \xxx < {\2}^{\nnn})
\ra
(\exists ! \nnn \ {\2}^{\nnn \hik \1} \leq \xxx + \1 < {\2}^{\nnn})
).
$$

The induction axiom of PA implies
\begin{eqnarray*}  
\PA \vdash & & \
({\2}^{\0} \leq \1 < {\2}^{\1}) \ \land \ 
\forall \xxx \ (
\exists ! \nnn \
{\2}^{\nnn \hik \1} \leq \xxx < {\2}^{\nnn}
\ra
\exists ! \nnn \ {\2}^{\nnn \hik \1} \leq \xxx + \1 < {\2}^{\nnn}
)
\\
& & \mbox{  \ \ } \ 
\ra 
\forall \xxx > \0 \ 
\exists ! \nnn \ \ 
{\2}^{\nnn \hik \1} \leq \xxx < {\2}^{\nnn}.
\end{eqnarray*}
Hence we obtain finally
$$ 
\PA \vdash \forall \xxx > \0 \ 
\exists ! \nnn \ \ 
{\2}^{\nnn \hik \1} \leq \xxx < {\2}^{\nnn}.
$$

\begin{flushright}
$\dashv$
\end{flushright}
\end{proof}

Following the claim above, 
we will introduce a function symbol, $| \cdot |$,  in PA, 
which is intended to denote the binary expression length of numeral 
$\xxx$, such that
\begin{equation}\label{eq:length} 
\PA \vdash \forall \xxx > \0 \ \forall \nnn \
(
{\2}^{\nnn \hik \1} \leq \xxx < {\2}^{\nnn}
\lra 
\nnn = | \xxx |
).
\end{equation}

\begin{claim}\label{claim2}
$$ 
\PA \vdash 
\forall \xxx > \0 \ \forall \nnn \
(
{\2}^{\nnn \hik \1} \leq \xxx < {\2}^{\nnn}
\ra
( 
({\2}^{\nnn \hik \1} \leq \xxx < {\2}^{\nnn \hik \1} + {\2}^{\nnn \hik \2}) 
\lor
({\2}^{\nnn \hik \1} + {\2}^{\nnn \hik \2} \leq \xxx < {\2}^{\nnn}) 
)
$$
\end{claim}

We omit the proof since it is similarly obtained.

\begin{claim}\label{claim3}
\begin{eqnarray*}   
\PA \vdash & & \ 
\forall \xxx > \1 \ \ 
\forall \iii < | \xxx | \ \ 
\exists ! \xxx_i < \2 \ \ 
\exists ! \yyy < \2^{\iii} \ \ 
\exists ! \zzz < \2^{| \xxx | \hik \iii \hik \1} \
\\
& & 
(
\xxx = \yyy + \xxx_i \cdot {\2}^{\iii} + \zzz \cdot {\2}^{\iii+\1}
).
\end{eqnarray*}
\end{claim}

This claim can be proven by applying the previous claims
repeatedly.

Based on the claim above,
we will introduce a function symbol, $\Bit(\cdot)$, in PA, 
which is intended to denote the $i$-th rightmost value of
the binary expression of a numeral 
\begin{eqnarray}    
\PA \vdash  & & \ 
\forall \xxx > \1 \ \ 
\forall \iii < | \xxx | \ \
\forall \xxx_i < \2 \ \ 
\exists ! \yyy < \2^{\iii} \ \ 
\exists ! \zzz < \2^{| \xxx | \hik \iii \hik \1} \
\nonumber
\\
& & 
( \ 
\xxx = \yyy + \xxx_i \cdot {\2}^{\iii} + \zzz \cdot {\2}^{\iii+\1}
\ \lra \
\xxx_i = \Bit(\xxx,\iii)
\ )
.
\label{eq:bit} 
\end{eqnarray}

Hereafter, we will also denote the binary representation

of variable $\xxx$ by
$[\xxx] =$ $\Bit(\xxx,\nnn \hik \1)$ $\Bit(\xxx,\nnn \hik \2)$ 
$\cdots$ $\Bit(\xxx,\0)$, where 
$\nnn = | \xxx |$.

In order to construct a formula, $\rho_R(\xxx)$, 
in PA which polynomial-time represents
relation $R$,
we will employ the approach of constructing
a family of polynomial size 
Boolean circuits that represents relation $R$,
which is introduced in Proposition \ref{circuit}
(Theorem 9.25 in \cite{Sipser97}).

Since $R$ is polynomial-time computable relation, 
there exists a PTM, $\U_{\PTM}(e_R, \cdot)$,
that computes relation $R$ correctly.
Then, $R$ can be decided by a family of Boolean circuits,
$\{ B_{n} \mid n \in \N \}$, that are polynomial size in $n$.

Before showing formula $\rho_R(\xxx)$,
we will show how to construct a family of Boolean circuits,
$\{ B_{n} \mid n \in \N \}$, based on the description of
Theorem 9.25 in \cite{Sipser97}.

Let the size of input $x$ be $n$ bits, and the computation time 
be $t(n) = n^c$ steps ($c$ is a constant determined by each PTM). 
The circuit is constructed by $(n^c)^2 k$ nodes. 
The value of each node is F (false/0/off) or T (true/1/on),
and each value is denoted by $light[i,j,s]$
($0\leq i < n^c$, $0\leq j < n^c$, $0\leq s < k$). 

$light[i,j,s]$ = T ($light[i,j,s]$ is on) denotes 
the element of $cell[i,j]$ (i.e., in the $i$-th computation step and
at the $j$-th leftmost tape square) is the $s$-th element,
where there are $k ( = 3 +3\ell)$ elements, $\Gamma \cup \Gamma \times Q$,
$\Gamma \equiv \{0,1,\sqcup \}$ is the set of tape alphabets, 
and $Q \equiv \{q_0$ (initial state), $q_1, \ldots,$ 
$q_{\ell-2}$ (reject state), $q_{\ell-1}$ (accept state) $\}$ 
is the set of states of the underlying PTM
to decide $x \in R$. 
$light[i,j,s]$ = F ($light[i,j,s]$ is off) denotes that 
the element of $cell[i,j]$ is not the $s$-th element.
The set of the elements is 
$\{ (0), (1), (\sqcup),$ $(q_0, 0), (q_0, 1), (q_0, \sqcup),$ $\ldots, 
(q_i, 0), (q_i, 1), (q_i, \sqcup),$ $\ldots, 
(q_{\ell}, 0),$ $(q_{\ell-1}, 1),$ $(q_{\ell-1}, \sqcup)\}$.
So, for each $cell[i,j]$, only one $light[i,j,s]$ (i.e., only one $s$)
is T (true/1/on) and the others are F (false/0/off).
Each node is connected to $3k$ nodes 
through $\land$ and $\lor$ gates.
More precisely, 
for all $i (1 \leq i < n^c)$,
for all $j (0 \leq j < n^c)$,
for all $s (0 \leq s < k)$,
$$light[i,j,s] = 
\bigvee_{(a,b,c) \in  A_s}
(light[i-1,j-1,a] \land light[i-1,j,b] \land light[i-1,j+1,c],
$$
where subset $A_s \equiv \{ (a_0, b_0, c_0), \ldots, (a_t,b_t,c_t) \}$ 
($ t < k^3)$ is uniquely determined for each $s$ 
based on the transition function $\delta$ of the underlying PTM
to decide $x \in R$.
For example, 
\begin{itemize}
\item
$A_1 \equiv \{ (1,1,1), (1,1,0), (2,3+3i-1,1), \ldots \}$,
where $\delta(q_i,1) = (q_j,0,$L).
\item
$A_2 \equiv \{ (1,2,1), (2,3+3i-2,1), \ldots \}$,
where $\delta(q_i,0) = (q_j,1,$R).
\item
$A_{3+3i-1} \equiv \{ (0,1,3+3j-2), \ldots \}$,
where $\delta(q_j,0) = (q_i,1,$L).
\end{itemize}

The values of $light[0,j,s]$, for $0\leq j < n^c$ and
$0\leq s < k$,  are determined by the input 
$[x] = `` {x_{n-1} x_{n-2} \cdots x_0 } ''$,
i.e., 
\begin{eqnarray*}   
& &  
\left\{
\begin{array}{ll}
light[0,0,3] = 1 &  \mbox{iff} x_{n-1}=0,\\
light[0,0,4] = 1 &  \mbox{iff} x_{n-1}=1, \\
light[0,0,s] = 0 &  \mbox{for all $s$ with $s \not= 3$ and $s\not=4$.}
\end{array} 
\right. 
\\
& &  
\left\{
\begin{array}{ll}
light[0,1,0] = 1 &  \mbox{iff} x_{n-2}=0,\\
light[0,1,1] = 1 &  \mbox{iff} x_{n-2}=1, \\
light[0,1,s] = 0 &  \mbox{for all $s$ with $s \not= 0$ and $s\not=1$.}
\end{array} 
\right. 
\\
& &  
\mbox{ \ \ } \cdots 
\\
& &  
\left\{
\begin{array}{ll}
light[0,n-1,0] = 1 &  \mbox{iff} x_{0}=0,\\
light[0,n-1,1] = 1 &  \mbox{iff} x_{0}=1, \\
light[0,n-1,s] = 0 &  \mbox{for all $s$ with $s \not= 0$ and $s\not=1$.}
\end{array} 
\right. 
\\
& &  
\left\{
\begin{array}{ll}
light[0,j,2] = 1 &  \mbox{for all $j \geq n$.} \\ 
light[0,j,s] = 0 &  \mbox{for all $j \geq n$ and for all $s$ with $s \not=2$.} 
\end{array} 
\right. 
\end{eqnarray*}

The input $[w]$ ($n^c k$ bit string) 
to Boolean circuit $B_{n}$ is
$$[w] = `` { light[0,1,1], light[0,1,2], \ldots, light[0,n^c-1,k-1] }'',$$
The output of the circuit is the value of node 
$light[n^c - 1,1,k-6]$, $light[n^c - 1,1,k-5]$, $light[n^c - 1,1,k-4]$ 
(reject) or 
$light[n^c - 1,1,k-3]$, $light[n^c - 1,1,k-2]$,
$light[n^c - 1,1,k-1]$ (accept).

We now show formula $\rho_R(\xxx)$ in PA
based on the above construction of 
Boolean circuit $B_{n}$.

First we define three formulas: $\ISET(\xxx, \yyy)$
in which only $\xxx$, $\yyy$ occurs free,
$\TRANS(\yyy)$ in which only $\yyy$ occurs free,
and $\EVAL(\yyy)$ in which only $\yyy$ occurs free.
Formula $\ISET(\xxx, \yyy)$ denotes that the information of
$x$ is transformed/copied to the value of a part of $y$,
formula $\TRANS(\yyy)$ denotes that
the transition history of computing $R(x)$
is mapped to the value of the other part of $y$,
and  $\EVAL(\yyy)$ is true if and only if
the evaluation result of $R(x)$ is true.

\begin{eqnarray*}  
& & \ISET(\xxx,\yyy) \  \ \equiv
\\
& & \mbox{ \ \ } 
( \ 
(
\Bit(\xxx,\0) = \0 \ \ra \ 
( \Bit(\yyy, \3)= \1 \ 
  \land \
  \forall \sss (\0 < \sss < \3 \ \lor \ \3 < \sss < \kkk) \ \Bit(\yyy, \sss) = \0 ) 
) \ \land \   
\\
& & \mbox{ \ \ \ \ } 
(
\Bit(\xxx,\0) = \1 \ra \
( \Bit(\yyy, \4)= \1 \ 
  \land \
  \forall \sss (\0 < \sss < \4 \ \lor \ \4 < \sss < \kkk) \ 
  \Bit(\yyy, \sss) = \0  )
) \ 
) 
\\
& & \mbox{ \ }   
\land \
\\
& & \mbox{ \ \ } 
( \ 
(
\forall \jjj (\0 < \jjj < \nnn ) \ 
\land \
\Bit(\xxx, \jjj) = \0 \
\ra \ 
\Bit(\yyy, \jjj\cdot \kkk ) = \1 \
\land \ 
\forall \sss (\0 < \sss < \kkk ) 
\ \Bit(\yyy, \jjj\cdot \kkk + \sss) = \0 
) \land \ 
\\
& & \mbox{ \ \ \ \ } 
(
\forall \jjj < \nnn \  
\land \
\Bit(\xxx, \jjj) = \1 \
\ra \
\Bit(\yyy, \jjj\cdot \kkk + \1 ) = \1 \
\land \ 
\forall \sss (\sss = \0 \ \lor \ 
\1 < \sss < \kkk ) \ 
\Bit(\yyy, \jjj\cdot \kkk + \sss) = \0 
) \ 
)
\\
& & \mbox{ \ } 
\land \
\\
& & \mbox{ \ \ } 
( \ 
\forall \jjj \ (\nnn < \jjj < \nnn^{\ccc}) \ 
(
\Bit(\yyy, \jjj\cdot \kkk + \2) = \1 \
\land \
\forall \sss \ (\sss < \2 \ \land \ \2 < \sss < \kkk) \
\Bit(\yyy, \jjj\cdot \kkk + \sss) = \0 
) \ 
).
\end{eqnarray*}

\begin{eqnarray*}  
& & \TRANS(\yyy) \ \ \equiv
\\
& & \mbox{ \ \ } 
\forall \iii (\0 < \iii < \nnn^{\ccc}) \ \
\forall \jjj < \nnn^{\ccc} \ \
\forall \sss <\kkk
\\
& & \mbox{ \ \ \ \ \ } 
( \ 
(
\exists \aaa<\kkk \ \ \exists \bbb<\kkk \ \ \exists \ccc<\kkk \ \ \
( \eta(\yyy, \iii \hik \1,\jjj,\aaa,\bbb,\ccc,\sss) \
\ra \ \Bit(\yyy,\iii\cdot \nnn^{\ccc}\cdot \kkk + \jjj\cdot\kkk + \sss) = \1
) 
\\
& & \mbox{ \ \ \ \ \ } 
\ \land
\\
& & \mbox{ \ \ \ \ \ \ } \ 
(
\forall \aaa<\kkk \ \ \forall \bbb<\kkk \ \ \forall \ccc<\kkk \ \ \
\neg \eta(\yyy, \iii \hik \1,\jjj,\aaa,\bbb,\ccc,\sss) \
\ra \ \Bit(\yyy,\iii\cdot \nnn^{\ccc}\cdot \kkk + \jjj\cdot\kkk + \sss) = \0
) \
).
\end{eqnarray*}

Here, formula $\eta(\cdot)$
is uniquely fixed for each $s$ 
based on the transition function $\delta$ of the underlying PTM
to decide $x \in R$, and corresponds to subset $A_s$ ($0 \leq s < k$)
in the above-mentioned Boolean circuit $B_n$.
In more detail, $\eta(\cdot)$ is formulated as follows:
\begin{eqnarray*}  
\eta(\yyy, \iii \hik \1, \jjj,\aaa,\bbb,\ccc,\sss) 
\ \ \equiv 
& &
\eta_0(\aaa,\sss,
\Bit(\yyy, (\iii \hik \1)\cdot\nnn^{\ccc}\cdot\kkk
+ (\jjj \hik \1)\cdot\kkk + \aaa) )
\ \land \
\\
& & 
\eta_1(\bbb,\sss,\Bit(\yyy, (\iii \hik \1)\cdot\nnn^{\ccc}\cdot\kkk
+ \jjj\cdot\kkk + \bbb))
\ \land \
\\
& & 
\eta_2(\ccc,\sss,\Bit(\yyy, (\iii \hik \1)\cdot\nnn^{\ccc}\cdot\kkk
+ (\jjj + \1)\cdot\kkk + \ccc)).
\end{eqnarray*}
Remark: If $\jjj = \0$ (or $\jjj = \nnn^{\ccc} \hik \1$),
then $\aaa$ (or $\ccc$) is ignored.

$$ 
\EVAL(\yyy) \ \ \equiv \ \ 
\exists \sss \ ( \kkk \hik \4 <\sss <\kkk) \ \ \
\Bit(\yyy,(\nnn^{\ccc} \hik \1)\cdot \nnn^{\ccc}\cdot \kkk + \sss) = \1 .
$$

Finally 
$$
\rho_R(\xxx)  \ \ \equiv \ \    
\exists ! \yyy < {\2}^{\nnn^{\2 \ccc} \kkk} \
( \nnn = | \xxx | \ \land
\ISET(\xxx,\yyy) \ \land
\TRANS(\yyy) \ \land
\EVAL(\yyy)
).
$$

By the above-mentioned formula, $\rho_R(\xxx)$,
for any input $x \in \N$  
($[\xxx] = \xxx_{n-1} \cdots \xxx_{0}$),
numeral $\yyy$  ($[\yyy] =  \yyy_{(n^c)^2 k - 1} \cdots \yyy_{0}$),
is uniquely determined, and 
the truth or falsity of $\rho_R(\xxx)$
is also determined by the truth or falsity of
term $\EVAL(\yyy)$.

It is clear from Proposition \ref{circuit}
that formula $\rho_R(\xxx)$ represents $R$,
since each atomic formula of  $\rho_R(\xxx)$
represents the corresponding atomic execution of 
$\U_{\PTM}(e_R, x)$,
and such atomic execution is primitive recursive.
That is, for all $x \in \N$ 
$$\U_{\PTM}(e_R, x) \mbox{accepts}
\ \ \Ra \ \ 
\PA \vdash \ 
\exists ! \yyy < {\2}^{\nnn^{\2 \ccc} \kkk} \
( \nnn = | \xxx | \ \land
\ISET(\xxx,\yyy) \ \land
\TRANS(\yyy) \ \land
\EVAL(\yyy)
).
$$
$$\U_{\PTM}(e_R, x) \mbox{rejects}
\ \ \Ra \ \ 
\PA \vdash \ 
\exists ! \yyy < {\2}^{\nnn^{\2 \ccc} \kkk} \
( \nnn = | \xxx | \ \land
\ISET(\xxx,\yyy) \ \land
\TRANS(\yyy) \ \land
\neg\EVAL(\yyy)
).
$$

Since
$\aaa$ is represented by the binary form of numerals,
and the proof tree of the formula,
$${\2}^{\nnn \hik \1} \leq \aaa < {\2}^{\nnn}
\lra 
\nnn = | \aaa |,
$$
can be constructed in $O(|a|)$, 
there exists $e_L \in \N$ such that 
for every $a$ in $\N$,
$$
\PTM_{e_L}(a) 
\vdash_{\PA} 
\forall \nnn \  
(
{\2}^{\nnn \hik \1} \leq \aaa < {\2}^{\nnn}
\lra 
\nnn = | \aaa |
).
$$

Similarly, $\Bit(\cdot)$ can be also
functionally polynomial-time represented.

Given $a \in \N$, in order to evaluate formula $\rho_R(\aaa)$,
we need to evaluate function $| \cdot |$ once,
polynomially many repetitions of function $\Bit$,
and polynomially many repetitions of formula $\eta$,
where the size of formula $\eta$ is constant in $|a|$
and $\eta$ is $\Delta_1$-formula
(since all quantifiers in $\eta$ are bounded).

Therefore, in total,
formula $\rho_R(\aaa)$ can be 
functionally polynomial-time represented.

We now introduce formula $\widetilde{\rho_R}(\xxx,\yyy)$ 
that is defined by
$$
\nnn = | \xxx | \ \land
\ISET(\xxx,\yyy) \ \land
\TRANS(\yyy).
$$
Here, $\widetilde{\rho_R}(\xxx,\yyy)$ 
represents a polynomial-time function,
which, given $x$, computes $y$.
(Usually, a part of execution history, $y$, 
is output in a polynomial-time function.) 
 
Then,
$$
\rho_R(\xxx) =  \exists ! \yyy < {\2}^{\nnn^{\2 \ccc} \kkk} \
\ ( \ \widetilde{\rho_R}(\xxx,\yyy) \ \land \ \EVAL(\yyy) \ ).
$$


By repeatedly proving an atomic formula 
on a pair of $kn^c$ bit parts (pair of laws) 
of the binary expression of $\yyy$,
we obtain  
$$ 
\PA \vdash \ 
\forall \xxx \ 
\exists ! \yyy < {\2}^{\nnn^{\2 \ccc} \kkk} \ \ 
( \nnn = | \xxx | \ \land
\ISET(\xxx,\yyy) \ \land
\TRANS(\yyy) \ ), 
$$
where $| \xxx |$ and $\Bit(\xxx,\iii)$
for $\xxx < \2$ are defined additionally.
That is, we obtain
$$ \PA \vdash \forall \xxx \ 
\exists ! \yyy < {\2}^{\nnn^{\2 \ccc} \kkk} \ \ 
\widetilde{\rho_R}(\xxx,\yyy).$$

\begin{flushright}
$\dashv$
\end{flushright}

\end{proof}


\subsection{Formalization of Polynomial-Time Proofs}\label{formal-ptm}

Let $\Phi$ be a set of an infinite number of formulas, 
$\{ \phi(\aaa) \mid  a \in \N \}$. 

Let formula
$$\mbox{\PTM-Out}(\eee, \lc \Phi \rc, \aaa, \bbb)$$
polynomial-time represent
$$\U_{\PTM}(e,(p,{\#}\Phi,a))=b$$
over natural numbers, $(e, {\#}\Phi, a, b)$,
and
formula 
$$\mbox{\PTM-\Acpt}(\vvv_T, \lc \phi(\aaa) \rc, \bbb)$$
polynomial-time represent
$$\U_{\PTM}(v_T,({\#}\phi(a),b)) \mbox{accepts}$$
over natural numbers, $(v_T, {\#}\phi(a), b)$.
(For the definition of $v_T$, see Section \ref{PT-proofs}.)
Here, these formulas are constructed 
by following (the multiple-variable version of)
the method of constructing a formula that was shown in the proof
of the polynomial-time representability theorem
(Theorem \ref{representability}).

Then, 
$$
\mbox{\Prv}_{T}[\phi(\aaa)](\eee, \lc \Phi \rc, \aaa) 
\ \ \ \equiv \ \ \ 
\exists \bbb < \ 2^{{\Size}(\aaa)^{\ccc}} \ \ 
( \  
\mbox{\PTM-Out}(\eee, \lc \Phi \rc, \aaa, \bbb) \ \ \
\land \ \ \ 
\mbox{\PTM-\Acpt}(\vvv_T, \lc \phi(\aaa) \rc, \bbb)
\ ),
$$
where $c$ is uniquely determined by $e$ (i.e.,
there is a primitive recursive function $f$ such that $c=f(e)$.)

Clearly, $\mbox{\Prv}_{T}[\phi(\aaa)](\eee, \lc \Phi \rc, \aaa)$ 
represents the relation 
$$\PTM_{e}(a) \vdash_T  \phi(\aaa)$$
over natural numbers, $(e, {\#}\Phi, a) \in \N^4$
(for the definition, see Section \ref{PT-proofs}).
Then, for any $(e, {\#}\Phi, a) \in \N^3$,
\begin{eqnarray*}
& & 
\PTM_{e}(a) \vdash_T  \phi(\aaa)
\ \ \ \Ra \ \ \ 
\PA \vdash \ 
\mbox{\Prv}_{T}[\phi(\aaa)](e, \lc \Phi \rc, \aaa),
\\
& & 
\PTM_{e}(a) \not\vdash_T  \phi(\aaa)
\ \ \ \Ra \ \ \ 
\PA \vdash \ 
\neg\mbox{\Prv}_{T}[\phi(\aaa)](e, \lc \Phi \rc, \aaa).
\end{eqnarray*}

Here, note that the above-mentioned relation
over $(e, {\#}\Phi, a)$ is polynomial-time
decidable in $a$ with a fixed value of $(e, {\#}\Phi)$, 
but that the asymptotic computational complexity of this relation 
in $(e, {\#}\Phi)$ is not explicitly specified.
However, the way of constructing a formula shown in 
the proof of Theorem \ref{representability}
can be applied to any primitive recursive relation.


Here it is worth noting that,
although formula $\mbox{\Prv}_{T}[\phi(\zzz)](\xxx, \yyy, \zzz)$,  
with free variables $\xxx$, $\yyy$ and $\zzz$, 
is specified by the construction shown in the proof
of Theorem \ref{representability},
there still exists ambiguity with regard to details of the formula.
However, notation $\mbox{\Prv}_{T}[\phi(\zzz)](\xxx, \yyy, \zzz)$
means a fixed formula selected from among the possible formulas. 
The difference of a formula selected from them does not
affect the results in this paper. It is important to note
that the fixed formula of $\mbox{\Prv}_{T}[\phi(\zzz)](\xxx, \yyy, \zzz)$  
is assumed throughout this paper. 

Informally, formula (sentence) 
$\mbox{\Prv}_{T}[\phi(\aaa)](\eee, \lc \Phi \rc, \aaa)$
is true if and only if 
$\U_{\PTM}(e,\cdot)$, on input $(p,{\#}\Phi,a) \in \N^3$,
outputs a proof tree of formula $\phi(\aaa) \in \Phi$ 
in theory $T$.
Here, note that $[ \ \cdot \  ]$ 
does not mean a variable part of the formula,
but just implies the target for $\U_{\PTM}(e,(p,{\#}\Phi,a))$ to prove,
while $( \ \cdot, \ \cdot, \ \cdot \ )$ means 
a variable part of the formula.
Therefore,
the part of 
$\mbox{\Prv}_{T}$
in formula 
$\mbox{\Prv}_{T}[\phi(\aaa)](\eee, \lc \Phi \rc, \aaa)$
identifies the form of the formula 
(like $\rho$ in $\rho(\eee, \lc \Phi \rc, \aaa)$).
The part of $[\phi(\aaa)]$ in the formula
is perfectly redundant and is not necessary to identify the formula,
but helps readers in understanding the meaning of the formula.
(Note that $\mbox{\Prv}_{T}[X](\cdot, \cdot, \cdot)$
is a single formula, regardless of $X$.)




\section{Incompleteness Theorems of Polynomial-Time Proofs}\label{sec:rbicthm}

This section shows the {\it polynomial-time proof} version of 
the (second) G{\"o}del incompleteness theorem.
First, we introduce the G{\"o}del sentences of polynomial-time proofs, 
and the first incompleteness theorem of polynomial-time proofs.
We then present the second incompleteness theorem of polynomial-time proofs,
based on the the first incompleteness theorem and 
the derivability conditions of polynomial-time proofs.

\subsection{Derivability Conditions of Polynomial-Time Proofs}\label{DC}

This section introduces several properties, 
the {\it derivability conditions} of polynomial-time proofs.
(They correspond to the derivability conditions regarding conventional
incompleteness theorems.) 
These properties are used to prove the (first and second)
incompleteness theorems of polynomial-time proofs in this paper.

\begin{lemma}\label{D1}
(D.1 of PTPs) \ \ \
Let $\Phi \equiv
\{ \phi(\aaa) \mid  a \in \N \}$
be a set of an infinite number of formulas. 
Suppose that $T$ is a PT-extension of PA.
Then the following holds for all $e,T$:

For any $e \in \N$ and any $a \in \N$
there exists $e^* \in \N$
such that
\begin{eqnarray*}   
& & \PTM_{e}(a) \vdash_T  \phi(\aaa)
\\ 
 \Ra \  
& & 
\PTM_{e^*}(a) \vdash_{\PA} 
\mbox{\Prv}_{T}[\phi(\aaa)](\eee, \lc \Phi \rc, \aaa).
\end{eqnarray*}
\end{lemma}
\begin{proof}
Since $$ 
\PTM_{e}(a) \vdash_T  \phi(\aaa)$$
is a polynomial-time relation computed by 
$\U_{\PTM}(e,({\#}\Phi,\cdot))$
and $\U_{\PTM}(v_{T},(\cdot, \cdot))$,
given $a \in \N$, such that 
$$\U_{\PTM}(e,({\#}\Phi,a))={\#}\pi  \  \land  \  
\U_{\PTM}(v_{T},({\#}\phi(\aaa), {\#}\pi)) \ \mbox{accepts}.$$
Therefore, 
this result is obtained immediately from 
Theorem \ref{representability}.

\begin{flushright}
$\dashv$
\end{flushright}
\end{proof}

\begin{lemma}\label{D2}
(D.2 of PTPs) \ \ \ 
Let 
$\Phi \equiv 
\{ \phi(\aaa) \mid  a \in \N \}$,
$\Omega \equiv 
\{ \phi(\aaa) \ra \psi(\aaa) \mid  \ a \in \N \}$, and
$\Psi \equiv 
\{ \psi(\aaa) \mid  a \in \N \}$.
Suppose that $T$ is a PT-extension of PA.

For all $e_1 \in \N$ and for all $e_2 \in \N$,
there exists $e_3 \in \N$ such that 
\begin{eqnarray*}   
\PA \vdash \ \ \
\forall \xxx \  
( \ \ & & 
 \mbox{\Prv}_{T}[\phi(\xxx)](\eee_1, \lc \Phi \rc, \xxx) \ \land
 \mbox{\Prv}_{T}[\phi(\xxx) \ra \psi(\xxx)]
(\eee_2, \lc {\Omega} \rc, \xxx) 
\\
& &
\ra \
 \mbox{\Prv}_{T}[\psi(\xxx)](\eee_3, \lc \Psi \rc, \xxx)
\ \ ).
\end{eqnarray*}

\end{lemma}

\begin{proof}

First, we introduce a two-place polynomial-time function, 
$h$, over $\N^2$ such that 
$$ 
h(s,t) \equiv \left\{
\begin{array}{ll}
 {\#}\pi & \mbox{\quad if there exist proof trees 
                       $\pi_1$ and $\pi_2$ in $T$} \\
          & \mbox{\quad such that $s = {\#}\pi_1$, $t = {\#}\pi_2$. Here } \\
          & \mbox{\quad 
                 $\pi \equiv <\psi, \mbox{\rm Modus Ponens}>  [\pi_1, \pi_2].$ } \\
 0        & \mbox{\quad otherwise.}          
\end{array}\right. 
$$ 
(Given $s \in \N$, it is polynomial-time (in $|s|$)
computable to check whether $u$ is the G{\"o}del number of 
a proof tree in $T$ in a syntactic sense as a symbol sequence.)

PTM $\U_{\PTM}(e_3, \cdot)$ is constructed by using
two PTMs,  
$\U_{\PTM}(e_1, (p,{\#}\Phi,\cdot))$ and 
$\U_{\PTM}(e_2, (p,{\#}\Omega,\cdot))$, and function $h$
as follows:

\begin{enumerate}
\item
(Input: ) $(p,{\#}\Psi,x) \in \N^3$.
\item
(Output: ) G{\"o}del number of a proof tree of $\psi(\xxx)$
or 0.
\item
Run the following computation 
$$\U_{\PTM}(e_1, (p,{\#}\Phi,x))=s, $$
$$\U_{\PTM}(e_2, (p,{\#}\Omega,x))=t. $$
\item
Compute $h(s,t)$ and output the result.
\end{enumerate}

Since function $h$ is primitive recursive,
there exists a $\Delta_1$-formula, $\mu(\sss,\ttt,\uuu)$, in PA 
which represents function $h$ 
such that (from Proposition \ref{c-representability}) 
$$
\PA \vdash \ 
\forall \sss \ \forall \ttt \ \exists !\uuu \ \ \  
\mu(\sss,\ttt,\uuu).
$$

We now introduce function symbol $h$ in PA, then
$$
\PA \vdash \ 
\forall \sss \ \forall \ttt \ \forall \uuu \ \ \ 
\ ( \mu(\sss,\ttt,\uuu) \ \lra \ \uuu = h(\sss,\ttt) \  ).
$$
That is,
$$
\PA \vdash \ 
\forall \sss \ \forall \ttt \ \exists ! \uuu \ \  
\uuu = h(\sss,\ttt). 
$$

Therefore,
for all $e_1 \in \N$ and for all $e_2 \in \N$,
\begin{eqnarray*}  
\PA \vdash \ 
\forall \xxx \ \forall \sss \ \forall \ttt \ \exists ! \uuu \ \ 
( \ & & 
\mbox{\PTM-Out}(\eee_1, \lc \Phi \rc, \xxx, \sss)
 \ \land \ 
\mbox{\PTM-Out}(\eee_2, \lc \Omega \rc, \xxx, \ttt)
\ )
\\
& & 
\ra \ \ 
( \
\mbox{\PTM-Out}(\eee_1, \lc \Phi \rc, \xxx, \sss)
 \ \land \ 
\mbox{\PTM-Out}(\eee_2, \lc \Omega \rc, \xxx, \ttt)
\ \ \land \ \
\uuu = h(\sss,\ttt)
 \  ).
\end{eqnarray*}

See Section \ref{formal-ptm} for the definition of notation 
$\mbox{\PTM-Out}(\cdot)$.


Then, by the construction of
$\U_{\PTM}(e_3,\cdot)$,
for all $e_1 \in \N$ and for all $e_2 \in \N$,
there exists $e_3 \in \N$ such that 
\begin{eqnarray*}  
\PA \vdash \  & &  
\forall \xxx \ \forall  \uuu 
\\
& & 
( \ \exists ! \sss \ \exists !  \ttt \ 
( 
\mbox{\PTM-Out}(\eee_1, \lc \Phi \rc, \xxx, \sss)
 \ \land \ 
\mbox{\PTM-Out}(\eee_2, \lc \Omega \rc, \xxx, \ttt)
\ \land \ 
\uuu = h(\sss,\ttt) 
 \  )
\ )
\\
& & \mbox{ } \ 
\ra \ \ \ 
\mbox{\PTM-Out}(\eee_3, \lc \Psi \rc, \xxx, \uuu)
\ ).
\end{eqnarray*}

Therefore,
for all $e_1 \in \N$ and for all $e_2 \in \N$,
there exists $e_3 \in \N$ such that 
\begin{eqnarray}\label{eq:D2-1}  
\PA \vdash \ 
\forall \xxx \ \forall \sss \ \forall \ttt \ \exists ! \uuu
\ ( \ & & 
\mbox{\PTM-Out}(\eee_1, \lc \Phi \rc, \xxx, \sss)
 \ \land \ 
\mbox{\PTM-Out}(\eee_2, \lc \Omega \rc, \xxx, \ttt)
\nonumber 
\\
& & 
\ra \ \ 
\mbox{\PTM-Out}(\eee_3, \lc \Psi \rc, \xxx, \uuu).
\end{eqnarray}

On the other hand,
a polynomial-time computation (relation) 
of $\U_{\PTM}(v_T,({\#}\psi,u))$
over $(v_T, {\#}\psi, u)$
is composed of two computation parts as follows:
\begin{enumerate}
\item
If $u = 0$, reject. 
Otherwise, 
check whether there exists a proof tree $\pi$,  
in which $u = {\#}\pi$, and 
whether 
the inference of the root node of $\pi$ 
is correct.
If both of them are valid,
go to next step.
(For example,
if $\pi = <r_0, r_1>[\pi_1, \pi_2]$,
then check whether inference from $(\pi_1, \pi_2)$
to $r_0$ by the rule of inference $r_1$ is correct. 
If $u = h(s,t)$ and $h(s,t) \not= 0$,
then $s={\#}\pi_1$, $t={\#}\pi_2$,
$r_0 = \psi(\xxx)$ and $r_1$ is Modus Ponens.
Hence, if the inference is correct,  
$\pi_2$ should be $\pi_1 \ra \psi$.
)

\item
Let $\pi_1$ and $\pi_2$ be parent nodes of the root node
of $\pi$.
Check whether $\pi_1$ and $\pi_2$ are valid
proof trees in $T$. 
\end{enumerate}

Since the first computation part is primitive recursive,
we can construct formula $\nu(\uuu)$
to represent the first computation (relation) part of
$\U_{\PTM}(v_T,({\#}\psi,u))$,
over $u \in \N$. 
Then from the definition of $h$, we obtain  
$$
\PA \vdash \ 
\forall \sss \ \forall \ttt \ \exists ! \uuu \ \ 
( \ \nu(\uuu) \ \land \ \uuu = h(\sss,\ttt) \ ).
$$

Then,
\begin{eqnarray*}  
\PA \vdash \ \ \ 
\forall \xxx \ \forall \sss \ \forall \ttt \ \exists ! \uuu \ 
( \ 
& & 
( 
\mbox{\PTM-\Acpt}(\vvv_T, {\#}\phi(\xxx), \sss)
 \ \land \  
\mbox{\PTM-\Acpt}(\vvv_T, 
{\#}\phi(\xxx) \ra \psi(\xxx), \ttt)
 )
\\
& & 
\ra \ \ 
( \ 
\mbox{\PTM-\Acpt}(\vvv_T, {\#}\phi(\xxx), \sss)
 \ \land \  
\mbox{\PTM-\Acpt}(\vvv_T, 
{\#}\phi(\xxx) \ra \psi(\xxx), \ttt)
\\
& & \mbox{ \ \ }
\ \land \
( \ \nu(\uuu) \ \land \ \uuu = h(\sss,\ttt) \ )
\ ) .
\end{eqnarray*}
See Section \ref{formal-ptm} for the definition of notation,
$\mbox{\PTM-\Acpt}(\cdot)$.

Here,
\begin{eqnarray*}  
\PA \vdash \ \ & & 
\forall \xxx \ \forall \sss \ \forall \ttt  \ \exists ! \uuu \ \ \ 
\\
& & 
( \  
\mbox{\PTM-\Acpt}(\vvv_T, {\#}\phi(\xxx), \sss)
 \ \land \ \  
\mbox{\PTM-\Acpt}(\vvv_T, 
{\#}\phi(\xxx) \ra \psi(\xxx), \ttt)
\ \ \land \ \
( \ \nu(\uuu) \ \land \ \uuu = h(\sss,\ttt) \ )
\  )
\\
& & \mbox{ } 
\ra \ \ \ 
\mbox{\PTM-\Acpt}(\vvv_T, {\#}\psi(\xxx), \uuu ) 
\ ).
\end{eqnarray*}

Therefore, 
\begin{eqnarray}\label{eq:D2-2}  
\PA \vdash \ \ 
\forall \xxx \ \forall \sss \ \forall \ttt \ \exists ! \uuu \ \ \ 
( \ 
& & 
\mbox{\PTM-\Acpt}(\vvv_T, {\#}\phi(\xxx), \sss)
 \ \land \  
\mbox{\PTM-\Acpt}(\vvv_T, {\#}\phi(\xxx) \ra \psi(\xxx), \ttt)
\nonumber 
\\
& & 
\ra
\mbox{\PTM-\Acpt}(\vvv_T, {\#}\psi(\xxx), \uuu ) 
\ ).
\end{eqnarray}

Hence, combining Eqs. (\ref{eq:D2-1}) and (\ref{eq:D2-2}),
for all $e_1 \in \N$ and for all $e_2 \in \N$,
there exists $e_3 \in \N$ such that 
\begin{eqnarray*}  
\PA \vdash \ \ & & 
\forall \xxx \ \forall \sss \ \forall \ttt \ \exists ! \uuu \ \ \ 
\\
& & 
( \ 
( \ 
\mbox{\PTM-Out}(\eee_1, \lc \Phi \rc, \xxx, \sss)
 \ \land \ 
\mbox{\PTM-\Acpt}(\vvv_T, {\#}\phi(\xxx), \sss)
\ ) 
\\
& & 
\land \ \ 
( \ 
\mbox{\PTM-Out}(\eee_2, \lc \Omega \rc, \xxx, \ttt)
\ \land \  
\mbox{\PTM-\Acpt}(\vvv_T, 
{\#}\phi(\xxx) \ra \psi(\xxx), \ttt)
\ )
\\
& & 
\ra \ \ \ 
\mbox{\PTM-Out}(\eee_3, \lc \Psi \rc, \xxx, \uuu)
\ \ \land \ \ 
\mbox{\PTM-\Acpt}(\vvv_T, {\#}\psi(\xxx), \uuu ) 
\ ).
\end{eqnarray*}

Therefore, finally we obtain that 
for all $e_1 \in \N$ and for all $e_2 \in \N$,
there exists $e_3 \in \N$ such that 
$$
\PA \vdash \ \ 
\forall \xxx \ \ 
( \ 
 \mbox{\Prv}_{T}[\phi(\xxx)](\eee_1, \lc \Phi \rc, \xxx) \ \land
 \mbox{\Prv}_{T}[\phi(\xxx) \ra \psi(\xxx)]
(\eee_2, \lc {\Omega} \rc, \xxx) 
\ \ \ra \ \ 
 \mbox{\Prv}_{T}[\psi(\xxx)](\eee_3, \lc \Psi \rc, \xxx) \
).
$$

\begin{flushright}
$\dashv$
\end{flushright}

\end{proof}

\begin{corollary}\label{D-col1}
Let 
$\Phi \equiv 
\{ \phi(\aaa) \mid  a \in \N \}$ and 
$\Psi \equiv 
\{ \psi(\aaa) \mid  a \in \N \}$.
Suppose that $T$ is a consistent PT-extension of PA.
We assume
$$ 
T \vdash \ 
\forall \xxx \ (\phi(\xxx) \ra \psi(\xxx)).
$$
Then, for all $e_1 \in \N$ there exists $e_2 \in \N$ such that  
$$  
\PA \vdash \ \ \
\forall \xxx  \
( \
 \mbox{\Prv}_{T}[\phi(\xxx)](\eee_1, \lc \Phi \rc, \xxx) \ 
\ra \
 \mbox{\Prv}_{T}[\psi(\xxx)](\eee_2, \lc \Psi \rc, \xxx) \ 
).
$$

\end{corollary}

\begin{proof}
From the first derivability condition (D.1)
of a traditional proof theory \cite{Barwise77}
and the assumption of this lemma,
we obtain
$$ 
\PA \vdash \ 
\mbox{\Prv}_{T}(\lc \forall \xxx \ (\phi(\xxx) \ra \psi(\xxx)) \rc).
$$

Then,
PTM $\U_{\PTM}(e_2, \cdot )$ is constructed by using
PTM $\U_{\PTM}(e_1, (p, {\#}\Phi,\cdot))$
as follows:

\begin{enumerate}
\item
(Input :) \ 
$(p, {\#}\Psi, x)$ 
\item
(Output: )
G{\"o}del number of a proof tree of $\psi(\xxx)$
or 0.
\item
Run the following computation 
$$\U_{\PTM}(e_1, (p,{\#}\Phi,x))=z, \ \  
\U_{\PTM}(v_{T},({\#}\phi(\xxx),z)).$$
\item
Compute the proof (say $\pi_2$) of 
$\forall \yyy \ (\phi(\yyy) \ra \psi(\yyy))$,
since there exists a proof for the predicate from the assumption.
(The computation time is finite i.e., constant in 
$\Size_{\Phi}(x)$ and $\Size_{\Psi}(x)$.) 
\item
Check whether $\U_{\PTM}(v_{T},({\#}\phi(\xxx),z))$
accepts or rejects.
If it rejects, output 0 and halt.
If it accepts, then combine $\pi_1$ ($z = {\#}\pi_1$) and
$\pi_2$ and make a new proof tree, $\pi_3$, for 
$\psi(\xxx)$,
as follows: 
$$
\pi_3 \ \ \equiv \ \ 
<\psi(\xxx),\mbox{Modus Ponens}> 
[\pi_1, <\phi(\xxx) \ra \psi(\xxx), \mbox{Modus Ponens}> 
[\pi_2, \mbox{Axiom X}] ], 
$$
where Axiom X is a logical axiom, 
``$\forall \yyy \ (\phi(\yyy) \ra \psi(\yyy))
\ \ra \ (\phi(\xxx) \ra \psi(\xxx))$''.
\item
Output $\pi_3$ for the proof tree of formula 
$\psi(\xxx)$.
\end{enumerate}

The other part of the proof can be completed in an analogous 
manner to that in Lemma \ref{D2} except for the constructions of 
functions $h$ and $g$ to meet the above-mentioned 
construction of $\U_{\PTM}(e_2, \cdot)$) in this proof.

\begin{flushright}
$\dashv$
\end{flushright}

\end{proof}

\begin{corollary}\label{D-col2}
Let 
$\Phi \equiv 
\{ \phi(\aaa) \mid  a \in \N \}$,
$\Psi \equiv 
\{ \psi(\aaa) \mid  a \in \N \}$, and
$\Omega \equiv 
\{ \phi(\aaa) \land \psi(\aaa) \mid  \ a \in \N \}$.
Suppose that $T$ is a PT-extension of PA.
For all $e_1 \in \N$ and all $e_2 \in \N$,
there exists $e_3 \in \N$ such that 
$$ 
T \vdash \ \ \
\forall \xxx \ \ 
( \ 
 \mbox{\Prv}_{T}[\phi(\xxx)](\eee_1, \lc \Phi \rc, \xxx) \ \land
 \mbox{\Prv}_{T}[\psi(\xxx)](\eee_2, \lc \Psi \rc, \xxx) 
$$
$$
\ra \
 \mbox{\Prv}_{T}[\phi(\xxx) \land \psi(\xxx)]
(\eee_3, \lc {\Omega} \rc, \xxx) \ 
).
$$
\end{corollary}

\begin{proof}

By using the following logical axiom of first order logic:  
$$ \phi \ra (\psi \ra (\phi \land \psi)), $$
and the derivability condition D.1. 
of the standard proof theory \cite{Barwise77},
we can obtain
$$ 
T \vdash \ 
\mbox{\Prv}_{T}(\lc \forall \yyy \ 
(\phi(\yyy) \ra (\psi(\yyy) \ra (\phi(\yyy) \land \psi(\yyy))) \rc).
$$
By applying Corollary \ref{D-col1}, we obtain
\begin{eqnarray*}   
T \vdash \ \ \
\forall \xxx  
( \ 
( & & 
 \mbox{\Prv}_{T}[\phi(\xxx)](\eee_1, \lc \Phi \rc, \xxx) \ \land
 \mbox{\Prv}_{T}[\psi(\xxx)](\eee_2, \lc \Psi \rc, \xxx) 
)
\\
& & 
\ra \
(
 \mbox{\Prv}_{T}
[(\psi(\xxx) \ra (\phi(\xxx) \land \psi(\xxx))]
(\eee_3', \lc {\Omega'} \rc, \xxx) 
\ \land
 \mbox{\Prv}_{T}[\psi(\xxx)](\eee_2, \lc \Psi \rc, \xxx) 
)
\\
& & 
\ra \
 \mbox{\Prv}_{T}[\phi(\xxx) \land \psi(\xxx)]
(\eee_3, \lc {\Omega} \rc, \xxx) \ 
).
\end{eqnarray*}

\begin{flushright}
$\dashv$
\end{flushright}

\end{proof}

\begin{lemma}\label{D3}
(D.3 of PTPs) \ \ \ 
Let $R$ be a polynomial-time relation over $\N$.
Let formula $\rho_R(x)$ (in which only $x$ occurs free)
polynomial-time represent
relation $R$ in theory $T$, and 
the concrete form of formula $\rho_R(\xxx)$ 
follow the construction given in the proof of Theorem \ref{representability}.
Let ${\cal R} \equiv
\{ \rho_R(\aaa) \mid  a \in \N \}$.
Suppose that $T$ is a consistent PT-extension of PA.
Then, there exists $e \in \N$ such that 
$$ 
\PA \vdash \
\forall \xxx \ 
( \
\rho_R(\xxx) \ \ra \ 
\mbox{\Prv}_{T}[\rho_R(\xxx)](\eee, \lc {\cal R} \rc, \xxx)
\ ).
$$
\end{lemma}
\begin{proof}

Here we will follow the notations employed in 
the proof of Theorem \ref{representability}.  

Formula $\rho_R(\xxx)$ has two atomic functions,
$| \cdot |$ and $\Bit(\cdot)$, and three atomic formulas 
$\eta_0(\cdot)$, $\eta_1(\cdot)$ and $\eta_2(\cdot)$.
Since $\rho_R(\xxx)$ is a $\Delta_1$-formula,
these atomic functions and formulas are composed by
a finite number of
logical symbols, $\land$, $\lor$, $\ra$, $\neg$,
and bounded quantifiers. 
Here bounded quantifiers can be replaced by
a finite number of $\land$ and $\lor$.

Hence, by applying Lemma \ref{D2} and Corollaries
\ref{D-col1} and \ref{D-col2},
formula $\mbox{\Prv}_{T}[\rho_R(\xxx)](\eee, \lc {\cal R} \rc, \xxx)$
can be deduced from a logical composition of 
the corresponding atomic formulas,
\begin{eqnarray*}  
&  & \mbox{\Prv}_{T}[\www = | \zzz | ]
(\eee_3, \lc {\cal L} \rc, (\zzz,\www)),
 \ \ \
\mbox{\Prv}_{T}[\www = \Bit(\zzz)]
(\eee_4, \lc {\cal B} \rc, (\zzz,\www)), 
\\
& & 
\mbox{\Prv}_{T}[\eta_i(\zzz,\www,\vvv)]
(\eee_i, \lc {{\cal E}_i} \rc, (\zzz,\www,\vvv)), \ \
\\
& & 
\mbox{\Prv}_{T}[\neg( \www = | \zzz | )]
(\eee_3^*, \lc {\cal L}^* \rc, (\zzz,\www)),
 \ \ \
\mbox{\Prv}_{T}[\neg( \www = \Bit(\zzz) )]
(\eee_4^{*}, \lc {\cal B} \rc, (\zzz,\www)),
\\
& & 
\mbox{\Prv}_{T}[\neg \eta_i(\zzz,\www,\vvv)]
(\eee^*_i, \lc {{\cal E}_i}^* \rc, (\zzz,\www,\vvv)), \ \
\end{eqnarray*}
where $i = 0,1,2$.

Therefore, to prove this Lemma
it is sufficient to prove the following
atomic formulas:
\begin{eqnarray*}  
& &  
\exists \eee \in \N \ \ \  
\PA \vdash \ 
\forall \zzz \ \forall \www \
( \ 
(\www = | \zzz | ) \ 
\ra \ 
\mbox{\Prv}_{T}[\www = | \zzz | ]
(\eee, \lc {\cal L} \rc, (\zzz,\www))
\ ),
\\
& & 
\exists \eee \in \N \ \ \  
\PA \vdash \
\forall \zzz \ \forall \www \
( \
(\www = \Bit(\zzz)) \ 
\ra \ 
\mbox{\Prv}_{T}[\www = \Bit(\zzz)]
(\eee, \lc {\cal B} \rc, (\zzz,\www))
\ ),
\\
& & 
\exists \eee \in \N \ \ \  
\PA \vdash \
\forall \zzz \ \forall \www \ \forall \vvv \
( \
\eta_i(\zzz) \
\ra \ 
\mbox{\Prv}_{T}[\eta_i(\zzz,\www,\vvv)]
(\eee, \lc {{\cal E}_i} \rc, (\zzz,\www,\vvv))
\ ),
\\
& & 
\exists \eee \in \N \ \ \   
\PA \vdash \ 
\forall \zzz \ \forall \www \
( \ 
\neg( \www = | \zzz | )  \ 
\ra \ 
\mbox{\Prv}_{T}[\neg( \www = | \zzz | ]
(\eee, \lc {\cal L}^* \rc, (\zzz,\www))
\ ),
\\
& & 
\exists \eee \in \N \ \ \  
\PA \vdash \
\forall \zzz \ \forall \www \
( \
\neg(\www = \Bit(\zzz)) \ 
\ra \ 
\mbox{\Prv}_{T}[\neg(\www = \Bit(\zzz))]
(\eee, \lc {\cal B}^* \rc, (\zzz,\www))
\ ),
\\
& & 
\exists \eee \in \N \ \ \   
\PA \vdash \
\forall \zzz \ \forall \www \ \forall \vvv \
( \
\neg\eta_i(\zzz,\www,\vvv) \
\ra \ 
\mbox{\Prv}_{T}[\neg\eta_i(\zzz,\www,\vvv)]
(\eee, \lc {{\cal E}_i}^* \rc, (\zzz,\www,\vvv))
\ ),
\end{eqnarray*}
where $i = 0,1,2$.

We will then show a construction of 
$\U_{\PTM}(e, \cdot )$ that outputs a proof tree of
each atomic formula.

First, $\U_{\PTM}(e, \cdot )$ for atomic function 
$| \cdot |$ is as follows: 
\begin{enumerate}
\item (Input:) $(p, {\#}{\cal L}, z, w)$, where
${\cal L} \equiv \{  \aaa = | \bbb | \ \mid \ a \in \N, b \in \N \}$
\ \ \
\item (Output:) ${\#}\pi_L$ or 0,  
where $\pi_L$ is a proof tree 
of formula $\www = | \zzz |$ in PA,
and 0 means ``Fail''.
Note that $\zzz$ and $\www$ are given in binary form such as
\begin{eqnarray*}  
& & 
\zzz_0 + \zzz_1\cdot\2 + \cdots + \zzz_{w-1}\cdot \2^{\www \hik \1}
\\
& & \mbox{ \ \ }
(\mbox{more precisely,} \  
\zzz_0 + \zzz_1\cdot\Se\Se\mbox{\bm $0$} + \cdots +
\zzz_{n-1}\cdot 
\Se\Se\mbox{\bm $0$}\cdot\Se\Se\mbox{\bm $0$} 
\cdots \cdot\Se\Se\mbox{\bm $0$}),
\\
& & 
\www_0 + \www_1\cdot\2 + \cdots + \www_{\ttt-1}\cdot \2^{\ttt \hik \1}.
\end{eqnarray*}  

\item
Check whether $\2^{\www \hik \1} \leq \zzz < \2^{\www}$
or not.
If it is false, output 0.
Otherwise, go to next step.

\item
Make a proof tree, $\pi_L$, of
$\2^{\www \hik \1} \leq \zzz < \2^{\www}$
by showing 
$\zzz' = \zzz_0 + \zzz_1\cdot\2 + \cdots + \zzz_{w-2}\cdot \2^{\www \hik \2}$
and
$\zzz'' = \ov{\zzz_0} + \ov{\zzz_1}\cdot\2 + \cdots + 
\ov{\zzz_{w-2}}\cdot \2^{\www \hik \2}$,
along with the proof tree of
$\zzz = \2^{\www \hik \1} + \zzz'$
and 
$\zzz + \zzz'' = \2^{\www}$,
where $\ov{\zzz_i}$ denotes the complement of $\zzz_i$
(e.g., if $\zzz_i=\0$, $\ov{\zzz_i}=\1$).

(Note that $\www$ in the above equations is expressed in
binary form.)

\end{enumerate}
 
$\mbox{\PTM-Out}(\eee, \lc {\cal L} \rc, \zzz, \www, \yyy)$
represents the above-mentioned computation (function) of 
$\U_{\PTM}(e,$ $(p,$ ${\#}{\cal L},$ $z, w) )$
to output $y$ such that $y = {\#}\pi_L$ or $y=0$.
(For $\mbox{\PTM-Out}$, see Section \ref{formal-ptm}).
From the definition of $| \cdot |$,
$$\PA \vdash \ \forall \zzz \forall \www \
(\www = | \zzz | \lra 
(\2^{\www \hik \1} \leq \zzz < \2^{\www}) ).$$
In addition, from the construction of
$\U_{\PTM}(e, (p, {\#}{\cal L}, z, w) )$,
$$ \PA \vdash \ \forall \zzz \forall \www \ 
(\2^{\www \hik \1} \leq \zzz < \2^{\www}) \ 
\ra \
\mbox{\PTM-Out}(\eee, \lc {\cal L} \rc, \zzz, \www, \lc \pi_L \rc).
$$
Since 
$\mbox{\PTM-\Acpt}(\vvv_T, \lc \zzz = | \www | \rc, \lc \pi_L \rc )$
represents 
computation $\U_{\PTM}(v_T, ({\#}\zzz = | \www |,{\#}\pi_L))$
(see Section \ref{formal-ptm}),
$$  
\PA \vdash \ \forall \zzz \ \forall \www \ \exists ! \yyy \ \   
(\www = | \zzz | \ \land \  
\mbox{\PTM-Out}(\eee, \lc {\cal L} \rc, \zzz, \www, \yyy) )
\ra \ 
\mbox{\PTM-\Acpt}(\vvv_T, \lc \www = | \zzz | \rc, \yyy )
\ ) .
$$
Namely,
$$ 
\PA \vdash \
\forall \zzz \ \forall \www \
( \
(\www = | \zzz |) \ 
\ra \ 
\mbox{\Prv}_{T}[\www = | \zzz |]
(\eee, \lc {\cal L} \rc, (\zzz,\www))
\ ).
$$

We can also prove similar results on 
$\neg( \www = | \zzz | )$,
$(\www = \Bit(\zzz))$, and 
$\neg(\www = \Bit(\zzz))$ 
in a manner similar to that on $( \www = | \zzz | ) $.

We will now prove the results on 
formulas $\eta_i$ $(i=0,1,2)$.
 
Since the values of variables 
of formula $\eta_i$ $(i=0,1,2)$ are bounded by constant 
$\kkk$ and the number of variables is also
bounded by 3,
all possible evaluation values of $\eta_i$ $(i=0,1,2)$
with possible values of variable are 
bounded by a constant.
This means that a proof of each possibility of formula  
$\eta_i$ $(i=0,1,2)$ can be created ahead of time and
stored by PTM $\U_{\PTM}(e, \cdot)$.
So, the role of PTM $\U_{\PTM}(e, \cdot)$
is just pattern matching
against the value of the input variables.
   
Given an input value,
$\U_{\PTM}(e, \cdot)$ outputs the G{\"o}del number of 
a proof tree of formula $\eta_i$ $(i=0,1,2)$
as follows:
\begin{enumerate}
\item (Input:) $(p, \lc {\cal E}_i \rc, z,w,v)$, where
${{\cal E}_i}  \equiv 
\{ \eta_i(\aaa,\bbb,\ccc) \ \mid \ a \in \N, b \in \N, c \in \N \}$
\item (Output:) ${\#}\pi_{E,i}(\zzz,\www,\vvv)$ or 0, where 
$\pi_{E,i}(\zzz,\www,\vvv)$ is a proof tree
of formula $\eta_i(\zzz,\www,\vvv)$ in theory PA.
\item (Preprocessing Phase before getting Input)
List up all input values of $(z,w,v)$
for which $\eta_i(\zzz,\www,\vvv)$ is true
(say the list ``TList'').
Make (the G{\"o}del number of) 
a proof tree, $\pi_{E,i}(\zzz,\www,\vvv)$, 
of $\eta_i(\zzz,\www,\vvv)$ 
for all values of $(z,w,v) \in$ TList.
Make a list of (the G{\"o}del number of) the proof trees along with TList,
which is retrieved by entry $(z,w,v)$
(say PList; $\{((z,w,v), {\#}\pi_{E,i}(\zzz,\www,\vvv)) \mid (z,w,v) \in 
\mbox{TList} \}$ ).
Note that the size of PList is finite and constant
in the size of input $\xxx$ to $\rho_R(\cdot)$.
\item 
Given input $(z,w,v)$,
search PList by the input.
If entry $(z,w,v)$ is found in PList,
output the corresponding ${\#}\pi_{E,i}(\zzz,\www,\vvv)$.
Otherwise, output 0.
\end{enumerate}

Let $\mbox{\PTM-Out}(\eee, \lc {\cal E}_i \rc, \zzz, \www, \vvv, \yyy)$
be a formula to represent the computation,
$\U_{\PTM}(e, (p, \lc {\cal E}_i \rc,$ $z,$ $w,v))=y$,
where 
$y=\pi_{E,i}(\zzz,\www,\vvv)$ or $y=0$.

Since the computation is just pattern matching, 
the formula should be effectively 
equivalent to the following form:
\begin{eqnarray*}  
\forall \zzz \ \forall \www \ \forall \vvv \ \exists ! \yyy \ \ 
& & 
( \ 
( (\zzz,\www,\vvv) = (\zzz_0,\www_0,\vvv_0) \
\ra \ \yyy=\pi_0)
\\
& & 
\land \
( (\zzz,\www,\vvv) = (\zzz_1,\www_1,\vvv_1) \
\ra \ \yyy=\pi_1)
\\
& & \mbox{ } \ \ \
\ldots
\\
& & \mbox{ } \ \ \
\ldots
\\
& & 
\land \
( (\zzz,\www,\vvv) = (\zzz_K,\www_K,\vvv_K) \
\ra \ \yyy=\pi_K),
\\
& & 
\land \
( (\zzz,\www,\vvv) \not= (\zzz_0,\www_0,\vvv_0) \
\land \ (\zzz,\www,\vvv) \not= (\zzz_1,\www_1,\vvv_1) \
\cdots \ 
\\
& & \mbox{  \ \ \  \ \  \ } \ 
\land \ (\zzz,\www,\vvv) \not= (\zzz_K,\www_K,\vvv_K) \
\ra \ \yyy=\0) 
\ ),
\end{eqnarray*}
where
TList $\equiv \{ (z_0,w_0,v_0), (z_1,w_1,v_1), \ldots,
(z_K,w_K,v_K) \}$.

From the construction,
\begin{eqnarray*}  
\PA \vdash \ \ 
& & 
\forall \zzz \ 
\forall \www \ \forall \vvv \
( \ 
\eta_i(\zzz,\www,\vvv) \ \
\\
& & 
\lra \ \ ( \
(\zzz,\www,\vvv) = (\zzz_0,\www_0,\vvv_0) \
\lor \ (\zzz,\www,\vvv) = (\zzz_1,\www_1,\vvv_1) \ \cdots \ 
\lor \ (\zzz,\www,\vvv) = (\zzz_K,\www_K,\vvv_K) 
\ ) 
\ ). 
\end{eqnarray*}

For all $i$ ($0 \leq i  \leq K$),
$$ 
\PA \vdash \ \ \forall \zzz \ \forall \www \ \forall \vvv \
( \ 
(\zzz,\www,\vvv) = (\zzz_i,\www_i,\vvv_i) \
\ra \ \ 
\mbox{\PTM-Out}(\eee, \lc {\cal E}_i \rc, \zzz, \www, \vvv, \lc \pi_i \rc) 
\ ).
$$
For all $i$ ($0 \leq i  \leq K$),
$$ 
\PA \vdash \ \ \forall \zzz \ \forall \www \ \forall \vvv \
( \ 
(\zzz,\www,\vvv) = (\zzz_i,\www_i,\vvv_i) \
\ra \ \ 
\mbox{\PTM-\Acpt}(\vvv_T, \lc \eta_i(\zzz,\www,\vvv) \rc, \lc \pi_i \rc)
\ ).
$$

Hence, 
$$
\PA \vdash \ \forall \zzz \ \forall \www \ \forall \vvv \ \exists ! \yyy \ \ 
( \ 
\eta_i(\zzz,\www,\vvv) \
\ra \ 
\mbox{\PTM-Out}(\eee, \lc {\cal E}_i \rc, \zzz, \www, \vvv, \yyy) 
 \ \land \
\mbox{\PTM-\Acpt}(\vvv_T, \lc \eta_i(\zzz,\www,\vvv) \rc, \yyy)
\ ) .
$$
Namely,
$$ 
T \vdash \
\forall \zzz \ \forall \www \
( \
\eta_i(\zzz,\www,\vvv) \
\ra \ 
\mbox{\Prv}_{T}[\eta_i(\zzz,\www,\vvv)]
(\eee, \lc {{\cal E}_i} \rc, \zzz,\www,\vvv)
\ ).
$$

We can also prove a similar result on 
$\neg\eta_i(\zzz,\www,\vvv)$
in a manner 
similar to that on $\eta_i(\zzz,\www,\vvv)$.

\begin{flushright}
$\dashv$
\end{flushright}
\end{proof}

\subsection{Recursion Theorem of Polynomial-Time Proofs}

\begin{proposition}\label{recursion}
(Recursion Theorem)
Let $\U(t,(\cdot,\cdot))$ be a Turing machine that computes 
a two-place function: $\N \times \N \ra \N$.
There exists a Turing machine $\U(k,\cdot)$ (i.e., there exists $k \in \N$) 
that computes a function: $\N \ra \N$, where for every $w \in \N$,
$$\U(k,w) = \U(t, (k,w)).$$
\end{proposition}  
(Note: for notation $\U(\cdot,\cdot)$, see Section \ref{TM}.) 

For the proof of this proposition,
see \cite{Sipser97}(Section 6.1).
The point is that we can construct a Turing machine
$\U_{\PTM}(k,\cdot)$ 
that can read its own code, $k$.
Note that the computational complexity
of reading its own code is constant in (independent from) input size, $|w|$.
$\U(k,\cdot)$, on input $w$, first reads $k$, and then 
simulates $\U(t, (\cdot,\cdot))$ on input $(k,w)$. 

By using this proposition,
we can obtain the PTM version of 
the recursion theorem.

\begin{lemma}\label{ptm-recursion}
(PTM and formula version of Recursion Theorem) \
Given $t \in \N$, let formula $\xi_t(\kkk,\www)$, in which
only $\kkk$ and $\www$ occur free, polynomial-time represent
function $\U_{\PTM}(t, (k,w))$
on $(k,w) \in \N^2$.
Then, for any $t \in \N$, there exists $k \in \N$ 
and formula $\rho_k$
such that 
formula $\rho_k(\www)$, in which
only $\www$ occurs free, polynomial-time represents
function $\U_{\PTM}(k,w)$
on $w \in \N$, and
$$ \PA \vdash 
\forall \www \ \
( \rho_k(\www) \ \lra \ \xi_t(\kkk,\www) )$$

\end{lemma}

\begin{proof}
From the recursion theorem (Proposition \ref{recursion}),
for any $t \in \N$, there exists $k \in \N$ such that
for any $w \in \N$,
$$\U_{\PTM}(k,w) = \U_{\PTM}(t, (k,w)).$$
Here, PTM $\U_{\PTM}(k,\cdot)$ runs  
as follows:
\begin{enumerate}
\item (Input:) \ $w \in \N$ 
\item (Output:) \ accept/reject
\item
First, read its own code, $k \in \N$ via the recursion 
theorem (Proposition \ref{recursion}).
\item
Simulate PTM $\U_{\PTM}(t, (\cdot, \cdot) )$
on input $(k,w)$.
\item
Accept if and only if 
$\U_{\PTM}(t, (k,w))$ accepts.
\end{enumerate}

Therefore, the difference between
$\U_{\PTM}(k,w)$ and $\U_{\PTM}(t,(k,w))$ 
is the step in which $\U_{\PTM}(k,w)$ reads its own code, $k$,
while $\U_{\PTM}(t,(k,w))$ obtains $k$ as an input.

Let $\rho_k(\www)$ polynomial-time represent $\U_{\PTM}(k,w)$. 
Let $\theta(\kkk)$ represent the computation of the step 
in which $\U_{\PTM}(k,w)$ reads its own code, $k$.

Since $\U_{\PTM}(k,w)$ can always reads its own code, $k$,
clearly from Proposition \ref{c-representability}
there exists $k \in \N$ such that
$$ \PA \vdash \ \theta(\kkk).$$

Then, 
there exists $k \in \N$ such that
$$ \PA \vdash \
\forall \www \
( \ 
\theta(\kkk) \land \xi_t(\kkk,\www) 
 \lra \ \xi_t(\kkk,\www) 
\ ).
$$
Since 
formula $(\theta(\kkk) \land \xi_t(\kkk,\www) )$
polynomial-time represents $\U_{\PTM}(k,w)$
from Proposition \ref{composition},
we identify it by $\rho_k(\www)$.
Then, there exists $k \in \N$ and $\rho_k(\www)$ such that
$$ \PA \vdash \
\forall \www \
( \ \rho_k(\www) \ \lra \ \xi_t(\kkk,\www) \ ),
$$

\begin{flushright}
$\dashv$
\end{flushright}
\end{proof}

\subsection{G{\"o}del Sentences of Polynomial-Time Proofs}

\begin{lemma}\label{def-goedel}
Let $T$ be a consistent PT-extension of PA.
Then, for any $e \in \N$, there exists a set of formulas,
${\cal G} \equiv \{ \rho_{e,T}(\aaa) \mid  a \in \N \}$,
such that 
$$ \PA \vdash 
\forall \xxx \ \
( \rho_{e,T}(\xxx) \ \lra \ 
\neg\mbox{\Prv}_{T}[\rho_{e,T}(\xxx)](\eee, \lc \cal G \rc, \xxx) \
).
$$
For all $x$,
$\rho_{e,T}(\xxx)$ is called a ``G{\"o}del sentence'' with respect to
$\mbox{\PTM}$.
\end{lemma}

\begin{proof}
Given $e \in \N$ and theory $T$, 
PTM $\U_{\PTM}(t,(k,x))$ in Lemma \ref{ptm-recursion}
is specialized to this lemma as follows:
\begin{enumerate}
\item (Input:) \ $(k,x) \in \N^2$ 
\item (Output:) \ accept/reject
\item 
Construct formula $\rho_k(\xxx)$
that polynomial-time represents the computation
of $\U_{\PTM}(k,x)$ via the polynomial-time representability theorem
(Theorem \ref{representability}).
Let ${\cal G}_k \equiv \{ \rho_k(\aaa) \mid a \in \N \}$.
\item
Construct PTM $\U_{\PTM}(e, (p, {\#}{\cal G}_k,x))$ to produce
a proof of formula $\rho_k(\xxx)$.
Then, check whether
it outputs a valid proof tree of the input by using 
$\U_{\PTM}(v_T, \cdot)$. 
That is, verify whether the following holds or not:
$$ \PTM_{e}(x) \vdash_T \rho_k(\xxx),$$ 
i.e.,, check 
$$\U_{\PTM}(e, (p, {\#}{\cal G}_k,x))=y  \  \land  \  
\U_{\PTM}(v_{T},({\#}\rho_k(\xxx), y)) \ \mbox{accepts},$$
\item
Accept if and only if the above-mentioned relation ``does not'' 
hold.  
\end{enumerate}

It is clear from the definition of formula
$\mbox{\Prv}_{T}[\cdot](\cdot, \cdot, \cdot)$
in Section \ref{formal-ptm}
that  
$\neg\mbox{\Prv}_{T}[\rho_k(\xxx)](\eee, \lc {\cal G}_k \rc, \xxx)$
represents the above-mentioned relation that 
$\U_{\PTM}(t,(k,x))$ accepts.

Therefore, from Lemma \ref{ptm-recursion},
for any $t \in \N$ (i.e., for any $e \in \N$), 
there exists $k \in \N$ 
and formula $\rho_k$
such that 
$$ \PA \vdash 
\forall \xxx \ \
( \rho_k(\xxx) \ \lra \ 
\neg\mbox{\Prv}_{T}[\rho_k(\xxx)](\eee, \lc {\cal G}_k \rc, \xxx) \ ).
$$
We rename $\rho_k$ as $\rho_{e,T}$,
which is a special symbol for a ``G{\"o}del sentence''
with respect to PTM,
in this paper. (We also rename ${\cal G}_k$ as ${\cal G}$.)

\begin{flushright}
$\dashv$
\end{flushright}
\end{proof}

\subsection{The First Incompleteness Theorem
of Polynomial-Time Proofs}\label{sec:1-rbicthm}

\begin{theorem}\label{rbic-1}
Let $T$ be a consistent PT-extension of PA. 
Let $\rho_{e,T}(\aaa)$ be a G{\"o}del sentence
with respect to PTM, where $a \in \N$.


For all $e \in \N$ and all $x \in \N$,
$$ {\PTM}_{e}(x) \not\vdash_T \ \rho_{e,T}(\xxx).$$ 


\end{theorem}

\begin{proof}

Assuming 
\begin{equation}\label{eq:1-rbicthm-1} 
\exists e \in \N \ \ \exists  x \in \N \ \ \  
{\PTM}_{e}(x) \vdash_T \ \rho_{e,T}(\xxx),
\end{equation}
then
\begin{equation}\label{eq:1-rbicthm-2} 
\exists e \in \N \ \ \exists  x \in \N \ \ \  
\PA  \vdash \ 
\mbox{\Prv}_{T}[\rho_{e,T}(\xxx)](\eee, \lc \cal G \rc, \xxx), \
\end{equation} 
from Lemma \ref{D1}.

On the other hand, 
Eq. (\ref{eq:1-rbicthm-1}) implies
$$ 
\exists e \in \N \ \ \exists  x \in \N \ \ \  
T \vdash \ \rho_{e,T}(\xxx).
$$ 
According to the property of 
the G{\"o}del sentence with respect to PTM (Lemma \ref{def-goedel}),
for any $e \in \N$
$$ 
\PA \vdash 
\forall \xxx \ \
( \rho_{e,T}(\xxx) \ \lra \ 
\neg\mbox{\Prv}_{T}[\rho_{e,T}(\xxx)](\eee, \lc \cal G \rc, \xxx) \
).
$$
Therefore,
\begin{equation}\label{eq:1-rbicthm-3} 
\exists e \in \N \ \ \exists  x \in \N \ \ \  
T \vdash \ 
\neg\mbox{\Prv}_{T}[\rho_{e,T}(\xxx)](\eee, \lc \cal G \rc, \xxx).
\end{equation}

Since $T$ is a consistent PT-extension of PA,
Eq. (\ref{eq:1-rbicthm-2}) contradicts 
Eq. (\ref{eq:1-rbicthm-3}).

Thus,
$$ {\PTM}_{e}(x) \not\vdash_T \ \rho_{e,T}(\xxx).$$

\begin{flushright}
$\dashv$
\end{flushright}
\end{proof}

\subsection{The Second Incompleteness Theorem
of Polynomial-Time Proofs}\label{sec:2-rbicthm}

\begin{theorem}\label{rbic-2nd}
Let $T$ be a consistent PT-extension of PA.
For any $e \in \N$ and any set of formulas
$\Psi \equiv \{ \psi(\aaa) \mid  a \in \N \}$, 
there exists $e^* \in \N$ such that
for any $x \in \N$ 
\begin{eqnarray}\label{eq:rbic-2nd-0}    
{\PTM}_{e}(x) \not\vdash_{T} \
\neg
\mbox{\Prv}_{T}
[\psi(\xxx)]
(\eee^*, \lc \Psi \rc, \xxx).
\end{eqnarray}
\end{theorem}

\begin{proof}

Let ${\cal G} \equiv \{ \rho_{e,T}(\aaa) \mid  a \in \N \}$
be a set of G{\"o}del sentences with respect to $\mbox{\PTM}$.
Let 
${\cal G}^{+} \equiv 
\{ \mbox{\Prv}_{T}[\rho_{e,T}(\aaa)](\eee, \lc {\cal G} \rc, \aaa)
\mid  a \in \N \}$,
${\cal G}^{++} \equiv \{ \neg\rho_{e,T}(\aaa) \mid  a \in \N \}$, and
${\cal G}^{+++} \equiv 
\{ \rho_{e,T}(\aaa) \land \neg\rho_{e,T}(\aaa) \mid  a \in \N \}$.

For any $e \in \N$, there exist
$e^+ \in \N$, $e^{++} \in \N$ and $e^{+++} \in \N$
such that
\begin{eqnarray}\label{eq:rbic-2nd-1}  
\PA \vdash \
\forall \xxx \ \ 
( & & \ \
\mbox{\Prv}_{T}
[\rho_{e,T}(\xxx)]
(\eee, \lc {\cal G} \rc, \xxx)
\nonumber \\ 
&\ra& \ \
\mbox{\Prv}_{T}
[\mbox{\Prv}_{T}[\rho_{e,T}(\xxx)](\eee, \lc {\cal G} \rc, \xxx)]
(\eee^+, \lc {\cal G}^+ \rc, \xxx)
\ \ \
\mbox{(by Lemma \ref{D3})}
\nonumber \\
&\ra& \ \
\mbox{\Prv}_{T}
[\neg\rho_{e,T}(\xxx)]
(\eee^{++}, \lc {{\cal G}^{++}} \rc, \xxx)
\ \ \
\mbox{(by Lemma \ref{def-goedel} and Corollary \ref{D-col1})}
\nonumber \\
&\ra& \ \
\mbox{\Prv}_{T}
[\rho_{e,T}(\xxx)]
(\eee, \lc {\cal G} \rc, \xxx)
\land \ 
\mbox{\Prv}_{T}
[\neg\rho_{e,T}(\xxx)]
(\eee^{++}, \lc {\cal G}^{++} \rc, \xxx)
\nonumber \\ 
&\ra& \ \
\mbox{\Prv}_{T}
[\rho_{e,T}(\xxx) \land \neg\rho_{e,T}(\xxx)]
(\eee^{+++}, \lc {\cal G}^{+++} \rc, \xxx)
\ \ 
).
\ \ \
\mbox{(by Corollary \ref{D-col2})}
\end{eqnarray}

For any formula family $\Psi \equiv \{ \psi(\aaa) \mid  a \in \N \}$,
\begin{equation}\label{eq:rbic-2nd-1x} 
\PA \vdash \
\forall \xxx \
( \
\rho_{e,T}(\xxx) \land \neg\rho_{e,T}(\xxx) \ 
\ra \   \psi(\xxx) \
).
\end{equation}

Hence by Corollary \ref{D-col1},
for any $e^{+++} \in \N$, there exists $e^* \in \N$ 
such that
\begin{equation}\label{eq:rbic-2nd-1y}  
\PA \vdash \
\forall \xxx \
( \
\mbox{\Prv}_{T}
[\rho_{e,T}(\xxx) \land \neg\rho_{e,T}(\xxx)]
(\eee^{+++}, \lc {\cal G}^{+++} \rc, \xxx) \ 
\ra \ 
\mbox{\Prv}_{T}
[\psi(\xxx)]
(\eee^*, \lc \Psi \rc, \xxx) \
).
\end{equation}

Therefore,
for any $e \in \N$, there exists $e^* \in \N$ 
such that
$$ 
\PA \vdash \
\forall \xxx \ \ 
( \ 
\mbox{\Prv}_{T}
[\rho_{e,T}(\xxx)]
(\eee, \lc {\cal G} \rc, \xxx)
\ra \ 
\mbox{\Prv}_{T}
[\psi(\xxx)]
(\eee^*, \lc \Psi \rc, \xxx) \
).
$$
That is,
for any $e \in \N$, there exists $e^* \in \N$ 
such that
$$
\PA \vdash \
\forall \xxx \ \ 
( \ 
\neg\mbox{\Prv}_{T}[\psi(\xxx)](\eee^*, \lc \Psi \rc, \xxx) \
\ra \ 
\neg\mbox{\Prv}_{T}[\rho_{e,T}(\xxx)](\eee, \lc {\cal G} \rc, \xxx)
).
$$
Since $\rho_{e,T}(\xxx)$ is a ``G{\"o}del sentence'' with respect to
$\mbox{\PTM}$, from Lemma \ref{def-goedel}, 
$$ 
\PA \vdash 
\forall \xxx \ \
( \rho_{e,T}(\xxx) \ \lra \ 
\neg\mbox{\Prv}_{T}[\rho_{e,T}(\xxx)](\eee, \lc \cal G \rc, \xxx) \
).
$$
Hence,
for any $e \in \N$, there exists $e^* \in \N$ 
such that
\begin{equation}\label{eq:rbic-2nd-2}  
\PA \vdash \
\forall \xxx \ \ 
( \ 
\neg\mbox{\Prv}_{T}[\psi(\xxx)](\eee^*, \lc \Psi \rc, \xxx) \
\ra \ 
\rho_{e,T}(\xxx)
\ ).
\end{equation}

We now assume that
there exist $e \in \N$ and a formula set $\Psi$
such that 
\begin{equation}\label{eq:rbic-2nd-3}   
\forall e^* \in \N \ \ \exists x \in \N \ \ \ 
{\PTM}_{e}(x) \vdash_T \
\neg\mbox{\Prv}_{T}[\psi(\xxx)](\eee^*, \lc \Psi \rc, \xxx).
\end{equation}

Then, 
PTM $\U_{\PTM}(e',\cdot)$ is constructed 
using PTM $\U_{\PTM}(e,\cdot)$ as follows:
\begin{itemize}
\item
(Input: ) \ $(p, \#{\cal G}', a)  \in \N^3$,
where ${\cal G}' \equiv \{ \rho_{e',T}(\aaa) \mid a \in \N \}$. 
\item
(Output: ) \ G{\"o}del number of a proof tree of $\rho_{e',T}(\aaa)$
or 0.
\item
First, read its own code, $e' \in \N$, via the recursion 
theorem (Proposition \ref{recursion}).
\item
Syntactically check 
whether the input has the form of $(p, {\#}{\cal G}', a)$
and ${\#}{\cal G}' = {\#}\{ \rho_{e',T}(\aaa) \mid a \in \N \}$.
If it is not correct, output 0.
Otherwise, go to the next step.
\item
Find a proof, $\pi$, of formula 
$$
\forall \xxx \ \ 
( \ 
\neg\mbox{\Prv}_{T}[\psi(\xxx)](\eee^{* \prime}, \lc \Psi \rc, \xxx) \
\ra \ 
\rho_{e',T}(\xxx)
\ ),
$$
where there exists $e^{* \prime} \in \N$ such that
a proof of the formula exists, according to Eq. (\ref{eq:rbic-2nd-2}).
Here, the size of $\pi$ is constant in $|a|$.  
\item
Simulate $\U_{\PTM}(e,(p, {\#}\Phi[e^{* \prime}], a))$,
and check whether its output is the G{\"o}del number of a valid proof tree
of $\neg\mbox{\Prv}_{T}[\psi(\aaa)](\eee^{* \prime}, \lc \Psi \rc, \aaa)$ 
by using $\U_{\PTM}(v_{T},\cdot)$,
where
$\Phi[e^{* \prime}] \equiv \{ 
\neg\mbox{\Prv}_{T}[\psi(\aaa)](\eee^{* \prime}, \lc \Psi \rc, \aaa)
\mid  a \in \N \}$.
\item
If it is not a valid proof tree, then output 0.
\item
If it is a valid proof tree (say $\theta$), 
using proofs, $\theta$ and $\pi$,
construct the following proof tree of $\rho_{e',T}(\aaa)$:
$$
< \rho_{e',T}(\aaa), \mbox{Modus Ponens}> \ 
[ \ \theta, \ 
<\neg\mbox{\Prv}_{T}[\psi(\aaa)](\eee^{* \prime}, \lc \Psi \rc, \aaa) \
\ra \ 
\rho_{e',T}(\aaa)
, \mbox{Modus Ponens}> 
$$
$$
[ \ 
\pi, 
\forall \xxx \ \ 
( \ 
\neg\mbox{\Prv}_{T}[\psi(\xxx)](\eee^{* \prime}, \lc \Psi \rc, \xxx) \
\ra \ 
\rho_{e',T}(\xxx)
\ ) 
\ \ \ra \ \ 
( \ 
\neg\mbox{\Prv}_{T}[\psi(\aaa)](\eee^{*  \prime}, \lc \Psi \rc, \aaa) \
\ra \ 
\rho_{e',T}(\aaa)
\ ) 
\ ]
\ ].
$$
Output the G{\"o}del number of the proof tree.
\end{itemize}

The running time of $\U_{\PTM}(e',\cdot)$
is that of $\U_{\PTM}(e,\cdot)$ plus polynomial-time in $|a|$.

Since we assume that Eq. (\ref{eq:rbic-2nd-3})
holds, $\U_{\PTM}(e',(p, {\#}{\cal G}', a))$ 
outputs the G{\"o}del number of a valid proof tree
of $\rho_{e',T}(\aaa)$.
Thus,
$$
\exists x \in \N \ \ \ 
{\PTM}_{e'}(x) \vdash_T \
\rho_{e',T}(\xxx).
$$

This contradicts Theorem \ref{rbic-1}.
Therefore, Eq. (\ref{eq:rbic-2nd-3}) does not hold.
That is,
for any $e \in \N$ and any $\Psi$, there exists $e^* \in \N$ 
such that
for any $x \in \N$
$$ 
{\PTM}_{e}(x) \not\vdash_T \
\neg\mbox{\Prv}_{T}[\psi(\xxx)](\eee^*, \lc \Psi \rc, \xxx).
$$

\begin{flushright}
$\dashv$
\end{flushright}
\end{proof}


\section{Polynomial-Time Decisions}\label{sec:pt-dec}

In order to prove the (resource bounded)
unprovability of ${\ov{\mbox{P$\not=$NP}}}$,
this section introduces our formalization of a decision made  
by a polynomial-time Turing machine (polynomial-time decision: PTD).

\subsection{Polynomial-Time Decisions}\label{pt-dec}

This section introduces the formalization of a decision 
made by a polynomial-time Turing machine 
(polynomial-time decision: PTD).

Let $\Phi \equiv \{ \phi(\aaa) \mid a \in \N  \}$
be a set of an infinite number of formulas in PA.
If $\U_{\PTM}(e,(d,{\#}\Phi,a))$ accepts 
and $\NNN \models \phi(\aaa)$
(i.e., $\phi(\aaa)$ is true in the standard model
of natural numbers),
we then denote 
$$
\PTM_{e}^{\Phi}(a) \dec \phi(\aaa).  
$$ 
(This can be interpreted as 
``$\U_{\PTM}(e,\cdot)$ {\it correctly accepts} $\phi(\aaa)$.)
Here $d$ denotes a natural number (e.g., 1), which
indicates $\U_{\PTM}(e,\cdot)$ that the output target
is a decision on the formula's truth.
In other words, 
\begin{eqnarray*} 
& & \PTM_{e}^{\Phi}(a) \dec \phi(\aaa) \ \ \  
\\
& & \mbox{ \ \ }  \  
\LRa   \ \ \ 
\U_{\PTM}(e,(d,{\#}\Phi,a)) \mbox{accepts} \  
\land  \ \NNN \models \phi(\aaa).
\end{eqnarray*} 

If $\U_{\PTM}(e,(d,{\#}\Phi,a))$ rejects 
and $\NNN \models \neg\phi(\aaa)$
(i.e., $\phi(\aaa)$ is false in the standard model
of natural numbers),
we then denote 
$\PTM_{e}^{\Phi}(a) \dec \neg\phi(\aaa)$.
(This can be interpreted as 
``$\U_{\PTM}(e,\cdot)$ {\it correctly rejects} $\phi(\aaa)$.)
In other words, 
\begin{eqnarray*}
& & 
\PTM_{e}^{\Phi}(a) \dec \neg\phi(\aaa) \ \ \  
\\
& & \mbox{ \ \ }  \  
\LRa   \ \ \ 
\U_{\PTM}(e,(d,{\#}\Phi,a)) \mbox{rejects} \  
\land  \ \NNN \models \neg\phi(\aaa).
\end{eqnarray*} 
Here note that 
$\PTM_{e}^{\Phi}(a) \dec \neg\phi(\aaa)$
is different from $\PTM_{e}^{\Omega}(a) \dec \neg\phi(\aaa)$,
where $\Omega \equiv \{ \neg\phi(\aaa) \mid a \in \N  \}$.
\footnote{
In the notation of polynomial-time proofs,
$$\PTM_{e}(a) \vdash \phi(\aaa),$$  
we omit 
$\Phi \equiv \{ \phi(\aaa) \mid a \in \N  \}$
in a place of $\PTM_{e}(a) \vdash \phi(\aaa)$
(e.g., the upper right position of $\PTM_{e}$),
since ${\Phi}$ is uniquely determined by
the object of the proof, $\phi(\aaa)$.
However, in polynomial-time decisions,
we have two different types of decisions as 
follows, as described above:
\begin{eqnarray*}
& & 
\PTM_{e}^{\Phi}(a) \dec \phi(\aaa), \\ 
& & 
\PTM_{e}^{\Omega}(a) \dec \phi(\aaa), \\ 
\end{eqnarray*}
where 
$\Omega \equiv \{ \neg\phi(\aaa) \mid a \in \N  \}$.
In the former notation, $\phi(\aaa)$
is {\it correctly accepted}, while,
in the latter notation, $\neg\phi(\aaa)$
is {\it correctly rejected}.
Therefore,
in the notation of polynomial-time decisions,
$$\PTM_{e}^{\Phi}(a) \dec \phi(\aaa),$$ 
we {\it cannot} omit ${\Phi}$ in the upper right position 
of $\PTM_{e}$.
}

In addition,
we use the notation 
$\PTM_{e}^{\Phi}(a) \not\dec \phi(\aaa)$
if and only if 
$ \neg( \ \PTM_{e}^{\Phi}(a) \dec \phi(\aaa) \ )$.

We now introduce a relaxed notion of
$\PTM_{e}^{\Phi}(a) \dec \phi(\aaa)$. 
We denote 
$$\PTM_{e}^{\Phi}(a) \dec_v \phi(\aaa)$$ 
if and only if 
$$
\U_{\PTM}(e,(d,{\#}\Phi,a)) \mbox{accepts} \  
\land  \ \U(v,(d,{\#}\Phi,a)) \mbox{accepts}. 
$$

\begin{lemma}\label{lemma-dec-def}
Let $\Phi \equiv \{ \phi(\aaa) \mid a \in \N  \}$
be a set of an infinite number of $\Delta_1$-formulas in PA, and
$\phi$ represent a primitive recursive relation $R_{\Phi}$, 
where there exists a Turing machine $\U(v_{\phi}, \cdot)$ such that,
for every $a \in \N$, 
$$
a \in R_{\Phi} 
\ \ \LRa \ \ 
\U(v_{\phi},(d,{\#}\Phi,a)) \mbox{accepts}.
$$
Then, 
$$ 
\PTM_{e}^{\Phi}(a) \dec_{v_{\phi}}  \phi(\aaa) \ \ \  \LRa \ \ \ 
\PTM_{e}^{\Phi}(a) \dec \phi(\aaa), 
$$
and
$$
\PTM_{e}^{\Phi}(a) \dec_{v_{\phi}}  \neg\phi(\aaa) \ \ \  \LRa \ \ \ 
\PTM_{e}^{\Phi}(a) \dec \neg\phi(\aaa). 
$$

\end{lemma}

\begin{proof}

Since $R_{\Phi}$ is a relation that formula $\phi$ represents,
for every $a \in \N$
$$ 
\U(v_{\phi},(d,{\#}\Phi,a)) \mbox{accepts} 
 \ \ \Ra \ \ 
a \in R_{\Phi}
\ \ \Ra \ \ \PA \vdash \phi(\aaa),
$$
$$
\U(v_{\phi},(d,{\#}\Phi,a)) \mbox{rejects} 
 \ \ \Ra \ \ 
a \not\in R_{\Phi}
\ \ \Ra \ \ \PA \vdash \neg\phi(\aaa).$$
Since $\NNN$ is a model of PA, from the soundness of PA,
for every $a \in \N$
$$ \U(v_{\phi},(d,{\#}\Phi,a)) \mbox{accepts} 
\ \ \Ra \ \ \NNN \models \phi(\aaa),$$
$$\U(v_{\phi},(d,{\#}\Phi,a)) \mbox{rejects} 
\ \ \Ra \ \ \NNN \models \neg\phi(\aaa).$$

\begin{flushright}
$\dashv$
\end{flushright}
\end{proof}

\subsection{Formalization of Polynomial-Time Decisions}
\label{formal-pt-dec}

A formula to represent the relation on polynomial-time decisions,

$ \PTM_{e}^{\Phi}(a) \dec \phi(\aaa) $,
is obtained, in a manner similar to that shown 
in Section \ref{formal-ptm}. 

Let $\Phi \equiv \{ \phi(\aaa) \mid a \in \N  \}$
be a set of an infinite number of $\Delta_1$-formulas in PA.
Let formula 
$$\mbox{\PTM-\Acc}(\eee, \lc \Phi \rc, \aaa)$$
polynomial-time represent
$$\U_{\PTM}(e,(d,{\#}\Phi,a)) \mbox{accepts}$$
over natural numbers, $(e, {\#}\Phi, a)$.

Let formula
$$\Acc(\eee, \lc \Phi \rc, \aaa)$$
represent
$$\U(e,(d,{\#}\Phi,a)) \mbox{accepts}$$
over natural numbers, $(e, {\#}\Phi, a)$.
Here, if $a = (a_1, \ldots, a_k)$, 
we then denote 
$$\Acc(\eee, \lc \Phi \rc, \aaa_1, \ldots, \aaa_k)$$
to represent
$$\U(e,(d,{\#}\Phi,(a_1, \ldots, a_k))) \mbox{accepts}.$$
The method of constructing these formulas is the same
as that described in Section \ref{formal-ptm}.

We then define the following formulas:
for all $a \in \N$,
\begin{eqnarray*}
\CA_v[\phi(\aaa)](\eee, \lc \Phi \rc, \aaa) \ \ \ &\equiv& \ \ \
\mbox{\PTM-\Acc}(\eee, \lc \Phi \rc, \aaa) \ \land \ 
\Acc(\vvv, \lc \Phi \rc, \aaa),
\\
\CR_v[\phi(\aaa)](\eee, \lc \Phi \rc, \aaa) \ \ \ &\equiv& \ \ \
\neg\mbox{\PTM-\Acc}(\eee, \lc \Phi \rc, \aaa) \ \land \ 
\neg\Acc(\vvv, \lc \Phi \rc, \aaa),
\\
\CD_v[\phi(\aaa)](\eee, \lc \Phi \rc, \aaa) \ \ \ &\equiv& \ \ \
\CA_v[\phi(\aaa)](\eee, \lc \Phi \rc, \aaa) \ \lor \ 
\CR_v[\phi(\aaa)](\eee, \lc \Phi \rc, \aaa),
\end{eqnarray*}
(Here, CA, CR, and CD stand for `correctly accept',
`correctly reject', and `correctly decide', respectively.) 

We also define the following formulas:
for all $a \in \N$,
\begin{eqnarray*}
\CA[\phi(\aaa)](\eee, \lc \Phi \rc, \aaa) \ \ \ &\equiv& \ \ \
\mbox{\PTM-\Acc}(\eee, \lc \Phi \rc, \aaa) \ \land \ 
\phi(\aaa),
\\
\CR[\phi(\aaa)](\eee, \lc \Phi \rc, \aaa) \ \ \ &\equiv& \ \ \
\neg\mbox{\PTM-\Acc}(\eee, \lc \Phi \rc, \aaa) \ \land \ 
\neg\phi(\aaa),
\\
\CD[\phi(\aaa)](\eee, \lc \Phi \rc, \aaa) \ \ \ &\equiv& \ \ \
\CA[\phi(\aaa)](\eee, \lc \Phi \rc, \aaa) \ \lor \ 
\CR[\phi(\aaa)](\eee, \lc \Phi \rc, \aaa).
\end{eqnarray*}

\begin{lemma}\label{lem:f}
Let $\Omega \equiv \{ \omega(\aaa) \mid a \in \N  \}$
be a set of an infinite number of $\Delta_1$-formulas in PA.

Then, there exists a primitive recursive function $f$
such that 
\begin{eqnarray*}
& & 
\exists y < f({\#}\Omega,\Size_{\Omega}(a)) \ \ \ 
\U_{\PTM}(v_{\PA},({\#}\omega(\aaa),y)) \mbox{accepts}
\ \ \ \LRa \ \ \
\PA \vdash \ \omega(\aaa), \ \ \ \mbox{and}
\\
& & 
\exists z < f({\#}\Omega,\Size_{\Omega}(a)) \ \ \ 
\U_{\PTM}(v_{\PA},({\#}\neg\omega(\aaa),z)) \mbox{accepts}
\ \ \ \LRa \ \ \
\PA \vdash \ \neg\omega(\aaa).
\end{eqnarray*}

\end{lemma}

\begin{proof}
\ \ 
Since $\omega(\aaa)$ is a $\Delta_1$-formula,
there exists a TM $\U(e_0, \cdot )$
such that
$$ 
\forall a \in \N  \  \
( \
\PA \vdash \ \omega(\aaa) \ \Ra \ 
\TM_{e_0}^{\Omega}(a) \vdash_{\PA} \ \omega(\aaa)
\ \ \lor \ \ 
\PA \vdash \ \neg\omega(\aaa) \ \Ra \ 
\TM_{e_0}^{\Omega'}(a) \vdash_{\PA} \ \neg\omega(\aaa)
\ ),
$$ 
where 
${\Omega'} \equiv \{ \neg\omega(\aaa) \mid a \in \N \}$,
and $\Size_{\Omega}(a) = \Size_{\Omega'}(a)$ for all $a \in \N$.

Therefore, there exists another TM $\U(e_1, \cdot )$
such that 
$$ 
\forall a \in \N  \  \
\U(e_1, ({\#}\Omega,a)) = |\pi| \ \land \
( \ 
\U(e_0, (p,{\#}\Omega,a)) = {\#}\pi \ \lor 
\U(e_0, (p,{\#}\Omega',a)) = {\#}\pi .
$$ 
(That is, $\pi$ is a proof tree of $\omega(\aaa)$ or $\neg\omega(\aaa)$,
generated by $\U(e_0, \cdot)$.)

Hence, there exists a TM $\U(e_2, \cdot )$
such that
\begin{eqnarray*}
& & \U(e_2, ({\#}\Omega,n)) \ =    
\max\{ \ \  |\pi(a)| \ \mid \
\\
& & \mbox { \ \ \  } \ \  
a \in \N \ \ \land \ \ n=\Size_{\Omega}(a) \ \ \land \ \ 
( \U(e_0, (p,{\#}\Omega,a)) = {\#}\pi \ \lor \ 
\U(e_0, (p,{\#}\Omega',a)) = {\#}\pi \ ) \ \ \}.
\end{eqnarray*} 
(That is, $\U(e_2, ({\#}\Omega,n))$ computes
the maximum length of proofs that $\U(e_0, \cdot)$
outputs where the input size is $n$.)

Thus, there exists the above-mentioned primitive recursive function 
$f$ that 
is computed by TM $\U(e_2, \cdot)$.

\begin{flushright}
$\dashv$
\end{flushright}
\end{proof}

\begin{definition}\label{def-v-{Omega}*}
Let $\Omega \equiv \{ \omega(\aaa) \mid a \in \N  \}$ and 
$\Phi \equiv \{ \phi(\aaa) \mid a \in \N  \}$
be sets of an infinite number of $\Delta_1$-formulas in PA.

Let 
$\U(v_{\Omega}^A,\cdot)$ be a TM 
as follows:
\begin{itemize}
\item
(Input: ) \ $(d,{\#}{\Phi},a)$
\item
(Output: ) \ accept or reject
\item
Let $\Phi_1 \equiv \{ \phi_1(\aaa) \mid a \in \N  \}$ and 
$\Phi_2 \equiv \{ \phi_2(\aaa) \mid a \in \N  \}$. 
If $\phi(\aaa) \equiv \phi_1(\aaa) \land \phi_2(\aaa)$,
then let 
$\Phi \equiv \{ \phi_1(\aaa) \land \phi_2(\aaa) 
\mid a \in \N^2  \}$, where
$\Size_{\Phi}(a) = \Size_{\Phi_1}(a) + \Size_{\Phi_2}(a)$.       
Then, simulate 
$\U(v_{\Omega}^A,(d,{\#}{\Phi}_1,a))$
and $\U(v_{\Omega}^A,(d,{\#}{\Phi}_2,a))$.
(Here, whether $\phi(\aaa)$ is the form of 
$\phi_1(\aaa) \land \phi_2(\aaa)$
is syntactically checked by some rule, and
is uniquely decided. For example,
search a formula from left to right and
syntactically check the form based on the leftmost 
$\land$, and if it is not the form then move to the 
right direction to find another $\land$, etc.)   

Accept if and only if both of them accept.

\item
If theorem in PA, 
$$
\PA \vdash \ 
\forall \xxx
( \ \psi(\xxx) \ \ra \ \phi(\xxx) \ ),
$$
is installed in 
$\U(v_{\Omega}^A,\cdot)$,
then 
simulate 
$\U(v_{\Omega}^A,(d,{\#}{\Psi},a))$, 
where $\Psi \equiv \{ \psi(\aaa) \mid  a \in \N \}$
and $\Size_{\Phi}(a) =  \Size_{\Psi}(a) + c$ ($c$: constant).

Accept if and only if $\U(v_{\Omega}^A,(d,{\#}{\Psi},a))$ accepts.

A finite number of the theorems explicitly shown in this paper
are installed in $\U(v_{\Omega}^A,\cdot)$.
\item
Let $\Psi \equiv 
\{ \psi(\aaa) \mid  a \in \N \}$
be a set of an infinite number of $\Delta_1$-formulas in PA.
If $\Phi \equiv 
\{ \mbox{\CA}_{\Omega}[\psi(\aaa)](\eee, \lc \Psi \rc, \aaa)  
\mid  a \in \N \}$, then
simulate $\U_{\PTM}(e, (d, {\#}\Psi, a))$
and $\U(v_{\Omega}^A,(d,{\#}{\Psi},a))$.
Here $\Size_{\Phi}(a) = 2 \cdot\Size_{\Psi}(a)$.

Accept if and only if both of them accept.
 
\item
Unless the above-mentioned cases occur,
check (by exhaustive search for $y < f({\#}\Omega,\Size_{\Phi}(a))$)
whether 
\begin{equation}\label{eq:def-v-{Omega}*-1}
\exists y < f({\#}\Omega,\Size_{\Phi}(a)) \ \ \ 
\U_{\PTM}(v_{\PA},({\#}\phi(\aaa),y)) \mbox{accepts},
\end{equation} 
where $\U_{\PTM}(v_{\PA},\cdot)$ is defined in Section \ref{PT-proofs}, and
$f$ is a primitive recursive function defined in Lemma \ref{lem:f}.

Accept if and only if Eq. (\ref{eq:def-v-{Omega}*-1}) holds.
\end{itemize}

Let 
$\U(v_{\Omega}^R,\cdot)$ be a TM 
as follows:
\begin{itemize}
\item
(Input: ) \ $(d,{\#}{\Phi},a)$
\item
(Output: ) \ accept or reject
\item
Let $\Phi_1 \equiv \{ \phi_1(\aaa) \mid a \in \N  \}$ and 
$\Phi_2 \equiv \{ \phi_2(\aaa) \mid a \in \N  \}$. 
If $\phi(\aaa) \equiv \phi_1(\aaa) \lor \phi_2(\aaa)$,
then let 
$\Phi \equiv \{ \phi_1(\aaa) \lor \phi_2(\aaa) 
\mid a \in \N^2  \}$, where
$\Size_{\Phi}(a) = \Size_{\Phi_1}(a) + \Size_{\Phi_2}(a)$.
Then, simulate 
$\U(v_{\Omega}^R,(d,{\#}{\Phi}_1,a))$
and $\U(v_{\Omega}^R,(d,{\#}{\Phi}_2,a))$.
(Here, whether $\phi(\aaa)$ is the form of 
$\phi_1(\aaa) \lor \phi_2(\aaa)$
is syntactically checked by some rule, and is
uniquely decided.)

Reject if and only if both of them reject.

\item
If theorem in PA, 
$$
\PA \vdash \ 
\forall \xxx
( \ \neg\psi(\xxx) \ \ra \ \neg\phi(\xxx) \ ),
$$
is installed in 
$\U(v_{\Omega}^R,\cdot)$,
then 
simulate 
$\U(v_{\Omega}^R,(d,{\#}{\Psi},a))$, 
where $\Psi \equiv \{ \psi(\aaa) \mid  a \in \N \}$
and $\Size_{\Phi}(a) =  \Size_{\Psi}(a) + c$ ($c$: constant).

Reject if and only if $\U(v_{\Omega}^R,(d,{\#}{\Psi},a))$ rejects.

A finite number of the theorems explicitly shown in this paper
are installed in $\U(v_{\Omega}^R,\cdot)$.

\item
Let $\Psi \equiv 
\{ \psi(\aaa) \mid  a \in \N \}$
be a set of an infinite number of $\Delta_1$-formulas in PA.
If $\Phi \equiv 
\{ \neg\mbox{\CR}_{\Omega}[\psi(\aaa)](\eee, \lc \Psi \rc, \aaa)  
\mid  a \in \N \}$, then
simulate $\U_{\PTM}(e, (d, {\#}\Psi, a))$
and $\U(v_{\Omega}^R,(d,{\#}{\Psi},a))$.
Here $\Size_{\Phi}(a) = 2 \cdot\Size_{\Psi}(a)$.

Reject if and only if both of them reject`.

\item
Unless the above-mentioned cases occur,
check (by exhaustive search for $y < f({\#}\Omega,\Size_{\Phi}(a))$)
whether 
\begin{equation}\label{eq:def-v-{Omega}*-2}
\exists y < f({\#}\Omega,\Size_{\Phi}(a)) \ \ \ 
\U_{\PTM}(v_{\PA},({\#}\neg\phi(\aaa),y)) \mbox{accepts},
\end{equation}
where $\U_{\PTM}(v_{\PA},\cdot)$ is defined in Section \ref{PT-proofs}, and
$f$ is a primitive recursive function defined in Lemma \ref{lem:f}.

Reject if and only if Eq. (\ref{eq:def-v-{Omega}*-2}) holds. 
\end{itemize}

\end{definition}

If $\U(v_{\Omega}^A, \cdot)$ and 
$\U(v_{\Omega}^R, \cdot)$ are TMs defined in Definition 
\ref{def-v-{Omega}*}, 
we simply denote 
by 
\begin{eqnarray*}
& & \CA_{\Omega}[\phi(\aaa)](\eee, \lc \Phi \rc, \aaa)
\ \ \equiv \ \  
\CA_{\vvv_{\Omega}^A}[\phi(\aaa)](\eee, \lc \Phi \rc, \aaa),
\\
& & 
\CR_{\Omega}[\phi(\aaa)](\eee, \lc \Phi \rc, \aaa) 
\ \ \equiv \ \  
\CR_{\vvv_{\Omega}^R}[\phi(\aaa)](\eee, \lc \Phi \rc, \aaa),
\\
& & 
\CD_{\Omega}[\phi(\aaa)](\eee, \lc \Phi \rc, \aaa)
\ \ \equiv \ \  
\CA_{\vvv_{\Omega}^A}[\phi(\aaa)](\eee, \lc \Phi \rc, \aaa)
\ \lor \
\CR_{\vvv_{\Omega}^R}[\phi(\aaa)](\eee, \lc \Phi \rc, \aaa),
\end{eqnarray*}

\begin{definition}\label{def-sound-acc}
We say $\U(v,\cdot)$ ``soundly accepts'' if, 
for any $\Phi \equiv \{ \phi(\aaa) \mid a \in \N \}$,
for all $a \in \N$,
\begin{equation}\label{eq-def-v-{Omega}*-5}
\U(v,(d,{\#}\Phi,a)) \mbox{accepts}
\ \ \  \Ra \ \ \
\NNN \models \phi(a).  
\end{equation}

We say $\U(v,\cdot)$ ``soundly rejects'' if, 
for any $\Phi \equiv \{ \phi(\aaa) \mid a \in \N \}$,
for any $a \in \N$,
\begin{equation}\label{eq-def-v-{Omega}*-6}
\U(v,(d,{\#}\Phi,a)) \mbox{rejects}
\ \ \  \Ra \ \ \
\NNN \models \neg\phi(a).  
\end{equation}

The following lemma is obtained from Definitions \ref{def-v-{Omega}*}
and \ref{def-sound-acc},
and Lemma \ref{lem:f}.

\end{definition}

\begin{lemma}\label{lem-v-{Omega}*}


Let $\U(v_{\Phi}^A,\cdot)$ soundly accept, and
$\U(v_{\Phi}^R,\cdot)$ soundly reject.


For all $a \in \N$,
\begin{equation}\label{eq-def-v-{Omega}*-7} 
\U(v_{\Phi}^A,(d,{\#}\Phi,a)) \mbox{accepts}
\ \ \  \LRa \ \ \ 
\NNN \models \phi(a).
\end{equation} 
\begin{equation}\label{eq-def-v-{Omega}*-8} 
\U(v_{\Phi}^R,(d,{\#}\Phi,a)) \mbox{rejects}
\ \ \  \LRa \ \ \ 
\NNN \models \neg\phi(a).
\end{equation}

\begin{equation}\label{eq-def-v-{Omega}*-1}
\PA \vdash \  
\forall \xxx \ \ ( \
\Acc(\vvv_{\Omega}^A, \lc \Phi \rc, (\xxx,\xxx))
\ \lra \ 
\Acc(\vvv_{\Omega}^A, \lc \Phi_1 \rc, \xxx) \ \land \ 
\Acc(\vvv_{\Omega}^A, \lc \Phi_2 \rc, \xxx),
\end{equation} 
where
$\Phi \equiv \{ \phi_1(\aaa) \land \phi_2(\aaa) 
\mid (a,a) \in \N^2  \}$.
\begin{equation}\label{eq-def-v-{Omega}*-2}
\PA \vdash \  
\forall \xxx \ \ ( \
\neg\Acc(\vvv_{\Omega}^R, \lc \Phi' \rc, (\xxx,\xxx))
\ \lra \ 
\neg\Acc(\vvv_{\Omega}^R, \lc \Phi_1 \rc, \xxx) \ \land \ 
\neg\Acc(\vvv_{\Omega}^R, \lc \Phi_2 \rc, \xxx),
\end{equation} 
where
$\Phi' \equiv \{ \phi_1(\aaa) \lor \phi_2(\aaa) 
\mid (a,a) \in \N^2  \}$.

\begin{equation}\label{eq-def-v-{Omega}*-3}
\PA \vdash \ \ \
\forall \xxx \  
( \ \ 
\mbox{\CA}_{\Omega}[\psi(\xxx)](\eee, \lc \Psi \rc, \xxx) 
\ \lra \ \
\Acc(\vvv_{\Omega}^A, \lc {\cal CA}[e,\Omega] \rc, \xxx)
\ ),
\end{equation} 
where 
$\Psi \equiv 
\{ \psi(\aaa) \mid  a \in \N \}$
is a set of an infinite number of $\Delta_1$-formulas in PA, 
and
${\cal CA}[e,\Omega] \equiv 
\{ \mbox{\CA}_{\Omega}[\psi(\aaa)]
(\eee, \lc \Psi \rc, \aaa)  \mid  a \in \N \}$.

\begin{equation}\label{eq-def-v-{Omega}*-4}
\PA \vdash \ \ \
\forall \xxx \  
( \ \ 
\mbox{\CR}_{\Omega}[\psi(\xxx)](\eee, \lc \Psi \rc, \xxx) 
\ \lra \ \
\neg\Acc(\vvv_{\Omega}^R, \lc {\cal CR}[e,\Omega] \rc, \xxx)
\ ),
\end{equation} 
where 
${\cal CR}[e,\Omega] \equiv 
\{ \neg\mbox{\CR}_{\Omega}[\psi(\aaa)]
(\eee, \lc \Psi \rc, \aaa)  \mid  a \in \N \}$.

\end{lemma}

\noindent
{\bf Remark:} \ 
If  
$$
\exists x \in \N  \ \ \
\NNN \models \ \neg\phi(\xxx),
$$
then
$$
\PA \not\vdash \  
\forall \xxx \ \ ( \
\Acc(\vvv_{\Phi}^*, \lc \Phi \rc, \xxx)
\ \ra \ \phi(\xxx) \ ),
$$
since 
if 
$\PA \vdash 
\forall \xxx \ \ ( \
{\Prv}_{\PA}(\lc \phi(\xxx)\rc)
\ \ra \ \phi(\xxx) \ )$,
then 
$\exists x \in \N \ \ \ 
\PA \vdash 
\neg{\Prv}_{\PA}(\lc \phi(\xxx)\rc)
$, which implies 
$\PA \vdash \ \mbox{\rm Con}(\PA)$
and contradicts the second G{\"o}del Incompleteness Theorem.

\vspace{5pt}


\begin{lemma}\label{lem-ptd1}
Let 
$\Phi \equiv \{ \phi(\aaa) \mid a \in \N  \}$
be a set of an infinite number of $\Delta_1$-formulas in PA.

For all $e \in \N$, and for all $a \in \N$,
$$
\PTM_{e}^{\Phi}(a) \dec_{v_{\Phi}^A} \phi(\aaa)
\ \ \
\LRa \ \ \
\PTM_{e}^{\Phi}(a) \dec \phi(\aaa),
$$

For all $e \in \N$, and for all $a \in \N$,
$$
\PTM_{e}^{\Phi}(a) \dec_{v_{\Phi}^R} \neg\phi(\aaa)
\ \ \
\LRa \ \ \
\PTM_{e}^{\Phi}(a) \dec \neg\phi(\aaa).
$$

\end{lemma}

\begin{proof} 
For all $e \in \N$, and for all $a \in \N$,
$$
\PTM_{e}^{\Phi}(a) \dec_{v_{\Phi}^A} \phi(\aaa)
\ \ \
\LRa \ \ \
\U_{\PTM}(e,(d,{\#}\Phi,a)) \mbox{accepts} \  
\land  \ \U(v_{\Phi}^A,(d,{\#}\Phi,a)) \mbox{accepts}. 
$$
As shown in Eq. (\ref{eq-def-v-{Omega}*-7}),
$$ 
\U(v_{\Phi}^A,(d,{\#}\Phi,a)) \mbox{accepts}
\ \ \  \LRa \ \ \ 
\NNN \models \phi(a).
$$
Hence,
\begin{eqnarray*}
\PTM_{e}^{\Phi}(a) \dec_{v_{\Phi}^A} \phi(\aaa)
\ \ \
& & 
\LRa \ \ \
\U_{\PTM}(e,(d,{\#}\Phi,a)) \mbox{accepts} \  
\land  \ \NNN \models \phi(a)
\\ 
& & 
\LRa \ \ \
\PTM_{e}^{\Phi}(a) \dec \phi(\aaa).
\end{eqnarray*} 

Similarly, from Eq. (\ref{eq-def-v-{Omega}*-8}),
we obtain that
for all $e \in \N$, and for all $a \in \N$,
$$
\PTM_{e}^{\Phi}(a) \dec_{v_{\Phi}^R} \neg\phi(\aaa)
\ \ \
\LRa \ \ \
\PTM_{e}^{\Phi}(a) \dec \neg\phi(\aaa).
$$

\begin{flushright}
$\dashv$
\end{flushright}
\end{proof}

\begin{lemma}\label{lem-ptd2}
Let $\Omega \equiv \{ \omega(\aaa) \mid a \in \N  \}$ and 
$\Phi \equiv \{ \phi(\aaa) \mid a \in \N  \}$
be sets of an infinite number of $\Delta_1$-formulas in PA.

Then for all $e \in \N$, 
$$
\PA \vdash \ \
\forall \xxx \ \ ( \  
\CA_{\Omega}[\phi(\xxx)](\eee, \lc \Phi \rc, \xxx)
\ \ra \ 
\CA_{\Phi}[\phi(\xxx)](\eee, \lc \Phi \rc, \xxx)
\ ),
$$
$$
\PA \vdash \ \
\forall \xxx \ \ ( \  
\CR_{\Omega}[\phi(\xxx)](\eee, \lc \Phi \rc, \xxx)
\ \ra \ 
\CR_{\Phi}[\phi(\xxx)](\eee, \lc \Phi \rc, \xxx)
\ ).
$$

\end{lemma}

\begin{proof} 

For all $e \in \N$, 
\begin{eqnarray*}
\PA \vdash \ \
\forall \xxx \ \ ( \   & &  
\CA_{\Omega}[\phi(\xxx)](\eee, \lc \Phi \rc, \xxx)
\\
&\lra& \ \ \
\mbox{\PTM-\Acc}(\eee, \lc \Phi \rc, \xxx) \ \land \ 
\Acc(\vvv_{\Omega}^A, \lc \Phi \rc, \xxx)
\\
&\ra& \ \ \ 
\mbox{\PTM-\Acc}(\eee, \lc \Phi \rc, \xxx) \ \land \ 
\Acc(\vvv_{\Phi}^A, \lc \Phi \rc, \xxx)
\\
&\lra& \ \ \
\CA_{\Phi}[\phi(\aaa)](\eee, \lc \Phi \rc, \aaa)
\ ). 
\end{eqnarray*}

For all $e \in \N$, 
\begin{eqnarray*}
\PA \vdash \ \
\forall \xxx \ \ ( \   & &  
\CR_{\Omega}[\phi(\xxx)](\eee, \lc \Phi \rc, \xxx)
\\
&\lra& \ \ \
\neg\mbox{\PTM-\Acc}(\eee, \lc \Phi \rc, \xxx) \ \land \ 
\neg\Acc(\vvv_{\Omega}^R, \lc \Phi \rc, \xxx)
\\
&\ra& \ \ \ 
\neg\mbox{\PTM-\Acc}(\eee, \lc \Phi \rc, \xxx) \ \land \ 
\neg\Acc(\vvv_{\Phi}^R, \lc \Phi \rc, \xxx)
\\
&\lra& \ \ \
\CR_{\Phi}[\phi(\aaa)](\eee, \lc \Phi \rc, \aaa)
\ ). 
\end{eqnarray*}

\begin{flushright}
$\dashv$
\end{flushright}
\end{proof}

\section{Incompleteness Theorems of Polynomial-Time Decisions}
\label{sec:rbicthm-dec}

This section shows the {\it polynomial-time decision} version of 
the (second) G{\"o}del incompleteness theorem.
First, we introduce the G{\"o}del sentences of polynomial-time decisions, 
and the first incompleteness theorems of polynomial-time decisions.
We then present the second incompleteness theorem of polynomial-time decisions,
based on the the first incompleteness theorems and 
the derivability conditions of polynomial-time decisions.

\subsection{Derivability Conditions of Polynomial-Time Decisions}
\label{DC-dec}

\begin{lemma}\label{D1-ca}
(D.1-CA) \ \ \
Let $\Phi \equiv \{ \phi(\aaa) \mid a \in \N  \}$
be a set of an infinite number of $\Delta_1$-formulas in PA.

For any $e \in \N$, for any $v \in \N$, and for any $a \in \N$
$$ 
\PTM_{e}^{\Phi}(a) \dec_v  \phi(\aaa)
\ \ \ \Ra \ \ \ 
\PA \vdash 
\mbox{\CA}_v[\phi(\aaa)](\eee, \lc \Phi \rc, \aaa).
$$
\end{lemma}

\begin{proof} 
For all $e \in \N$, and for all $a \in \N$,
\begin{eqnarray*}
& & \ \ \PTM_{e}^{\Phi}(a) \dec_{v}  \phi(\aaa)
\\
&\LRa& \ \
\U_{\PTM}(e,(d,{\#}\Phi,a)) \mbox{accepts} \  
\land   \ \U(v,(d,{\#}\Phi,a)) \mbox{accepts} \
\\
&\Ra& \ \ 
\PA \vdash \ \ 
\mbox{\PTM-\Acc}(\eee, \lc \Phi \rc, \aaa) \ \land \ 
\Acc(\vvv, \lc \Phi \rc, \aaa) 
\ \ \ 
\mbox{(from $\Sigma_1$-Completeness Theorem of PA)}
\\
&\LRa& \ \
\PA \vdash \ \CA_{v}[\phi(\aaa)](\eee, \lc \Phi \rc, \aaa).
\end{eqnarray*}

\begin{flushright}
$\dashv$
\end{flushright}
\end{proof}

\begin{lemma}\label{D1-cr}
(D.1-CR) \ \ \
Let $\Phi \equiv \{ \phi(\aaa) \mid a \in \N  \}$
be a set of an infinite number of $\Delta_1$-formulas in PA.

For any $e \in \N$, for any $v \in \N$, and for any $a \in \N$
$$ 
\PTM_{e}^{\Phi}(a) \dec_v  \neg\phi(\aaa)
\ \ \ \Ra \ \ \ 
\PA \vdash 
\mbox{\CR}_v[\phi(\aaa)](\eee, \lc \Phi \rc, \aaa).
$$
\end{lemma}

\begin{proof} 
For all $e \in \N$, and for all $a \in \N$,
\begin{eqnarray*}
& & \ \ \PTM_{e}^{\Phi}(a) \dec_{v}  \neg\phi(\aaa)
\\
&\LRa& \ \
\U_{\PTM}(e,(d,{\#}\Phi,a)) \mbox{rejects} \  
\land   \ \U(v,(d,{\#}\Phi,a)) \mbox{rejects} \
\\
&\Ra& \ \ 
\PA \vdash \ \ 
\neg\mbox{\PTM-\Acc}(\eee, \lc \Phi \rc, \aaa) \ \land \ 
\neg\Acc(\vvv, \lc \Phi \rc, \aaa) 
\ \ \ 
\mbox{(from $\Sigma_1$-Completeness Theorem of PA)}
\\
&\LRa& \ \
\PA \vdash \ \CR_{v}[\phi(\aaa)](\eee, \lc \Phi \rc, \aaa).
\end{eqnarray*}

\begin{flushright}
$\dashv$
\end{flushright}
\end{proof}

\begin{lemma}\label{D2-ca}
(D.2-CA) \ \ \ 
Let 
$\Omega \equiv 
\{ \omega(\aaa) \mid  a \in \N \}$,
$\Phi \equiv 
\{ \phi(\aaa) \mid  a \in \N \}$,
$\Psi \equiv 
\{ \psi(\aaa) \mid  a \in \N \}$, and
$\Gamma \equiv 
\{ \phi(\aaa) \land \psi(\aaa) \mid  \ (a,a) \in \N^2 \}$
be sets of an infinite number of $\Delta_1$-formulas in PA.
 
For all $e_1 \in \N$ and for all $e_2 \in \N$,
there exists $e_3 \in \N$ such that 
\begin{eqnarray*} 
\PA \vdash \ \ \
\forall \xxx \  
( & &  
 \mbox{\CA}_{\Omega}[\phi(\xxx)](\eee_1, \lc \Phi \rc, \xxx) \ \land
 \mbox{\CA}_{\Omega}[\psi(\xxx)](\eee_2, \lc \Psi \rc, \xxx) 
\\
&\ra& \
 \mbox{\CA}_{\Omega}[\phi(\xxx) \ \land \ \psi(\xxx)]
(\eee_3, \lc \Gamma \rc, \xxx) \ 
\ ).
\end{eqnarray*}

\end{lemma}

\begin{proof}

PTM $\U_{\PTM}(e_3, \cdot)$ is constructed by using
two PTMs,  
$\U_{\PTM}(e_1, (d,{\#}\Phi,\cdot))$ and 
$\U_{\PTM}(e_2, (d,{\#}\Psi,\cdot))$
as follows:

\begin{enumerate}
\item
(Input: ) \ $(d,{\#}\Gamma,(x,x)) \in \N^4$.
\item
(Output: ) \ accept or reject
\item
Run the following computation 
$$\U_{\PTM}(e_1, (d,{\#}\Phi,x)), $$
$$\U_{\PTM}(e_2, (d,{\#}\Psi,x)). $$
\item
If both of them accept, then accept.
Otherwise reject.

\end{enumerate}

From the construction of $\U_{\PTM}(e_3, \cdot )$,
clearly 
$$
\PA \vdash \ \
\forall \xxx \ \ 
( \ \ 
\mbox{\PTM-\Acc}(\eee_1, \lc \Phi \rc, \xxx)
\ \land \
\mbox{\PTM-\Acc}(\eee_2, \lc \Psi \rc, \xxx)
\ \ \lra \ \ 
\mbox{\PTM-\Acc}(\eee_3, \lc \Gamma \rc, \xxx)
\ \ ).
$$

As shown in Eq. (\ref{eq-def-v-{Omega}*-1}),
\begin{equation}\label{eq-claim-D2-dec} 
\PA \vdash \ \ \
\forall \xxx \  
( \ \ 
\Acc(\vvv_{\Omega}^A, \lc \Phi \rc, \xxx)  \ \land
\Acc(\vvv_{\Omega}^A, \lc \Psi \rc, \xxx)
\ \ \ra \ \
\Acc(\vvv_{\Omega}^A, \lc \Gamma \rc, \xxx) 
\ \ ).
\end{equation} 

Then, 
for all $e_1 \in \N$ and for all $e_2 \in \N$,
there exists $e_3 \in \N$ such that 
\begin{eqnarray*}
\PA \vdash \ \ \ 
\forall \xxx \ \ \ 
( \ & & \   
 \mbox{\CA}_{\Omega}[\phi(\xxx)](\eee_1, \lc \Phi \rc, \xxx) \ \land
 \mbox{\CA}_{\Omega}[\psi(\xxx)](\eee_2, \lc \Psi \rc, \xxx) 
\\
&\lra&  \  
( \ 
\mbox{\PTM-\Acc}(\eee_1, \lc \Phi \rc, \xxx) \ \land \ 
\Acc(\vvv_{\Omega}^A, \lc \Phi \rc, \xxx)
\ )
\\
& & \mbox{ \ } \ \land \ \ 
( \ 
\mbox{\PTM-\Acc}(\eee_2, \lc \Psi \rc, \xxx) \ \land \ 
\Acc(\vvv_{\Omega}^A, \lc \Psi \rc, \xxx)
\ )
\\
&\ra& \  
( \ 
\mbox{\PTM-\Acc}(\eee_1, \lc \Phi \rc, \xxx) \ \land \ 
\mbox{\PTM-\Acc}(\eee_2, \lc \Psi \rc, \xxx)  
\ )
\\
& & \mbox{ \ } \ \land \ \ 
( \ 
\Acc(\vvv_{\Omega}^A, \lc \Phi \rc, \xxx) \ \land \ 
\Acc(\vvv_{\Omega}^A, \lc \Psi \rc, \xxx)
\ )
\\
&\lra& \ \ 
\mbox{\PTM-\Acc}(\eee_3, \lc \Gamma \rc, \xxx) \ \land \ 
\Acc(\vvv_{\Omega}^A, \lc \Gamma \rc, \xxx) 
\\
&\lra& \ \ 
\mbox{\CA}_{\Omega}[\phi(\xxx) \ \land \ \psi(\xxx)]
(\eee_3, \lc \Gamma \rc, \xxx) \ 
\ ).
\end{eqnarray*}

\begin{flushright}
$\dashv$
\end{flushright}

\end{proof}

\begin{lemma}\label{D2-cr}
(D.2-CR) \ \ \ 
Let 
$\Omega \equiv 
\{ \omega(\aaa) \mid  a \in \N \}$,
$\Phi \equiv 
\{ \phi(\aaa) \mid  a \in \N \}$,
$\Psi \equiv 
\{ \psi(\aaa) \mid  a \in \N \}$, and
$\Gamma \equiv 
\{ \phi(\aaa) \land \psi(\aaa) \mid  \ (a,a) \in \N^2 \}$
be sets of an infinite number of $\Delta_1$-formulas in PA.
 
For all $e_1 \in \N$ and for all $e_2 \in \N$,
there exists $e_3 \in \N$ such that 
\begin{eqnarray*}  
\PA \vdash \ \ \
\forall \xxx \  
( & &  
 \mbox{\CR}_{\Omega}[\phi(\xxx)](\eee_1, \lc \Phi \rc, \xxx) \ \land
 \mbox{\CR}_{\Omega}[\psi(\xxx)](\eee_2, \lc \Psi \rc, \xxx) 
\\
&\ra& \
 \mbox{\CR}_{\Omega}[\phi(\xxx) \ \lor \ \psi(\xxx)]
(\eee_3, \lc \Theta \rc, \xxx) \ 
\ ).
\end{eqnarray*}

\end{lemma}

\begin{proof}

PTM $\U_{\PTM}(e_3, \cdot)$ is constructed by using
two PTMs,  
$\U_{\PTM}(e_1, (d,{\#}\Phi,\cdot))$ and 
$\U_{\PTM}(e_2, (d,{\#}\Psi,\cdot))$
as follows:

\begin{enumerate}
\item
(Input: ) \ $(d,{\#}\Theta,(x,x)) \in \N^4$.
\item
(Output: ) \ accept or reject
\item
Run the following computation 
$$\U_{\PTM}(e_1, (d,{\#}\Phi,x)), $$
$$\U_{\PTM}(e_2, (d,{\#}\Psi,x)). $$
\item
If both of them reject, then reject.
Otherwise accept.
\end{enumerate}

From the construction of $\U_{\PTM}(e_3, \cdot )$,
clearly 
$$
\PA \vdash \ \
\forall \xxx \ \ 
( \ \ 
\neg\mbox{\PTM-\Acc}(\eee_1, \lc \Phi \rc, \xxx)
\ \land \
\neg\mbox{\PTM-\Acc}(\eee_2, \lc \Psi \rc, \xxx)
\ \ \lra \ \ 
\neg\mbox{\PTM-\Acc}(\eee_3, \lc \Theta \rc, \xxx)
\ \ ).
$$

As shown in Eq. (\ref{eq-def-v-{Omega}*-2}),
\begin{equation}\label{eq-claim-D2-dec-sr} 
\PA \vdash \ \ \
\forall \xxx \  
( \ \ 
\neg\Acc(\vvv_{\Omega}^R, \lc \Phi \rc, \xxx)  \ \land
\neg\Acc(\vvv_{\Omega}^R, \lc \Psi \rc, \xxx)
\ \ \ra \ \ 
\neg\Acc(\vvv_{\Omega}^R, \lc \Theta \rc, \xxx) 
\ \ ).
\end{equation}

Then, 
for all $e_1 \in \N$ and for all $e_2 \in \N$,
there exists $e_3 \in \N$ such that 
\begin{eqnarray*}
\PA \vdash \ \ \ 
\forall \xxx \ \ \ 
( \ & & \   
 \mbox{\CR}_{\Omega}[\phi(\xxx)](\eee_1, \lc \Phi \rc, \xxx) \ \land
 \mbox{\CR}_{\Omega}[\psi(\xxx)](\eee_2, \lc \Psi \rc, \xxx) 
\\
&\lra&  \ 
( \ 
\neg\mbox{\PTM-\Acc}(\eee_1, \lc \Phi \rc, \xxx) \ \land \ 
\neg\Acc(\vvv_{\Omega}^R, \lc \Phi \rc, \xxx)
\ )
\\
& & \mbox{ \ } \ \land \ \ 
( \
\neg\mbox{\PTM-\Acc}(\eee_2, \lc \Psi \rc, \xxx) \ \land \ 
\neg\Acc(\vvv_{\Omega}^R, \lc \Psi \rc, \xxx)
\ )
\\
&\ra& \  
( \ 
\neg\mbox{\PTM-\Acc}(\eee_1, \lc \Phi \rc, \xxx) \ \land \ 
\neg\mbox{\PTM-\Acc}(\eee_2, \lc \Psi \rc, \xxx)  
\ )
\\
& & \mbox{ \ } \ \land \ \ 
( \ 
\neg\Acc(\vvv_{\Omega}^R, \lc \Phi \rc, \xxx) \ \land \ 
\neg\Acc(\vvv_{\Omega}^R, \lc \Psi \rc, \xxx)
\ )
\\
&\lra& \ \ 
\neg\mbox{\PTM-\Acc}(\eee_3, \lc \Gamma \rc, \xxx) \ \land \ 
\neg\Acc(\vvv_{\Omega}^R, \lc \Gamma \rc, \xxx)
\\
&\lra& \ \ 
\mbox{\CR}_{\Omega}[\phi(\xxx) \ \land \ \psi(\xxx)]
(\eee_3, \lc \Gamma \rc, \xxx) \ 
\ ).
\end{eqnarray*}

\begin{flushright}
$\dashv$
\end{flushright}

\end{proof}

\begin{corollary}\label{D-col1-ca}
Let 
$\Omega \equiv 
\{ \omega(\aaa) \mid  a \in \N \}$,
$\Phi \equiv 
\{ \phi(\aaa) \mid  a \in \N \}$, and
$\Psi \equiv 
\{ \psi(\aaa) \mid  a \in \N \}$
be sets of an infinite number of $\Delta_1$-formulas in PA.

We assume that
$$ 
\PA \vdash \ 
\forall \xxx \ ( \phi(\xxx) \ra \psi(\xxx) )
$$
is installed in $\U(v_{\Omega}^A,\cdot)$.

Then, for all $e_1 \in \N$ there exists $e_2 \in \N$ such that 
$$  
\PA \vdash \ \ \
\forall \xxx  \
( \
\mbox{\CA}_{\Omega}[\phi(\xxx)](\eee_1, \lc \Phi \rc, \xxx) \ 
\ra \
 \mbox{\CA}_{\Omega}[\psi(\xxx)](\eee_2, \lc \Psi \rc, \xxx) \ 
).
$$

\end{corollary}

\begin{proof}

PTM $\U_{\PTM}(e_2, \cdot )$ is constructed by using
PTM $\U_{\PTM}(e_1, (d, {\#}\Phi,\cdot))$
as follows:

\begin{enumerate}
\item
(Input: ) \ 
$(d, {\#}\Psi, x)$ 
\item
(Output: )
accept or reject
\item
Run the following computation 
$$\U_{\PTM}(e_1, (d,{\#}\Phi,x)).$$
\item
Accept if and only $\U_{\PTM}(e_1, (d,{\#}\Phi,x))$ accepts. 
\end{enumerate}

From the construction of $\U_{\PTM}(e_2, \cdot )$,
clearly 
$$
\PA \vdash \ \
\forall \xxx \ \ 
( \
\mbox{\PTM-\Acc}(\eee_1, \lc \Phi \rc, \xxx)
\ \ \lra \ \ 
\mbox{\PTM-\Acc}(\eee_2, \lc \Psi \rc, \xxx)
\ \ ).
$$

Since $ \PA \vdash \ 
\forall \xxx \ ( \phi(\xxx) \ra \psi(\xxx) ) $
is installed in $\U(v_{\Omega}^A,\cdot)$,
$$
\PA \vdash \ \forall \xxx \ ( \
\Acc(\vvv_{\Omega}^A, \lc \Phi \rc, \xxx) \ \ \ra \ \  
\Acc(\vvv_{\Omega}^A, \lc \Psi \rc, \xxx) \ ).
$$

Then, for all $e_1 \in \N$ there exists $e_2 \in \N$ such that 
\begin{eqnarray*}
\PA \vdash \ \ \ 
\forall \xxx \ \ \ 
( \ & & \ 
\mbox{\CA}_{\Omega}[\phi(\xxx)](\eee_1, \lc \Phi \rc, \xxx)
\\
&\lra&  \ 
\mbox{\PTM-\Acc}(\eee_1, \lc \Phi \rc, \xxx) \ \land \ 
\Acc(\vvv_{\Omega}^A, \lc \Phi \rc, \xxx)
\\
&\ra& \  
\mbox{\PTM-\Acc}(\eee_2, \lc \Psi \rc, \xxx) \ \land \ 
\Acc(\vvv_{\Omega}^A, \lc \Psi \rc, \xxx)  
\\
&\lra&  \ 
\mbox{\CA}_{\Omega}[\psi(\xxx)](\eee_2, \lc \Psi \rc, \xxx)
\ \ ). 
\end{eqnarray*}

\begin{flushright}
$\dashv$
\end{flushright}

\end{proof}

\begin{corollary}\label{D-col1-cr}
Let 
$\Omega \equiv 
\{ \omega(\aaa) \mid  a \in \N \}$,
$\Phi \equiv 
\{ \phi(\aaa) \mid  a \in \N \}$, and
$\Psi \equiv 
\{ \psi(\aaa) \mid  a \in \N \}$
be sets of an infinite number of $\Delta_1$-formulas in PA.

We assume that 
$$ 
\PA \vdash \ 
\forall \xxx \ (\neg\phi(\xxx) \ra \neg\psi(\xxx))
$$
is installed in $\U(v_{\Omega}^R,\cdot)$.

Then, for all $e_1 \in \N$ there exists $e_2 \in \N$ such that 
$$  
\PA \vdash \ \ \
\forall \xxx  \
( \
\mbox{\CR}_{\Omega}[\phi(\xxx)](\eee_1, \lc \Phi \rc, \xxx) \ 
\ra \
\mbox{\CR}_{\Omega}[\psi(\xxx)](\eee_2, \lc \Psi \rc, \xxx) \ 
).
$$

\end{corollary}

\begin{proof}

PTM $\U_{\PTM}(e_2, \cdot )$ is constructed by using
PTM $\U_{\PTM}(e_1, (d, {\#}\Phi,\cdot))$
as follows:

\begin{enumerate}
\item
(Input: ) \ 
$(d, {\#}\Psi, x)$ 
\item
(Output: )
accept or reject
\item
Run the following computation 
$$\U_{\PTM}(e_1, (d,{\#}\Phi,x)).$$
\item
Reject if and only $\U_{\PTM}(e_1, (d,{\#}\Phi,x))$ rejects. 
\end{enumerate}

From the construction of $\U_{\PTM}(e_2, \cdot )$,
clearly 
$$
\PA \vdash \ \
\forall \xxx \ \ 
( \
\neg\mbox{\PTM-\Acc}(\eee_1, \lc \Phi \rc, \xxx)
\ \ \lra \ \ 
\neg\mbox{\PTM-\Acc}(\eee_2, \lc \Psi \rc, \xxx)
\ \ ).
$$

Since 
$ 
\PA \vdash \ 
\forall \xxx \ (\neg\phi(\xxx) \ra \neg\psi(\xxx))
$
is installed in $\U(v_{\Omega}^R,\cdot)$,
$$
\PA \vdash \ \forall \xxx \ ( \
\neg\Acc(\vvv_{\Omega}^R, \lc \Phi \rc, \xxx) \ \ \ra \ \  
\neg\Acc(\vvv_{\Omega}^R, \lc \Psi \rc, \xxx) \ ).
$$

Then, for all $e_1 \in \N$ there exists $e_2 \in \N$ such that 
\begin{eqnarray*}
\PA \vdash \ \ \ 
\forall \xxx \ \ \ 
( \ & & \ 
\mbox{\CR}_{\Omega}[\phi(\xxx)](\eee_1, \lc \Phi \rc, \xxx)
\\
&\lra&  \ 
\neg\mbox{\PTM-\Acc}(\eee_1, \lc \Phi \rc, \xxx) \ \land \ 
\neg\Acc(\vvv_{\Omega}^R, \lc \Phi \rc, \xxx)
\\
&\ra& \  
\neg\mbox{\PTM-\Acc}(\eee_2, \lc \Psi \rc, \xxx) \ \land \ 
\neg\Acc(\vvv_{\Omega}^R, \lc \Psi \rc, \xxx)  
\\
&\lra&  \ 
\mbox{\CR}_{\Omega}[\psi(\xxx)](\eee_2, \lc \Psi \rc, \xxx)
\ \ ). 
\end{eqnarray*}

\begin{flushright}
$\dashv$
\end{flushright}

\end{proof}

\begin{lemma}\label{D3-ca}
(D.3-CA) \ \ \ 
Let 
$\Omega \equiv 
\{ \omega(\aaa) \mid  a \in \N \}$
and 
$\Phi \equiv 
\{ \phi(\aaa) \mid  a \in \N \}$
be sets of an infinite number of $\Delta_1$-formulas in PA.
Let ${\cal CA}[e,\Omega] \equiv 
\{ \mbox{\CA}_{\Omega}[\phi(\aaa)](\eee, \lc \Phi \rc, \aaa)  \mid  a \in \N \}$.

For all $e_1 \in \N$,
there exists $e_2 \in \N$ such that 
$$ 
\PA \vdash \ \ \ 
\forall \xxx \ 
( \ 
\mbox{\CA}_{\Omega}[\phi(\xxx)](\eee_1, \lc \Phi \rc, \xxx) 
\ \ra \ 
\mbox{\CA}_{\Omega}
[\mbox{\CA}_{\Omega}[\phi(\xxx)](\eee_1, \lc \Phi \rc, \xxx) ]
(\eee_2, \lc {\cal CA}[e_1,{\Omega}] \rc, \xxx) \ ).
$$
\end{lemma}

\begin{proof}

PTM $\U_{\PTM}(e_2, \cdot )$ is constructed 
by using
PTM $\U_{\PTM}(e_1, (d, {\#}\Phi,\cdot))$
as follows:

\begin{enumerate}
\item
(Input: ) \ 
$(d, {\#}{\cal CA}[e_1,{\Omega}], x) \in \N^3$ 
\item
(Output: ) \ 
accept or reject
\item
Run the following computation 
$$\U_{\PTM}(e_1, (d,{\#}\Phi,x)).$$
\item
Accept if and only $\U_{\PTM}(e_1, (d,{\#}\Phi,x))$ accepts. 
\end{enumerate}

From the construction of $\U_{\PTM}(e_2, \cdot )$,
$$
\PA \vdash \ \
\forall \xxx \ \ 
( \
\mbox{\PTM-\Acc}(\eee_1, \lc \Phi \rc, \xxx)
\ \ \lra \ \ 
\mbox{\PTM-\Acc}(\eee_2, \lc \Psi \rc, \xxx)
\ \ ).
$$

As shown in Eq. (\ref{eq-def-v-{Omega}*-3}), 
$$
\PA \vdash \ \ \
\forall \xxx \  
( \ \ 
\mbox{\CA}_{\Omega}[\phi(\xxx)](\eee_1, \lc \Phi \rc, \xxx) 
\ \ra \ \
\Acc(\vvv_{\Omega}^A, \lc {\cal CA}[e_1,\Omega] \rc, \xxx)
\ ).
$$

Then, 
\begin{eqnarray*}
\PA \vdash \ \ \ 
\forall \xxx \ \ \ 
( \ & & \ 
\mbox{\CA}_{\Omega}[\phi(\xxx)](\eee_1, \lc \Phi \rc, \xxx) 
\\
&\lra&  \ 
\mbox{\PTM-\Acc}(\eee_1, \lc \Phi \rc, \xxx) \ \land \ 
\Acc(\vvv_{\Omega}^A, \lc \Phi \rc, \xxx)
\\
&\lra&  \ 
\mbox{\PTM-\Acc}(\eee_1, \lc \Phi \rc, \xxx) \ \land \ 
\mbox{\CA}_{\Omega}[\phi(\xxx)](\eee_1, \lc \Phi \rc, \xxx) 
\\
&\ra& \  
\mbox{\PTM-\Acc}(\eee_2, \lc {\cal CA}[e_1,\Omega] \rc, \xxx) \ \land \ 
\Acc(\vvv_{\Omega}^A, \lc {\cal CA}[e_1,\Omega] \rc, \xxx)
\\
&\ra& \  
\mbox{\CA}_{\Omega}
[ \mbox{\CA}_{\Omega}[\phi(\xxx)](\eee_1, \lc \Phi \rc, \xxx) ]
(\eee_2, \lc {\cal CA}[e_1,\Omega] \rc, \xxx)
\ \ ). 
\end{eqnarray*}

\begin{flushright}
$\dashv$
\end{flushright}

\end{proof}

\begin{lemma}\label{D3-cr}
(D.3-CR) \ \ \ 
Let 
$\Omega \equiv 
\{ \omega(\aaa) \mid  a \in \N \}$
and 
$\Phi \equiv 
\{ \phi(\aaa) \mid  a \in \N \}$
be sets of an infinite number of $\Delta_1$-formulas in PA.
Let ${\cal CR}[e,\Omega] \equiv 
\{ \neg\mbox{\CR}_{\Omega}[\phi(\aaa)](\eee, \lc \Phi \rc, \aaa)  \mid  a \in \N \}$.

For all $e_1 \in \N$,
there exists $e_2 \in \N$ such that 
$$ 
\PA \vdash \ \ \ 
\forall \xxx \ 
( \ 
\mbox{\CR}_{\Omega}[\phi(\xxx)](\eee_1, \lc \Phi \rc, \xxx) 
\ \ra \ 
\mbox{\CR}_{\Omega}
[\neg\mbox{\CR}_{\Omega}[\phi(\xxx)](\eee_1, \lc \Phi \rc, \xxx) ]
(\eee_2, \lc {\cal CR}[e_1,{\Omega}] \rc, \xxx) \ ).
$$
\end{lemma}

\begin{proof}

PTM $\U_{\PTM}(e_2, \cdot )$ is constructed 
by using
PTM $\U_{\PTM}(e_1, (d, {\#}\Phi,\cdot))$
as follows:

\begin{enumerate}
\item
(Input: ) \ 
$(d, {\#}{\cal CR}[e_1,{\Omega}], x) \in \N^3$ 
\item
(Output: ) \ 
accept or reject
\item
Run the following computation 
$$\U_{\PTM}(e_1, (d,{\#}\Phi,x)).$$
\item
Reject if and only $\U_{\PTM}(e_1, (d,{\#}\Phi,x))$ rejects. 
\end{enumerate}

From the construction of $\U_{\PTM}(e_2, \cdot )$,
$$
\PA \vdash \ \
\forall \xxx \ \ 
( \
\neg\mbox{\PTM-\Acc}(\eee_1, \lc \Phi \rc, \xxx)
\ \ \lra \ \ 
\neg\mbox{\PTM-\Acc}(\eee_2, \lc \Psi \rc, \xxx)
\ \ ).
$$

As shown in Eq. (\ref{eq-def-v-{Omega}*-4}), 
$$
\PA \vdash \ \ \
\forall \xxx \  
( \ \ 
\mbox{\CR}_{\Omega}[\phi(\xxx)](\eee_1, \lc \Phi \rc, \xxx) 
\ \ra \ \
\neg\Acc(\vvv_{\Omega}^R, \lc {\cal CR}[e_1,\Omega] \rc, \xxx)
\ ).
$$

Then, 
\begin{eqnarray*}
\PA \vdash \ \ \ 
\forall \xxx \ \ \ 
( \ & & \ 
\mbox{\CR}_{\Omega}[\phi(\xxx)](\eee_1, \lc \Phi \rc, \xxx) 
\\
&\lra&  \ 
\neg\mbox{\PTM-\Acc}(\eee_1, \lc \Phi \rc, \xxx) \ \land \ 
\neg\Acc(\vvv_{\Omega}^R, \lc \Phi \rc, \xxx)
\\
&\lra&  \ 
\neg\mbox{\PTM-\Acc}(\eee_1, \lc \Phi \rc, \xxx) \ \land \ 
\mbox{\CR}_{\Omega}[\phi(\xxx)](\eee_1, \lc \Phi \rc, \xxx) 
\\
&\ra& \  
\neg\mbox{\PTM-\Acc}(\eee_2, \lc {\cal CR}[e_1,\Omega] \rc, \xxx) \ \land \ 
\neg\Acc(\vvv_{\Omega}^R, \lc {\cal CR}[e_1,\Omega] \rc, \xxx)
\\
&\ra& \  
\mbox{\CR}_{\Omega}
[ \neg\mbox{\CR}_{\Omega}[\phi(\xxx)](\eee_1, \lc \Phi \rc, \xxx) ]
(\eee_2, \lc {\cal CR}[e_1,\Omega] \rc, \xxx)
\ \ ). 
\end{eqnarray*}

\begin{flushright}
$\dashv$
\end{flushright}

\end{proof}

\subsection{G{\"o}del  Sentences of Polynomial-Time Decisions}

\begin{lemma}\label{def-goedel-dec-ca}
For any $e \in \N$ and for any $v \in \N$,
there exists a set of formulas,
${\cal G}^{A} \equiv \{ \rho_{e,v}^{A}(\aaa) \mid  a \in \N \}$,
such that 
$$ \PA \vdash 
\forall \xxx \ \
( \rho_{e,v}^{A}(\xxx) \ \lra \ 
\neg\mbox{\CA}_v[\rho_{e,v}^{A}(\xxx)](\eee, \lc {\cal G}^{A} \rc, \xxx) \
).
$$
For all $x \in \N$,
$\rho_{e,v}^{A}(\xxx)$ is called a ``G{\"o}del sentence'' with respect to
CA. 
\end{lemma}

\begin{proof}
Let $e \in \N$ and $v \in \N$ be given.

Based on the recursion theorem (Proposition \ref{recursion}), 
TM $\U(k,\cdot)$ is constructed as follows:
\begin{enumerate}
\item (Input:) \ $(d,{\#}\N^3, (e,v,x)) \in \N^5$.,
\item (Output:) \ accept or reject
\item
First, read its own code, $k \in \N$.
\item 
Construct a formula, 
$$ \rho_{e,v}^{A}(\xxx)
\ \equiv \
\mbox{\Acc}(\kkk, \lc \N^3 \rc, \xxx, \eee, \vvv ),
$$
(i.e., $\rho_{e,v}^{A}(\xxx)$ 
represents the relation
that $\U_{\PTM}(k,(d,{\#}\N^3,(e,v,x)))$ accepts).
Let ${\cal G}^{A} \equiv \{ \rho_{e,v}^{A}(\aaa) \mid a \in \N \}$.
\item
Execute PTM $\U_{\PTM}(e, (d, {\#}{\cal G}^{A}, x))$ to decide
the truth of formula $\rho_{e,v}^{A}(\xxx)$.
\item
Execute TM $\U(v, (d, {\#}{\cal G}^{A}, x))$ to decide
the truth of formula $\rho_{e,v}^{A}(\xxx)$.
\item
Reject if and only if  
both of 
$\U_{\PTM}(e, (d, {\#}{\cal G}^{A}, x))$ 
and $\U(v, (d,{\#}{\cal G}^{A}, x))$ accepts.
\end{enumerate}

Here, note that when only $\xxx$ occurs free in formula 
$\rho_{e,v}^{A}(\xxx)$,
${\#}\rho_{e,v}^{A}(\xxx)$ is a finite number. 
For $a \in \N$,  $\rho_{e,v}^{A}(\aaa)$ is equivalent to 
$\rho_{e,v}'(\aaa) \equiv (\rho_{e,v}^{A}(\xxx) \land \xxx=\aaa)$,
and $|{\#}\rho_{e,v}'(\aaa)| = O(|a|)$. 

Then, in a manner similar to Lemma \ref{ptm-recursion},
we obtain
$$ 
\forall e \in \N \ \forall v \in \N \ \ \
\PA \vdash \ 
\forall \xxx \ \
( \rho_{e,v}^{A}(\xxx) \ \lra \ 
\neg\mbox{\CA}_v[\rho_{e,v}^{A}(\xxx)](\eee, \lc {\cal G}^{A} \rc, \xxx) \
).
$$

\begin{flushright}
$\dashv$
\end{flushright}
\end{proof}

\begin{lemma}\label{def-goedel-dec-cr}
For any $e \in \N$ and for any $v \in \N$,
there exists a set of formulas,
${\cal G}^{R} \equiv \{ \rho_{e,v}^{R}(\aaa) \mid  a \in \N \}$,
such that 
$$ \PA \vdash 
\forall \xxx \ \
( \rho_{e,v}^{R}(\xxx) \ \lra \ 
\mbox{\CR}_v[\rho_{e,v}^{R}(\xxx)](\eee, \lc {\cal G}^{R} \rc, \xxx) \
).
$$
For all $x \in \N$,
$\rho_{e,v}^{R}(\xxx)$ is called a ``G{\"o}del sentence'' with respect to
CR. 
\end{lemma}

\begin{proof}
Let $e \in \N$ and $v \in \N$ be given.

Based on the recursion theorem (Proposition \ref{recursion}), 
TM $\U(k,\cdot)$ is constructed as follows:
\begin{enumerate}
\item (Input:) \ $(d,{\#}\N^3, (e,v,x)) \in \N^5$.,
\item (Output:) \ accept or reject

\item
First, read its own code, $k \in \N$.
\item 
Construct a formula, 
$$ \rho_{e,v}^{R}(\xxx)
\ \equiv \
\mbox{\Acc}(\kkk, \lc \N^3 \rc, \xxx, \eee, \vvv ),
$$
(i.e., $\rho_{e,v}^{R}(\xxx)$ 
represents the relation
that $\U_{\PTM}(k,(d,{\#}\N^3,(e,v,x)))$ accepts).
Let ${\cal G}^{R} \equiv \{ \rho_{e,v}^{R}(\aaa) \mid a \in \N \}$.
\item
Execute PTM $\U_{\PTM}(e, (d, {\#}{\cal G}^{R}, x))$ to decide
on the truth of formula $\rho_{e,v}^{R}(\xxx)$.
\item
Execute TM $\U(v, (d, {\#}{\cal G}^{R}, x))$ to decide
on the truth of formula $\rho_{e,v}^{R}(\xxx)$.
\item
Accept if and only if  
both of 
$\U_{\PTM}(e, (d, {\#}{\cal G}^{R}, x))$ 
and $\U(v, (d,{\#}{\cal G}^{R}, x))$ reject.
\end{enumerate}

Then, in a manner similar to Lemma \ref{ptm-recursion},
we obtain
$$ 
\forall e \in \N \ \forall v \in \N \ \ \
\PA \vdash \ 
\forall \xxx \ \
( \rho_{e,v}^{R}(\xxx) \ \lra \ 
\mbox{\CR}_v[\rho_{e,v}^{R}(\xxx)](\eee, \lc {\cal G}^{R} \rc, \xxx) \
).
$$

\begin{flushright}
$\dashv$
\end{flushright}
\end{proof}

\subsection{The First Incompleteness Theorems
of Polynomial-Time Decisions}\label{sec:1-rbicthm-dec}

\begin{theorem}\label{rbic-1-thm-dec-ca}
Let $\rho_{e,v}^{A}(\aaa)$ be a G{\"o}del sentence
with respect to CA, where $a \in \N$.
Let 
${\cal G}^{A} \equiv
\{ \rho_{e,v}^{A}(\aaa) \mid  a \in \N \}$.
Let $\U(v,\cdot)$ soundly accept
(see Definition \ref{def-sound-acc}).

For all $e \in \N$, 
and for all $x \in \N$,
$$ {\PTM}_{e}^{{\cal G}^{A}}(x) \not\dec_v \ \rho_{e,v}^{A}(\xxx).$$

\end{theorem}

\begin{proof}
Assume that there exist $e \in \N$, 
$v \in \N$ 
and $x \in \N$ such that
\begin{equation}\label{rbic-1-thm-dec-ca-proof-eq1}
{\PTM}_{e}^{{\cal G}^{A}}(x) \dec_v \ \rho_{e,v}^{A}(\xxx).
\end{equation}

From Lemma \ref{D1-ca}
$$ 
\PA \vdash \  
\mbox{\CA}_v[\rho_{e,v}^{A}(\xxx)](\eee, \lc {\cal G}^{A} \rc, \xxx).
$$
Since PA has model $\NNN$,
\begin{equation}\label{rbic-1-thm-dec-ca-proof-eq2}
\NNN \models \  
\mbox{\CA}_v[\rho_{e,v}^{A}(\xxx)](\eee, \lc {\cal G}^{A} \rc, \xxx).
\end{equation}

On the other hand, from assumption of 
Eq. (\ref{rbic-1-thm-dec-ca-proof-eq1}),
$\U(v,(d,{\#}{\cal G}^{A},x))$ accepts.
Since $\U(v,\cdot)$ soundly accepts,
$$
\NNN \models \ \rho_{e,v}^{A}(\xxx).
$$
Applying Lemma \ref{def-goedel-dec-ca} to the above equation,
$$
\NNN \models \ 
\neg\mbox{\CA}_v[\rho_{e,v}^{A}(\xxx)](\eee, \lc {\cal G}^{A} \rc, \xxx).
$$
This contradicts Eq. (\ref{rbic-1-thm-dec-ca-proof-eq2}).
Thus, for all $e \in \N$, and for all $x \in \N$,
$$ {\PTM}_{e}^{{\cal G}^{A}}(x) \not\dec_v \ \rho_{e,v}^{A}(\xxx).$$

\begin{flushright}
$\dashv$
\end{flushright}
\end{proof}

\begin{theorem}\label{rbic-1-thm-dec-cr}
Let $\rho_{e,v}^{R}(\aaa)$ be a G{\"o}del sentence
with respect to CR, where $a \in \N$.
Let 
${\cal G}^{R} \equiv
\{ \rho_{e,v}^{R}(\aaa) \mid  a \in \N \}$.
Let $\U(v,\cdot)$ soundly reject
(see Definition \ref{def-sound-acc}).

For all $e \in \N$, 
and for all $x \in \N$,

$$ {\PTM}_{e}^{{\cal G}^{R}}(x) \not\dec_v \ \neg\rho_{e,v}^{R}(\xxx).$$ 

 
\end{theorem}

\begin{proof}
Assume that there exist $e \in \N$, 
$v \in \N$ 
and $x \in \N$ such that
\begin{equation}\label{rbic-1-thm-cr-proof-eq1}
{\PTM}_{e}^{{\cal G}^{R}}(x) \dec_v \ \neg\rho_{e,v}^{R}(\xxx).
\end{equation}

From Lemma \ref{D1-cr},
$$ 
\PA \vdash \  
\mbox{\CR}_v[\rho_{e,v}^{R}(\xxx)](\eee, \lc \cal G^{R} \rc, \xxx).
$$
Since PA has model $\NNN$,
\begin{equation}\label{rbic-1-thm-cr-proof-eq2}
\NNN \models \  
\mbox{\CR}_v[\rho_{e,v}^{R}(\xxx)](\eee, \lc \cal G^{R} \rc, \xxx).
\end{equation}

On the other hand, from assumption of Eq. (\ref{rbic-1-thm-cr-proof-eq1})
$\U(v,(d,{\#}{\cal G}^{R},x))$ rejects.
Since $\U(v,\cdot)$ soundly rejects,
$$
\NNN \models \ \neg\rho_{e,v}^{R}(\xxx).
$$
Applying Lemma \ref{def-goedel-dec-cr} to the above equation,
$$
\NNN \models \ 
\neg\mbox{\CR}_v[\rho_{e,v}^{R}(\xxx)](\eee, \lc {\cal G}^{R} \rc, \xxx).
$$
This contradicts Eq. (\ref{rbic-1-thm-cr-proof-eq2}).
Thus, for all $e \in \N$, and for all $x \in \N$,
$$ {\PTM}_{e}^{{\cal G}^{R}}(x) \not\dec_v \ \neg\rho_{e,v}^{R}(\xxx).$$ 


\begin{flushright}
$\dashv$
\end{flushright}
\end{proof}

The following Corollaries are immediately obtained 
from Theorems \ref{rbic-1-thm-dec-ca} and \ref{rbic-1-thm-dec-cr}, 
since $\U(v_{\Omega}^A,\cdot)$ soundly accepts and
$\U(v_{\Omega}^R,\cdot)$ soundly rejects,
as shown in Lemma \ref{lem-v-{Omega}*}.

\begin{corollary}\label{rbic-1-col-dec-ca}
Let $\Omega \equiv 
\{ \omega(\aaa) \mid  a \in \N \}$,
be a set of an infinite number of $\Delta_1$-formulas in PA.
Let $\U(v_{\Omega}^A,\cdot)$ be a TM as defined in
Definition \ref{def-v-{Omega}*}. 
Let $\rho_{e,{\Omega}}^{A}(\aaa)
\equiv \rho_{e,v_{\Omega}^A}^{A}(\aaa)
$ be a G{\"o}del sentence
with respect to CA, where $a \in \N$.
Let 
${\cal G}_{\Omega}^{A} \equiv
\{ \rho_{e,{\Omega}}^{A}(\aaa) \mid  a \in \N \}$.

For all $e \in \N$, 
and for all $x \in \N$,
$$ {\PTM}_{e}^{{\cal G}_{\Omega}^{A}}(x) \not\dec_{v_{\Omega}^A} \ 
\rho_{e,{\Omega}}^{A}(\xxx).$$ 

\end{corollary}

\begin{corollary}\label{rbic-1-col-dec-cr}
Let $\Omega \equiv 
\{ \omega(\aaa) \mid  a \in \N \}$,
be a set of an infinite number of $\Delta_1$-formulas in PA.
Let $\U(v_{\Omega}^R,\cdot)$ be a TM as defined in
Definition \ref{def-v-{Omega}*}. 
Let $\rho_{e,{\Omega}}^{R}(\aaa)
\equiv \rho_{e,v_{\Omega}^R}^{R}(\aaa)
$ be a G{\"o}del sentence
with respect to CR, where $a \in \N$.
Let 
${\cal G}_{\Omega}^{R} \equiv
\{ \rho_{e,{\Omega}}^{R}(\aaa) \mid  a \in \N \}$.

For all $e \in \N$, 
and for all $x \in \N$,
$$ {\PTM}_{e}^{{\cal G}_{\Omega}^{R}}(x) \not\dec_{v_{\Omega}^R} \ 
\neg\rho_{e,{\Omega}}^{R}(\xxx).$$ 

\end{corollary}

\subsection{The Second Incompleteness Theorem
of Polynomial-Time Decisions}\label{sec:2-rbicthm-dec}

\begin{lemma}\label{lem:2-rbicthm-dec}
For $a \in \N$,
let $\rho_{e,{\Omega}}^{A}(\aaa)
\equiv \rho_{e,v_{\Omega}^A}^{A}(\aaa)$
be a G{\"o}del sentence with respect to CA,
and $\rho_{e,{\Omega}}^{R}(\aaa)
\equiv \rho_{e,v_{\Omega}^R}^{R}(\aaa)
$ be a G{\"o}del sentence
with respect to CR.
(For the definition of $v_{\Omega}^A$ and $v_{\Omega}^R$, 
see Definition \ref{def-v-{Omega}*}.)

Then, there exists a primitive recursive function $h$
such that 
for any $\Delta_1$-formula sets 
$\Psi \equiv \{\psi(\aaa) \mid a \in \N \}$ and
$\Omega \equiv \{\omega(\aaa) \mid a \in \N \}$, and
for any $e \in \N$,
\begin{equation}\label{eq:rbic-2nd-dec-5a}    
\PA \vdash \
\forall \xxx \ \ 
( \ 
\neg\mbox{\CA}_{\Omega}[\psi(\xxx)](h(\eee), \lc \Psi \rc, \xxx) 
\ \ \
\ra \ \ \  
\rho_{e,{\Omega}}^{A}(\xxx) \ 
),
\end{equation}
and
\begin{equation}\label{eq:rbic-2nd-dec-5r}    
\PA \vdash \
\forall \xxx \ \ 
( \ 
\neg\mbox{\CR}_{\Omega}[\psi(\xxx)](h(\eee), \lc \Psi \rc, \xxx) 
\ \ \
\ra \ \ \  
\neg\rho_{e,{\Omega}}^{R}(\xxx) \ 
).
\end{equation}

In other words,
for any $e \in \N$, there exists $e^*$ such that 
\begin{equation}\label{eq:rbic-2nd-dec-5-1a}    
\PA \vdash \
\forall \xxx \ \ 
( \ 
\neg\mbox{\CA}_{\Omega}[\psi(\xxx)](\eee^*, \lc \Psi \rc, \xxx) 
\ \ \
\ra \ \ \  
\rho_{e,{\Omega}}^{A}(\xxx) \ 
),
\end{equation}
\begin{equation}\label{eq:rbic-2nd-dec-5-1r}    
\PA \vdash \
\forall \xxx \ \ 
( \ 
\neg\mbox{\CR}_{\Omega}[\psi(\xxx)](\eee^*, \lc \Psi \rc, \xxx) 
\ \ \
\ra \ \ \  
\neg\rho_{e,{\Omega}}^{R}(\xxx) \ 
),
\end{equation}

\end{lemma}

\begin{proof}
Let 
$$
{\cal G}_{\Omega}^{A} \equiv 
\{ \rho_{e,{\Omega}}^{A}(\aaa) \mid  a \in \N \}
$$
be a set of G{\"o}del sentences with respect to CA, and
$$
{\cal G}_{\Omega}^{R} \equiv 
\{ \rho_{e,{\Omega}}^{R}(\aaa) \mid  a \in \N \}
$$
be a set of G{\"o}del sentences with respect to CR.

Let 
\begin{eqnarray*}
& & {\cal G}^{A +} \equiv 
\{ \mbox{\CA}_{\Omega}[\rho_{e,{\Omega}}^{A}(\aaa)]
(\eee, \lc {\cal G}_{\Omega}^{A} \rc, \aaa) \mid  a \in \N \},\\
& & {\cal G}^{A ++} \equiv 
\{ \neg\rho_{e,{\Omega}}^{A}(\aaa) \mid  a \in \N \},\\ 
& & {\cal G}^{A +++} \equiv 
\{ \rho_{e,{\Omega}}^{A}(\aaa) \land 
\neg\rho_{e,{\Omega}}^{A}(\aaa) \mid  a \in \N \},\\
& & {\cal G}^{R +} \equiv 
\{ \neg\mbox{\CR}_{\Omega}[\rho_{e,{\Omega}}^{R}(\aaa)]
(\eee, \lc {\cal G}_{\Omega}^{R} \rc, \aaa) \mid  a \in \N \},\\
& & {\cal G}^{R ++} \equiv 
\{ \neg\rho_{e,{\Omega}}^{R}(\aaa) \mid  a \in \N \},\\ 
& & {\cal G}^{R +++} \equiv 
\{ \rho_{e,{\Omega}}^{R}(\aaa) \lor 
\neg\rho_{e,{\Omega}}^{R}(\aaa) \mid  a \in \N \}.
\end{eqnarray*}

For any $e \in \N$, there exist
$e^{+} \in \N$, $e^{++} \in \N$ and $e^{+++} \in \N$
such that
\begin{eqnarray}\label{eq:rbic-2nd-dec-1a}  
\PA \vdash \
\forall \xxx \ \ 
(  \  & & 
\mbox{\CA}_{\Omega}[\rho_{e,{\Omega}}^{A}(\xxx)]
(\eee, \lc {\cal G}_{\Omega}^{A} \rc, \xxx)
\nonumber \\ 
&\ra& \ \
\mbox{\CA}_{\Omega}
[\mbox{\CA}_{\Omega}[\rho_{e,{\Omega}}^{A}(\xxx)]
(\eee, \lc {\cal G} \rc, \xxx)](\eee^{+}, \lc {\cal G}^{A +} \rc, \xxx)
\ \ \ \mbox{(by Lemma \ref{D3-ca})}
\nonumber \\
&\ra& \ \
\mbox{\CA}_{\Omega}[\neg\rho_{e,{\Omega}}^{A}(\xxx)]
(\eee^{++}, \lc {{\cal G}^{A ++}} \rc, \xxx)
\ \ \ \mbox{(by Lemma \ref{def-goedel-dec-ca} and Corollary \ref{D-col1-ca} )} 
\nonumber \\ 
&\ra& \ \
\mbox{\CA}_{\Omega}[\rho_{e,{\Omega}}^{A}(\xxx)]
(\eee, \lc {\cal G}_{\Omega}^{A} \rc, \xxx)
\land \ 
\mbox{\CA}_{\Omega}[\neg\rho_{e,{\Omega}}^{A}(\xxx)]
(\eee^{++}, \lc {\cal G}^{A ++} \rc, \xxx)
\nonumber \\
&\ra& \ \
\mbox{\CA}_{\Omega}
[\rho_{e,{\Omega}}^{A}(\xxx) \ 
\land \ \neg\rho_{e,{\Omega}}^{A}(\xxx)]
(\eee^{+++}, \lc {\cal G}^{A +++} \rc, \xxx),
\ \ \
\mbox{(by Lemma \ref{D2-ca})}
\end{eqnarray}
and 
\begin{eqnarray}\label{eq:rbic-2nd-dec-1r}  
\PA \vdash \
\forall \xxx \ \ 
(  \  & & 
\mbox{\CR}_{\Omega}[\rho_{e,{\Omega}}^{R}(\xxx)]
(\eee, \lc {\cal G}_{\Omega}^{R} \rc, \xxx)
\nonumber \\ 
&\ra& \ \
\mbox{\CR}_{\Omega}
[\neg\mbox{\CR}_{\Omega}[\rho_{e,{\Omega}}^{R}(\xxx)]
(\eee, \lc {\cal G} \rc, \xxx)](\eee^{+}, \lc {\cal G}^{R +} \rc, \xxx)
\ \ \ \mbox{(by Lemma \ref{D3-cr})}
\nonumber \\
&\ra& \ \
\mbox{\CR}_{\Omega}[\neg\rho_{e,{\Omega}}^{R}(\xxx)]
(\eee^{++}, \lc {{\cal G}^{R ++}} \rc, \xxx)
\ \ \ \mbox{(by Lemma \ref{def-goedel-dec-cr} and Corollary \ref{D-col1-cr} )} 
\nonumber \\ 
&\ra& \ \
\mbox{\CR}_{\Omega}[\rho_{e,{\Omega}}^{R}(\xxx)]
(\eee, \lc {\cal G}_{\Omega}^{R} \rc, \xxx)
\land \ 
\mbox{\CR}_{\Omega}[\neg\rho_{e,{\Omega}}^{R}(\xxx)]
(\eee^{++}, \lc {\cal G}^{\CR ++} \rc, \xxx)
\nonumber \\
&\ra& \ \
\mbox{\CR}_{\Omega}
[\rho_{e,{\Omega}}^{R}(\xxx) \ 
\lor \ \neg\rho_{e,{\Omega}}^{R}(\xxx)]
(\eee^{+++}, \lc {\cal G}^{\CR +++} \rc, \xxx)
\ \ ). 
\ \ \
\mbox{(by Lemma \ref{D2-cr})}
\end{eqnarray}

For any formula set $\Psi \equiv \{ \psi(\aaa) \mid  a \in \N \}$,
\begin{eqnarray}
& & 
\PA \vdash \
\forall \xxx \
( \
\rho_{e,{\Omega}}^{A}(\xxx) \ \land \ 
\neg\rho_{e,{\Omega}}^{A}(\xxx)
\ra \   \psi(\xxx) \
),
\label{eq:rbic-2nd-dec-100a}
\\
& & 
\PA \vdash \
\forall \xxx \
( \
\neg( \ \rho_{e,{\Omega}}^{R}(\xxx) \ \lor \ 
\neg\rho_{e,{\Omega}}^{R}(\xxx) \ ) \ 
\ra \   \neg\psi(\xxx) \
).
\label{eq:rbic-2nd-dec-100r}    
\end{eqnarray}

Hence by Corollaries \ref{D-col1-ca} and \ref{D-col1-cr},
for any $e^{+++} \in \N$, there exists $e^* \in \N$ 
such that
\begin{eqnarray}
\PA \vdash \
\forall \xxx \
( \ & &
\mbox{\CA}_{\Omega}[ \rho_{e,{\Omega}}^{A}(\xxx) \ \land \ 
\neg\rho_{e,{\Omega}}^{A}(\xxx) ]
(\eee^{+++}, \lc {\cal G}^{A +++} \rc, \xxx) \ 
\nonumber \\
&\ra& \ 
\mbox{\CA}_{\Omega}[\psi(\xxx)]
(\eee^*, \lc \Psi \rc, \xxx) \ ),
\label{eq:rbic-2nd-dec-2a}  
\\
\PA \vdash \
\forall \xxx \
( \ & &
\mbox{\CR}_{\Omega}[ \rho_{e,{\Omega}}^{R}(\xxx) \ \land \ 
\neg\rho_{e,{\Omega}}^{R}(\xxx) ]
(\eee^{+++}, \lc {\cal G}^{R +++} \rc, \xxx) \ 
\nonumber \\
&\ra& \ 
\mbox{\CR}_{\Omega}[\psi(\xxx)]
(\eee^*, \lc \Psi \rc, \xxx) \ ).
\label{eq:rbic-2nd-dec-2r}  
\end{eqnarray}

Therefore,
for any $e \in \N$, there exists $e^* \in \N$ 
such that
\begin{eqnarray}
& & 
\PA \vdash \
\forall \xxx \ \ 
( \  
\mbox{\CA}_{\Omega}[\rho_{e,{\Omega}}^{A}(\xxx)]
(\eee, \lc {\cal G}_{\Omega}^{A} \rc, \xxx)  
\ \ra \ 
\mbox{\CA}_{\Omega}[\psi(\xxx)]
(\eee^*, \lc \Psi \rc, \xxx)  \ ),
\label{eq:rbic-2nd-dec-3a}   
\\
& & 
\PA \vdash \
\forall \xxx \ \ 
( \   
\mbox{\CR}_{\Omega}[\rho_{e,{\Omega}}^{R}(\xxx)]
(\eee, \lc {\cal G}_{\Omega}^{R} \rc, \xxx)
\ \ra \ 
\mbox{\CR}_{\Omega}[\psi(\xxx)]
(\eee^*, \lc \Psi \rc, \xxx) \ \ 
).
\label{eq:rbic-2nd-dec-3r}   
\end{eqnarray}

That is,
\begin{eqnarray}
& & 
\PA \vdash \
\forall \xxx \ \ 
( \ 
\neg\mbox{\CA}_{\Omega}[\psi(\xxx)]
(\eee^*, \lc \Psi \rc, \xxx) \
\ \ra \ 
\neg\mbox{\CA}_{\Omega}[\rho_{e,{\Omega}}^{A}(\xxx)]
(\eee, \lc {\cal G}_{\Omega}^{A} \rc, \xxx) 
\ ),
\label{eq:rbic-2nd-dec-4a}
\\
& & 
\PA \vdash \
\forall \xxx \ \ 
( \ 
\neg\mbox{\CR}_{\Omega}[\psi(\xxx)]
(\eee^*, \lc \Psi \rc, \xxx) \
 \ \ra \ 
\neg\mbox{\CR}_{\Omega}[\rho_{e,{\Omega}}^{R}(\xxx)]
(\eee, \lc {\cal G}_{\Omega}^{R} \rc, \xxx) \ \ 
\ ).
\label{eq:rbic-2nd-dec-4r}
\end{eqnarray}

In addition,
from the property of G{\"o}del sentences
(Lemmas \ref{def-goedel-dec-ca} and \ref{def-goedel-dec-cr}),
\begin{eqnarray*}
& & 
\PA \vdash 
\forall \xxx \ \
( \rho_{e,{\Omega}}^{A}(\xxx) \ \lra \ 
\neg\mbox{\CA}_{\Omega}[\rho_{e,{\Omega}}^{A}(\xxx)]
(\eee, \lc {\cal G}_{\Omega}^{A} \rc, \xxx) \
), 
\\
& & 
\PA \vdash 
\forall \xxx \ \
( \neg\rho_{e,{\Omega}}^{R}(\xxx) \ \lra \ 
\neg\mbox{\CR}_{\Omega}[\rho_{e,{\Omega}}^{R}(\xxx)]
(\eee, \lc {\cal G}_{\Omega}^{R} \rc, \xxx) \
).
\end{eqnarray*} 

Hence,
\begin{eqnarray}
& & 
\PA \vdash \
\forall \xxx \ \ 
( \ 
\neg\mbox{\CA}_{\Omega}[\psi(\xxx)](\eee^*, \lc \Psi \rc, \xxx) 
\ \ \
\ra \ \ \  
\rho_{e,{\Omega}}^{A}(\xxx) \ 
),
\\
& &
\PA \vdash \
\forall \xxx \ \ 
( \ 
\neg\mbox{\CR}_{\Omega}[\psi(\xxx)](\eee^*, \lc \Psi \rc, \xxx) 
\ \ \
\ra \ \ \  
\neg\rho_{e,{\Omega}}^{R}(\xxx) \ 
).
\end{eqnarray}

Since $e^*$ is computed from $e$ in a manner similar to those  
used in the lemmas and corollaries in Section \ref{DC-dec}, 
there exists a primitive recursive function $h$ 
such that 
for any formula sets 
$\Psi$ and $\Omega$, and
for any $e \in \N$,
\begin{eqnarray*}
& & 
\PA \vdash \
\forall \xxx \ \ 
( \ 
\neg\mbox{\CA}_{\Omega}[\psi(\xxx)](h(\eee), \lc \Psi \rc, \xxx) 
\ \ \
\ra \ \ \  
\rho_{e,{\Omega}}^{A}(\xxx) \ 
),
\\
& &
\PA \vdash \
\forall \xxx \ \ 
( \ 
\neg\mbox{\CR}_{\Omega}[\psi(\xxx)](h(\eee), \lc \Psi \rc, \xxx) 
\ \ \
\ra \ \ \  
\neg\rho_{e,{\Omega}}^{R}(\xxx) \ 
).
\end{eqnarray*} 

\begin{flushright}
$\dashv$
\end{flushright}
\end{proof}

\begin{lemma}\label{lem:2-rbicthm-dec-2}
Let $T$ be a consistent PT-extension of PA.
Let assume that there exist $e \in \N$, $e^* \in \N$, $x \in \N$
and a $\Delta_1$-formula set $\Psi\equiv \{ \psi(\aaa) \mid  a \in \N \}$
such that 
\begin{equation}\label{eq:rbic-2nd-dec-8}     
{\PTM}_{e}(x) \vdash_T \
\neg\mbox{\CD}[\psi(\xxx)](\eee^*, \lc \Psi \rc, \xxx).
\end{equation}

Then, there exists a primitive recursive function $s$
such that 
$\tilde{e} = s(e) \in \N$ and  
\begin{eqnarray}
& & 
\NNN \models \psi(\xxx) \ \ \ \Ra \ \ \
{\PTM}_{\tilde{e}}^{\Theta^A[e^*]}(x) \dec_{v_{\Theta^A[e^*]}^A} \
\neg\mbox{\CA}_{\Psi}[\psi(\xxx)](\eee^*, \lc \Psi \rc, \xxx),
\label{eq:rbic-2nd-dec-10a}
\\
& & 
\NNN \models \neg\psi(\xxx) \ \ \ \Ra \ \ \     
{\PTM}_{\tilde{e}}^{\Theta^R[e^*]}(x) \dec_{v_{\Theta^R[e^*]}^R} \
\neg\mbox{\CR}_{\Psi}[\psi(\xxx)](\eee^*, \lc \Psi \rc, \xxx),
\label{eq:rbic-2nd-dec-10r}
\end{eqnarray}
where 
$\Theta^A[e^*] \equiv \{ 
\neg\mbox{\CA}_{\Psi}[\psi(\aaa)](\eee^*, \lc \Psi \rc, \aaa) 
\mid a \in \N \}$, and
$\Theta^R[e^*] \equiv \{ 
\mbox{\CR}_{\Psi}[\psi(\aaa)](\eee^*, \lc \Psi \rc, \aaa) 
\mid a \in \N \}$.

\end{lemma}

\begin{proof}

PTM $\U_{\PTM}(\tilde{e},\cdot)$ is constructed 
using PTM $\U_{\PTM}(e,\cdot)$ as follows:
\begin{itemize}
\item
(Input: ) \ $(d, {\#}\Theta^A[e^*], x)  \in \N^3$ or
$(d, {\#}\Theta^R[e^*], x)  \in \N^3$.
\item
(Output: ) \ accept or reject
\item
Simulate $\U_{\PTM}(e,(p, {\#}\Phi[e^*], x))$,
and check whether its output is the G{\"o}del number of a valid proof tree
of $\neg\mbox{\CD}[\psi(\xxx)](\eee^*, \lc \Psi \rc, \xxx)$
by using $\U_{\PTM}(v_{T},\cdot)$. 
\item
Let input be $(d, {\#}\Theta^A[e^*], x)$.
Then, accept if and only if it is a valid proof tree.
\item
Let input be $(d, {\#}\Theta^R[e^*], x)$.
Then, reject if and only if it is a valid proof tree.
\end{itemize}

The running time of $\U_{\PTM}(\tilde{e},\cdot)$
is that of $\U_{\PTM}(e,\cdot)$ plus polynomial-time in 
$|x|$.
From the construction of $\U_{\PTM}(\tilde{e},\cdot)$,
there exists a primitive recursive function $s$
such that $\tilde{e} = s(e)$.

Since $T$ is a consistent PT-extension of PA,
if Eq. (\ref{eq:rbic-2nd-dec-8}) holds,
$$
\NNN \models \ \neg\mbox{\CD}[\psi(\xxx)](\eee^*, \lc \Psi \rc, \xxx).
$$
Then,
\begin{eqnarray*}
& & \mbox{ \     } \ 
\NNN \models \ \ \psi(\xxx) 
\\ 
\ \ \  &\Ra& \ \ \ 
\NNN \models \ \ \psi(\xxx) \ \land \ 
\neg\mbox{\CD}[\psi(\xxx)](\eee^*, \lc \Psi \rc, \xxx)
\\
\ \ \  &\LRa& \ \ \ 
\NNN \models \ \  
\neg\mbox{\PTM-\Acc}(\eee^*, \lc \Psi \rc, \xxx) \ \land \ 
\psi(\xxx) 
\\ 
\ \ \  &\LRa& \ \ \ 
\NNN \models \ \ \neg\mbox{\PTM-\Acc}(\eee^*, \lc \Psi \rc, \xxx) \ 
\land \ \NNN \models \ \ \psi(\xxx) 
\\
\ \ \  &\LRa& \ \ \ 
\NNN \models \ \ \neg\mbox{\PTM-\Acc}(\eee^*, \lc \Psi \rc, \xxx) \ 
\land \ \PA \vdash \  \psi(\xxx) 
\\
\ \ \  &\LRa& \ \ \ 
\NNN \models \ \ \neg\mbox{\PTM-\Acc}(\eee^*, \lc \Psi \rc, \xxx) \ 
\land \ \U_{\PTM}(v_{\Psi}^A,(d,{\#}\Psi,x)) \  \mbox{(soundly) accepts} 
\\
\ \ \  &\LRa& \ \ \ 
\NNN \models \ \ \neg\mbox{\PTM-\Acc}(\eee^*, \lc \Psi \rc, \xxx) \ 
\land \ \NNN \models \ \ \mbox{\Acc}(\vvv_{\Psi}^A, \lc \Psi \rc, \xxx) 
\\
\ \ \  &\LRa& \ \ \ 
\NNN \models \ \ \neg\mbox{\PTM-\Acc}(\eee^*, \lc \Psi \rc, \xxx) \ 
\land \ \mbox{\Acc}(\vvv_{\Psi}^A, \lc \Psi \rc, \xxx)
\\
\ \ \  &\Ra& \ \ \ 
\NNN \models \ \ 
\neg\mbox{\CA}_{\Psi}[\psi(\xxx)](\eee^*, \lc \Psi \rc, \xxx)
\end{eqnarray*}
and
\begin{eqnarray*}
& & \mbox{ \     } \ 
\NNN \models \ \ \neg\psi(\xxx) 
\\ 
\ \ \  &\Ra& \ \ \ 
\NNN \models \ \ \neg\psi(\xxx) \ \land \ 
\neg\mbox{\CD}[\psi(\xxx)](\eee^*, \lc \Psi \rc, \xxx)
\\
\ \ \  &\LRa& \ \ \ 
\NNN \models \ \ \mbox{\PTM-\Acc}(\eee^*, \lc \Psi \rc, \xxx) \ \land \ 
\neg\psi(\xxx) \ ) 
\\ 
\ \ \  &\LRa& \ \ \ 
\NNN \models \ \ \mbox{\PTM-\Acc}(\eee^*, \lc \Psi \rc, \xxx) \ 
\land \ \NNN \models \ \ \neg\psi(\xxx)  
\\
\ \ \  &\LRa& \ \ \ 
\NNN \models \ \ \mbox{\PTM-\Acc}(\eee^*, \lc \Psi \rc, \xxx) \ 
\land \ \PA \vdash \ \neg\psi(\xxx)
\\
\ \ \  &\LRa& \ \ \ 
\NNN \models \ \ \mbox{\PTM-\Acc}(\eee^*, \lc \Psi \rc, \xxx) \ 
\land \ \U_{\PTM}(v_{\Psi}^R,(d,{\#}\Psi,x)) \ \mbox{(soundly) rejects} 
\\
\ \ \  &\LRa& \ \ \ 
\NNN \models \ \ \mbox{\PTM-\Acc}(\eee^*, \lc \Psi \rc, \xxx) \ 
\land \ \NNN \models \ \ \neg\mbox{\Acc}(\vvv_{\Psi}^R, \lc \Psi \rc, \xxx)
\\
\ \ \  &\LRa& \ \ \ 
\NNN \models \ \ \mbox{\PTM-\Acc}(\eee^*, \lc \Psi \rc, \xxx) \ 
\land \ \neg\mbox{\Acc}(\vvv_{\Psi}^R, \lc \Psi \rc, \xxx)
\\
\ \ \  &\LRa& \ \ \ 
\NNN \models \ \ 
\neg\mbox{\CR}_{\Psi}[\psi(\xxx)](\eee^*, \lc \Psi \rc, \xxx).
\end{eqnarray*}

Therefore, if Eq. (\ref{eq:rbic-2nd-dec-8}) holds,
\begin{eqnarray*}
& &
\NNN \models \psi(\xxx) 
\ \ \  \Ra \ \ \ 
\NNN \models \ \ 
\neg\mbox{\CA}_{\Psi}[\psi(\xxx)](\eee^*, \lc \Psi \rc, \xxx),
\\
& &
\NNN \models \neg\psi(\xxx) 
\ \ \  \Ra \ \ \ 
\NNN \models \ \ 
\neg\mbox{\CR}_{\Psi}[\psi(\xxx)](\eee^*, \lc \Psi \rc, \xxx).
\end{eqnarray*} 

Since $\neg\mbox{\CA}_{\Psi}[\psi(\xxx)](\eee^*, \lc \Psi \rc, \xxx)$
and $\neg\mbox{\CR}_{\Psi}[\psi(\xxx)](\eee^*, \lc \Psi \rc, \xxx)$
are $\Delta_1$-formulas,
\begin{eqnarray*}
& &
\NNN \models \psi(\xxx) 
\ \ \  \Ra \ \ \ 
\PA \vdash \ \ 
\neg\mbox{\CA}_{\Psi}[\psi(\xxx)](\eee^*, \lc \Psi \rc, \xxx),
\\
& &
\NNN \models \neg\psi(\xxx) 
\ \ \  \Ra \ \ \ 
\PA \vdash \ \ 
\neg\mbox{\CR}_{\Psi}[\psi(\xxx)](\eee^*, \lc \Psi \rc, \xxx).
\end{eqnarray*} 

Then, from the definition of 
$\U(v_{\Theta[e^*]}^A,\cdot)$ 
and $\U(v_{\Theta[e^*]}^R,\cdot)$ (see Definition \ref{def-v-{Omega}*}),
$\U(v_{\Theta[e^*]}^A,(d, {\#}\Theta^A[e^*], x))$
accepts if $\NNN \models \psi(\xxx)$, 
and
$\U(v_{\Theta[e^*]}^A,(d, {\#}\Theta^R[e^*], x))$
rejects if $\NNN \models \neg\psi(\xxx)$.

On the other hand, 
from the construction of $\U_{\PTM}(\tilde{e},\cdot)$,
if Eq. (\ref{eq:rbic-2nd-dec-8}) holds,
$\U_{\PTM}(\tilde{e},(d, {\#}\Theta^A[e^*], x))$
accepts, and
$\U_{\PTM}(\tilde{e},(d, {\#}\Theta^R[e^*], x))$
rejects.
Thus,
\begin{eqnarray*}
& &
\NNN \models \psi(\xxx) \ \ \ \Ra \ \ \
{\PTM}_{\tilde{e}}^{\Theta^A[e^*]}(x) \dec_{v_{\Theta^A[e^*]}^A} \
\neg\mbox{\CA}_{\Psi}[\psi(\xxx)](\eee^*, \lc \Psi \rc, \xxx),
\\
& &
\NNN \models \neg\psi(\xxx) \ \ \ \Ra \ \ \     
{\PTM}_{\tilde{e}}^{\Theta^R[e^*]}(x) \dec_{v_{\Theta^R[e^*]}^R} \
\neg\mbox{\CR}_{\Psi}[\psi(\xxx)](\eee^*, \lc \Psi \rc, \xxx),
\end{eqnarray*}

\begin{flushright}
$\dashv$
\end{flushright}
\end{proof}

\begin{theorem}\label{rbic-2nd-dec}


Let $T$ be a consistent PT-extension of PA.
For any set of $\Delta_1$-formulas
$\Psi \equiv \{ \psi(\aaa) \mid  a \in \N \}$,
$$  
\forall e \in \N \ \ \exists e^* \in \N \ \ 
\forall x \in \N  \ \ \
{\PTM}_{e}(x) \not\vdash_T \
\neg\mbox{\CD}[\psi(\xxx)](\eee^*, \lc \Psi \rc, \xxx).$$

\end{theorem}

\begin{proof}

We assume that 

there exist $e \in \N$ and a formula set $\Psi$
such that 
\begin{equation}\label{eq:rbic-2nd-dec-7}     
\forall e^* \in \N \ \ \exists x \in \N \ \ \ 
{\PTM}_{e}(x) \vdash_T \
\neg\mbox{\CD}[\psi(\xxx)](\eee^*, \lc \Psi \rc, \xxx). 
\end{equation}

Then,
from Lemma \ref{lem:2-rbicthm-dec-2},
we can construct $\U_{\PTM}(\tilde{e},\cdot)$ 
using $\U_{\PTM}(e,\cdot)$ such that $\tilde{e} = s(e)$
and $\U_{\PTM}(\tilde{e},\cdot)$ satisfy
Eqs. (\ref{eq:rbic-2nd-dec-10a}) and (\ref{eq:rbic-2nd-dec-10r}).

Then, there exists a primitive recursive function $t$
such that $e' = t(\tilde{e}) \in \N$ and   
PTM $\U_{\PTM}(e',\cdot)$ is constructed 
using PTM $\U_{\PTM}(\tilde{e},\cdot)$ as follows:
\begin{itemize}
\item
(Input: ) \ 
$(d, {\#}{\cal G}_{\Theta^A[h(e')]}^{A}, a)  \in \N^3$
or $(d, {\#}{\cal G}_{\Theta^R[h(e')]}^{R}, a)  \in \N^3$
where ${\cal G}_{\Theta^A[h(e')]}^{A} 
\equiv \{ \rho_{e',{\Theta^A[h(e')]}}^{A}(\aaa) \mid a \in \N \}$
and ${\cal G}_{\Theta^R[h(e')]}^{R} 
\equiv \{ \rho_{e',{\Theta^R[h(e')]}}^{R}(\aaa) \mid a \in \N \}$. 
(For the definitions of $\Theta^A[\cdot]$ and 
$\Theta^R[\cdot]$, see Lemma \ref{lem:2-rbicthm-dec-2}.)
\item
(Output: ) \ accept or reject
\item
First, read its own code, $e' \in \N$ via the recursion 
theorem (Proposition \ref{recursion}).
\item
If input is $(d, {\#}{\cal G}_{\Theta^A[h(e')]}^{A}, a)$,
then 
simulate $\U_{\PTM}(\tilde{e},(d, {\#}\Theta^A[h(e')], a))$,
and accept if and only if 
$\U_{\PTM}(\tilde{e},(d, {\#}\Theta^A[h(e')], a))$ accepts.
\item
If input is $(d, {\#}{\cal G}_{\Theta^R[h(e')]}^{R}, a)$,
then 
simulate $\U_{\PTM}(\tilde{e},(d, {\#}\Theta^R[h(e')], a))$,
and reject if and only if 
$\U_{\PTM}(\tilde{e},(d, {\#}\Theta^R[h(e')], a))$ rejects.
\end{itemize}

The running time of $\U_{\PTM}(e',\cdot)$
is that of $\U_{\PTM}(e,\cdot)$ plus polynomial-time in 
$|a|$.

By substituting $\Theta[h(e')]$
for $\Omega$,
in Eqs. (\ref{eq:rbic-2nd-dec-5a}) and (\ref{eq:rbic-2nd-dec-5r}),
we obtain that
for any formula set 
$\Psi \equiv \{\psi(\aaa) \mid a \in \N \}$ and 
for any $e' \in \N$,
\begin{eqnarray}
& & 
\PA \vdash \
\forall \xxx \ \ 
( \ 
\neg\mbox{\CA}_{\Theta^A[h(e')]}[\psi(\xxx)](h(\eee'), \lc \Psi \rc, \xxx) 
\ \ \
\ra \ \ \  
\rho_{e',\Theta^A[h(e')]}^{A}(\xxx) \ 
),
\label{eq:rbic-2nd-dec-5-2a}    
\\
& &
\PA \vdash \
\forall \xxx \ \ 
( \ 
\neg\mbox{\CR}_{\Theta^R[h(e')]}[\psi(\xxx)](h(\eee'), \lc \Psi \rc, \xxx) 
\ \ \
\ra \ \ \  
\neg\rho_{e',\Theta^R[h(e')]}^{R}(\xxx) \ 
).
\label{eq:rbic-2nd-dec-5-2r}    
\end{eqnarray} 

For all $e \in \N$ and all $a \in \N$,
\begin{eqnarray*}
& & 
\NNN \models \ \psi(\xxx)
\ \ \ \Ra \ \ \
f({\#}\Psi, \Size_{\Psi}(a)) < 
f({\#}\Theta^A[e], \Size_{\Theta^A[e]}(a)), 
\\
& & 
\NNN \models \ \neg\psi(\xxx)
\ \ \ \Ra \ \ \ 
f({\#}\Psi, \Size_{\Psi}(a)) < 
f({\#}\Theta^R[e], \Size_{\Theta^R[e]}(a))
\end{eqnarray*} 
(for the definition of function $f$, 
see Definition \ref{def-v-{Omega}*}).

Therefore,
\begin{eqnarray}
& &
\NNN \models \ \psi(\xxx) \ \ \ \Ra 
\nonumber
\\
& & \mbox{ } \ \ \ \ \ 
\PA \vdash \
\forall \xxx \ \ 
( \ 
\neg\mbox{\CA}_{\Psi}[\psi(\xxx)](h(\eee'), \lc \Psi \rc, \xxx) 
\ \ \
\lra \ \ \  
\neg\mbox{\CA}_{\Theta^A[h(e')]}[\psi(\xxx)](h(\eee'), \lc \Psi \rc, \xxx) 
\ ),
\label{eq:rbic-2nd-dec-9a}
\\
& & 
\NNN \models \ \neg\psi(\xxx) \ \ \ \Ra 
\nonumber
\\
& & \mbox{ } \ \ \ \ \ 
\PA \vdash \
\forall \xxx \ \ 
( \ 
\neg\mbox{\CR}_{\Psi}[\psi(\xxx)](h(\eee'), \lc \Psi \rc, \xxx) 
\ \ \
\lra \ \ \  
\neg\mbox{\CR}_{\Theta^R[h(e')]}[\psi(\xxx)](h(\eee'), \lc \Psi \rc, \xxx) 
\ ).
\label{eq:rbic-2nd-dec-9r}    
\end{eqnarray}

Hence, if Eq. (\ref{eq:rbic-2nd-dec-7}) holds,
then
by applying Eqs. (\ref{eq:rbic-2nd-dec-10a}), (\ref{eq:rbic-2nd-dec-10r}),
(\ref{eq:rbic-2nd-dec-5-2a}), (\ref{eq:rbic-2nd-dec-5-2r}),
(\ref{eq:rbic-2nd-dec-9a}), and (\ref{eq:rbic-2nd-dec-9r}),
\begin{eqnarray*}
& & 
\NNN \models \ \psi(\xxx) \ \ \ \Ra \ \ \
\U(v_{\Theta^A[h(e')]}^A, 
(d, {\#}{\cal G}_{\Theta^A[h(e')]}^{A}, a)) \ 
\mbox{accepts},
\\
& & 
\NNN \models \ \neg\psi(\xxx) \ \ \ \Ra \ \ \
\U(v_{\Theta^R[h(e')]}^R, 
(d, {\#}{\cal G}_{\Theta^R[h(e')]}^{R}, a)) \ 
\mbox{rejects}.
\end{eqnarray*}

On the other hand, if Eq. (\ref{eq:rbic-2nd-dec-7}) holds, 
from the construction of $\U_{\PTM}(e',\cdot)$
and $\U_{\PTM}(\tilde{e},\cdot)$,
\begin{eqnarray*}
& &
\U_{\PTM}(e',(d, {\#}{\cal G}_{\Theta^A[h(e')]}^{A}, a)) \
\mbox{accepts},
\\
& & 
\U_{\PTM}(e',(d, {\#}{\cal G}_{\Theta^R[h(e')]}^{R}, a)) \
\mbox{rejects}.
\end{eqnarray*} 

Hence, if Eq. (\ref{eq:rbic-2nd-dec-7}) holds, 
for a formula set $\Psi$, 
\begin{eqnarray*}
& &
\NNN \models \ \psi(\xxx) \ \ \ \Ra \ \ \
\exists e' \in \N \ \
\exists x \in \N \ \ \ 
\PTM_{e'}^{{\cal G}_{\Phi[h(e')]}^{A}}(x) \ \dec_{v_{\Phi[h(e')]}^A} \  
\rho_{e',{\Phi[h(e')]}}^{A}(\xxx),
\\
& &
\NNN \models \ \neg\psi(\xxx) \ \ \ \Ra \ \ \
\exists e' \in \N \ \
\exists x \in \N \ \ \ 
\PTM_{e'}^{{\cal G}_{\Phi[h(e')]}^{R}}(x) \ \dec_{v_{\Phi[h(e')]}^R} \  
\neg\rho_{e',{\Phi[h(e')]}}^{R}(\xxx).
\end{eqnarray*} 

This contradicts Corollaries \ref{rbic-1-col-dec-ca} and
\ref{rbic-1-col-dec-cr}.
Therefore, Eq. (\ref{eq:rbic-2nd-dec-7}) does not hold
for $e^* = h(e') = h(t(s(e)))$.
That is,
there exists a primitive recursive function $g$
such that 
for any $e \in \N$, for any $\Psi$, and
for any $x \in \N$
$$ 
{\PTM}_{e}(x) \not\vdash_T \
\neg\mbox{\CD}[\psi(\xxx)](g(\eee), \lc \Psi \rc, \xxx), 
$$
where $g(e) = h(t(s(e)))$.

\begin{flushright}
$\dashv$
\end{flushright}
\end{proof}

\begin{corollary}\label{cor:rbic-2nd-dec}
Let $T$ be a consistent PT-extension of PA.
There exists a primitive recursive function
$g$ such that for any set of $\Delta_1$-formulas
$\Psi \equiv \{ \psi(\aaa) \mid  a \in \N \}$,
\begin{eqnarray*}
& &
\NNN \models \psi(\xxx) \ \ \ \Ra \ \ \
\forall e \in \N \ \ 
\forall x \in \N \ \ \ 
{\PTM}_{e}(x) \not\vdash_T \
\neg\mbox{\CA}[\psi(\xxx)](g(\eee), \lc \Psi \rc, \xxx),
\\
& &
\NNN \models \neg\psi(\xxx) \ \ \ \Ra \ \ \
\forall e \in \N \ \ 
\forall x \in \N \ \ \ 
{\PTM}_{e}(x) \not\vdash_T \
\neg\mbox{\CR}[\psi(\xxx)](g(\eee), \lc \Psi \rc, \xxx).
\end{eqnarray*} 

\end{corollary}


\section{Formalization of P$\not=$NP and 
a Super-Polynomial-Time Lower Bound}\label{formaldef-pnp}
We now introduce the notations and definitions 
necessary to consider the P$\not=$NP problem
in this paper.  
We omit the fundamental concepts and definitions
regarding P and NP (see \cite{Sipser97} for them). 
 
\subsection{${\ov{\mbox{P$\not=$NP}}}$}\label{subsec:formaldef-pnp}

\begin{definition}\label{def-sat}
Let $R_{\TSAT} \subset \N$ be a relation
such that
$x \in R_{\TSAT}$ if and only if 
there exists a satisfiable 3CNF formula $\phi$
and $x = {\#}\phi$. 
Let 
$\SAT(\xxx)$ be a formula in \PA
and $\SAT(\xxx)$ represent relation 
$R_{\TSAT}$ in \PA. 
(see Section \ref{convention} for representability.) 
That is,
for every $a \in \N$
\begin{eqnarray*}
& & 
a \in R_{\TSAT} \ \ \Ra \ \ 
\PA \vdash \SAT(\aaa),
\\
& & 
a \not\in R_{\TSAT}  \ \ \Ra \ \ 
\PA \vdash \neg\SAT(\aaa).
\end{eqnarray*}
Let $\CSAT$ be a set of formulas in \PA,
$\{ \SAT(\aaa) \mid a \in \N \}$,
and 
$\cCSAT$ be a set of formulas in \PA,
$\{ \neg\SAT(\aaa) \mid a \in \N \}$.
Let $\Size_{\CSAT}(a)$ and $\Size_{\cCSAT}(a)$ be
$|a|$.
Let ${\cal DS}$ be $\CSAT \cup \cCSAT$.
\end{definition}

\begin{definition}
Let theory $T$ be a PT-extension of PA. 
For $e \in \N$,
let 
$\U_{\PTM}(e,(d,{\#}\CSAT, \cdot))$, 
on input $x \in \N$, 
output 
one bit decision, whether $x \in R_{\TSAT}$ or $x \not\in R_{\TSAT}$;
in other words, $\SAT(\xxx)$ is true or false.
\end{definition}

We then define a formula that characterizes
the fact that a PTM, $\U_{\PTM}(e,(\cdot, \cdot))$,
given $x \in \N$, 
can solve the problem of deciding the truth/falsity of formula $\SAT(\xxx)$.
\begin{definition}\label{decsat-solvesat}
Let theory $T$ be a PT-extension of \PA.
$$\DecSAT(\eee,\xxx)$$ 
denotes a $\Delta_1$-formula in $\PA$, which 
represents the following primitive recursive relation
on $(e,x) \in \N^2$ 
such that
$$ 
\U_{\PTM}(e, (d,{\#}\CSAT, x)) \  \mbox{\rm accepts} \ 
\LRa \
x \in R_{\TSAT}. 
$$
More precisely, let 
$$
\DecSAT(\eee,\xxx)
\ \ \equiv \ \ 
\CD[\SAT(\xxx)](\eee, \lc \CSAT \rc, \xxx)
$$ 
(For the definition of this notation,
see Section \ref{formal-pt-dec}).
This primitive recursive relation on $(e,x)$ means whether
the decision (on $x \in R_{\TSAT}$)
of PTM $\U_{\PTM}(e, (d,{\#}\CSAT, \cdot))$
is correct or not.

\end{definition}

We now introduce the Cook-Levin Theorem \cite{Sipser97}, which 
characterizes the P vs NP problem by the satisfiability problem, 
3SAT (an NP-complete problem). 

\begin{proposition}\label{cook-levin}
(Cook-Levin Theorem) \
$$
\exists e \in \N \ \ \exists n \in \N \ \ \forall x \geq n  \ \ \ 
( \ 
\U_{\PTM}(e, (d,{\#}\CSAT, x)) \  \mbox{\rm accepts} \ 
\LRa \
x \in R_{\TSAT} \ 
) 
$$
if and only if \ \ $\mbox{P = NP}$.
\end{proposition}


\begin{lemma}\label{solvesat}
Let theory $T$ be a consistent PT-extension of PA.  
$$
\forall e \in \N \ \ \forall n \in N \ \ \exists x \geq n \ \ \
T \vdash \  \neg\DecSAT(\eee,\xxx),
$$
if and only if \ \ $\mbox{P$\not=$NP}$.

\end{lemma}

\begin{proof} 
From the representability theorem 
(Proposition \ref{c-representability}) regarding formula \\
$\DecSAT(\eee,\xxx)$,
the statement of this lemma is equivalent to 
$$
\exists e \in \N \ \ \exists n \in N \ \ \forall x \geq n \ \ \
T \vdash \  \DecSAT(\eee,\xxx),
$$
if and only if \ \ $\mbox{P$=$NP}$.

Thus, we obtain the statement of this lemma from
the definitions of formula $\DecSAT(\eee,\xxx)$, and 
Proposition \ref{cook-levin}.

\begin{flushright}
$\dashv$
\end{flushright}
\end{proof}

\begin{lemma}\label{solvesatY}
Let theory $T$ be a PT-extension of \PA \ and
$\omega$-consistent.  
$$
\forall e \in \N \ \ \forall n \in N \ \ \ 
T \vdash \  \exists \xxx \geq \nnn \ \ \
\neg\DecSAT(\eee,\xxx),
$$
if and only if \  P$\not=$NP.

\end{lemma}

\begin{proof}  \ \  

\noindent
(If:)

When
$$
\exists x \in \N \ \ \
T \vdash \  \neg\DecSAT(\eee,\xxx),
$$
the following holds
$$
T \vdash \ \exists \xxx \ \ \neg\DecSAT(\eee,\xxx).
$$

\noindent
(Only if:)

The following claim is obtained from 
$\omega$-consistency.  
\begin{claim}\label{solvesatX}
Let theory $T$ be a PT-extension of \PA \ and
$\omega$-consistent.  
\begin{eqnarray*}  
\exists e \in \N \ \ \exists n \in N \ \ \forall x \geq n \ \ \
T &\vdash& \  \DecSAT(\eee,\xxx).
\\
\Ra \ \ \
\exists e \in \N \ \ \exists n \in N \ \ \ 
T &\not\vdash& \  \exists \xxx \geq \nnn \ \ \
\neg\DecSAT(\eee,\xxx).
\end{eqnarray*}

\end{claim}

%

From Definition \ref{vartheta},
if P=NP, 
$$
\exists e \in \N \ \ \exists n \in N \ \ \forall x \geq n \ \ \
T \vdash \  \DecSAT(\eee,\xxx).
$$
We then have the following equation from the above-mentioned claim,
$$
\exists e \in \N \ \ \exists n \in N \ \ \ 
T \not\vdash \  \exists \xxx \geq \nnn \ \ \
\neg\DecSAT(\eee,\xxx).
$$
Hence, if
$$
\forall e \in \N \ \ \forall n \in N \ \ \ 
T \vdash \  \exists \xxx \geq \nnn \ \ \
\neg\DecSAT(\eee,\xxx),
$$
then P$\not=$NP.

\begin{flushright}
$\dashv$
\end{flushright}
\end{proof}

\noindent
{\bf Note:} \ \ 
This lemma implies
\begin{eqnarray*}  
\forall e \in \N \ \ \forall n \in N \ \ \ 
T &\vdash& \  \exists \xxx \geq \nnn \ \ \
\neg\DecSAT(\eee,\xxx)
\\
\LRa \ \ \
\forall e \in \N \ \ \forall n \in N \ \ \exists x \geq n \ \ \ 
T &\vdash& \  
\neg\DecSAT(\eee,\xxx).
\end{eqnarray*}

\begin{definition}\label{vartheta}
Let ${\ov{\mbox{P$\not=$NP}}}$ be 
a formula (sentence) in \PA \ such that
$$
{\ov{\mbox{P$\not=$NP}}} \ \ \equiv \ \
\forall \eee \ \ \forall \nnn \ \ \exists  \xxx \geq \nnn \ \ \
\neg\DecSAT(\eee,\xxx).
$$

\end{definition}

\begin{lemma}\label{pnp-formalization-and-standard-model}
$$
\NNN \models \ {\ov{\mbox{P$\not=$NP}}}, 
$$
if and only if {P$\not=$NP}.
\end{lemma}

\begin{proof} 

\begin{eqnarray*}
& & \mbox{P$\not=$NP}
\\
\LRa & & 
\forall e \in \N \ \ \forall n \in N \ \ \exists x \geq n \ \ \
\PA \vdash \  \neg\DecSAT(\eee,\xxx) \ \ \ \ \ (\mbox{from Lemma \ref{solvesat}})
\\
\LRa & & 
\forall e \in \N \ \ \forall n \in N \ \ \exists x \geq n \ \ \
\NNN \models \  \neg\DecSAT(\eee,\xxx) \ \ \ \ \ (\mbox{since $\neg\DecSAT(\eee,\xxx)$ is $\Delta_1$-formula})
\\
\LRa & & 
\NNN \models \ {\ov{\mbox{P$\not=$NP}}}.
\end{eqnarray*}

\begin{flushright}
$\dashv$
\end{flushright}
\end{proof} 

\begin{lemma}\label{solvesatV}
Let theory $T$ be a PT-extension of \PA \ and
$\omega$-consistent.  
If 
$$
T \vdash \ {\ov{\mbox{P$\not=$NP}}}, 
$$
then {P$\not=$NP}.
\end{lemma}

\begin{proof} 
If  
$$
T \vdash \  
\forall \eee \ \ \forall \nnn \ \ \exists \xxx \geq \nnn \ \ \
\neg\DecSAT(\eee,\xxx),
$$
then
$$ \forall e \in \N \ \ \forall n \in N \ \ \ 
T \vdash \  \exists \xxx \geq \nnn \ \ \
\neg\DecSAT(\eee,\xxx).
$$
We then obtain {P$\not=$NP} by Lemma \ref{solvesatY}.

\begin{flushright}
$\dashv$
\end{flushright}
\end{proof}

\subsection{Formalization of a Super-Polynomial-Time Lower Bound}
\label{subsec:super-poly}

This section shows a formalization of a super-polynomial-time 
lower bound in PA in a manner similar to ${\ov{\mbox{P$\not=$NP}}}$.

\begin{definition}\label{def-L}
Let $L$ be a language (a set of binary strings)
in PSPACE.
Let $R_L \subset \N$ be a relation
such that
$x \in R_L$ if and only if
$[x] \in L$.
Let 
$\phi_L(\xxx)$ be a formula in \PA
and $\phi_L(\xxx)$ represent relation 
$R_L$ in \PA. 
(see Section \ref{convention} for representability.) 
That is,
for every $a \in \N$
\begin{eqnarray*}
& & 
a \in R_L \ \ \Ra \ \ 
\PA \vdash \phi_L(\aaa),
\\
& & 
a \not\in R_L  \ \ \Ra \ \ 
\PA \vdash \neg\phi_L(\aaa).
\end{eqnarray*}
Let $\Phi_L$ be a set of formulas in \PA,
$\{ \phi_L(\aaa) \mid a \in \N \}$,
and 
$\Size_{\Phi_L}(a)$ be $|a|$.

\end{definition}
 
\begin{definition}\label{def:super-poly}
$$
\forall e \in \N \ \ \forall n \in \N \ \ \exists x \geq n  \ \ \ 
\neg( \ 
\U_{\PTM}(e, (d,{\#}\Phi_L, x)) \  \mbox{\rm accepts} \ 
\LRa \
x \in R_L \ 
) 
$$
if and only if \ \ $L$ has a {\it super-polynomial-time 
computational lower bound}.
\end{definition}

\begin{lemma}\label{lem:super-poly-1}
Let theory $T$ be a consistent PT-extension of PA.  
$$
\forall e \in \N \ \ \forall n \in N \ \ \exists x \geq n \ \ \
T \vdash \  \neg\CD[\phi_L(\xxx)](\eee, \lc \Phi_L \rc, \xxx),
$$
if and only if \ $L$ has a super-polynomial-time 
computational lower bound.
\end{lemma}

\begin{proof}
This is obtained from   
the representability theorem 
(Proposition \ref{c-representability}) regarding formula \\
$\CD[\phi_L(\xxx)](\eee, \lc \Phi_L \rc, \xxx)$,
and the definition of this formula notation, $\CD$ (see 
Section \ref{formal-pt-dec}).

\begin{flushright}
$\dashv$
\end{flushright}
\end{proof}

The following lemmas can be proven
in a manner similar to those used in Lemmas
\ref{solvesatY} and \ref{solvesatV}.

\begin{lemma}\label{lem:super-poly-2}
Let theory $T$ be a PT-extension of \PA \ and
$\omega$-consistent.  
$$
\forall e \in \N \ \ \forall n \in N \ \ \ 
T \vdash \  
\exists \xxx \geq \nnn \ \ \
\neg\CD[\phi_L(\xxx)](\eee, \lc \Phi_L \rc, \xxx),
$$
if and only if \ $L$ has a super-polynomial-time 
computational lower bound.

\end{lemma}

\begin{lemma}\label{lem:super-poly-3}
Let theory $T$ be a PT-extension of PA and
$\omega$-consistent.  
If 
$$
T \vdash \ 
\forall \eee \ \ \forall \nnn \ \ \exists \xxx \geq \nnn \ \ \
\neg\CD[\phi_L(\xxx)](\eee, \lc \Phi_L \rc, \xxx),
$$
then $L$ has a super-polynomial-time 
computational lower bound.

\end{lemma}

\section{Unprovability of P$\not=$NP and Super-Polynomial-Time Lower Bounds}
\label{sec:unprove-dec}

This section shows 
that there exists no formal proof of $\ov{\mbox{P$\not=$\NP}}$ 
in $T$, if $T$ is a consistent PT-extension of PA 
and \PTM-$\omega$-consistent for $\Delta_2^P$.
This result is based on the second incompleteness theorem 
of polynomial-time decisions, Theorem \ref{rbic-2nd-dec}.  

\subsection{PTM-$\omega$-Consistency}
\label{sec:ptm-omega-con}

\begin{definition}\label{delta-p-2}
Formula $\phi(\xxx)$ in PA is called 
$\Sigma^P_i$ ($i = 1,2, \ldots$) if 
there exists a formula $\psi(\xxx)$ in PA 
such that 
$$
\PA \vdash \ \forall \xxx
\ ( \phi(\xxx) \ \lra \ \psi(\xxx) \ ),  
$$
$$
\psi(\xxx) \equiv \
\exists \www_1 < \2^{|\xxx|^{\ccc_1}} \ 
\forall \www_2 < \2^{|\xxx|^{\ccc_2}} \ \cdots \ 
Q_i \www_i < \2^{|\xxx|^{\ccc_i}} \ \psi_0(\xxx,\www_1, \ldots,\www_i),
$$
where 
$Q_i$ is $\forall$ or $\exists$,
$\psi_0(\xxx,\www_1, \ldots,\www_i)$ 
is a formula that represents a polynomial-time relation
over $(x,w_1,\ldots,w_i)$, 
$c_j$ ($0 \leq j \leq i$) is a constant (in $|x|$). 

Similarly, 
formula $\phi(\xxx)$ in PA is called 
$\Pi^P_i$ ($i = 1,2, \ldots$) if 
there exists a formula $\psi(\xxx)$ in PA 
such that 
$$
\PA \vdash \ \forall \xxx
\ ( \phi(\xxx) \ \lra \ \psi(\xxx) \ ),  
$$
$$
\psi(\xxx) \equiv \
\forall \www_1 < \2^{|\xxx|^{\ccc_1}} \ 
\exists \www_2 < \2^{|\xxx|^{\ccc_2}} \ \cdots \ 
Q_i \www_i < \2^{|\xxx|^{\ccc_i}} \ \psi_0(\xxx,\www_1, \ldots,\www_i).
$$

Formula $\phi(\xxx)$ in PA is called 
$\Delta^P_i$ ($i = 1,2, \ldots$) if 
$\phi(\xxx)$ is $\Sigma^P_i$ and $\Pi^P_i$.

Formula $\phi(\xxx)$ in PA is called 
QBF if 
there exists a formula $\psi(\xxx)$ in PA 
such that 
$$
\PA \vdash \ \forall \xxx
\ ( \phi(\xxx) \ \lra \ \psi(\xxx) \ ),  
$$
$$
\psi(\xxx) \equiv \
\forall \www_1 < \2^{|\xxx|^{\ccc_1}} \ 
\exists \www_2 < \2^{|\xxx|^{\ccc_2}} \ \cdots \ 
Q_i \www_i < \2^{|\xxx|^{\ccc_k}} \ \psi_0(\xxx,\www_1, \ldots,\www_k),
$$ 
where $k \equiv |x|^c$ for a constant $c$.

\end{definition}

\begin{definition}\label{PTM-omega-con}
(\PTM-$\omega$-consistency) \ 
Let theory $S$ be a PT-extension of theory $T$.
$S$ is \PTM-$\omega$-{\it inconsistent} for $\Delta_1$-formula 
$\phi(\eee^*,\xxx)$ over $T$, 
if the following two conditions hold simultaneously. 
\begin{eqnarray} 
& & \forall e \in \N \ \exists e^* \in \N 
\ \exists \ell  \in \N \ \forall n \geq \ell 
\ \forall c \in \N \ \ \
\PTM_{e}(n) \ 
\not\vdash_T  
\exists \xxx \ (\nnn \leq \xxx < \nnn+|\nnn|^{\ccc}) \ \ 
\phi(\eee^*,\xxx),
\label{eq:PTM-omega-con-2} 
\\
& & \exists e \in \N \ \forall e^* \in \N \ 
\forall \ell \in \N 
\ \exists n \geq \ell \ \ \ \PTM_{e}(n) \ 
\vdash_S  
\exists \xxx \geq \nnn \ \ \phi(\eee^*,\xxx).
\label{eq:PTM-omega-con-1}
\end{eqnarray} 

Here,
$\Size_{\Phi[c]}(n) = |n|^{c+1}$, and
$
\Phi[c] \equiv 
\{ \exists \xxx \ (\aaa \leq \xxx < \nnn+|\nnn|^{\ccc}) \ \ 
\phi(\eee^*,\xxx)
\mid  a \in \N \}
$.

Theory $S$ is PTM-$\omega$-{\it consistent} for $\phi(\eee^*,\xxx)$ 
over $T$,
if theory $S$ is not PTM-$\omega$-inconsistent for 
$\phi(\eee^*,\xxx)$ over $T$.

Theory $S$ is PTM-$\omega$-consistent for $\Sigma^P_i$
($\Pi^P_i$, $\Delta^P_i$, resp.) over $T$,
if $S$ is PTM-$\omega$-consistent for any $\Sigma^P_i$ 
($\Pi^P_i$, $\Delta^P_i$, resp.) formula 
$\phi(\eee^*,\xxx)$ over $T$.

Theory $T$ is PTM-$\omega$-consistent for $\phi(\eee^*,\xxx)$
($\Sigma^P_i$, $\Pi^P_i$, $\Delta^P_i$, resp.),
if $T$ is PTM-$\omega$-consistent for $\phi(\eee^*,\xxx)$
($\Sigma^P_i$, $\Pi^P_i$, $\Delta^P_i$, resp.) over $T$.

\end{definition}

The following definition is equivalent to the above:
Theory $S$ is \PTM-$\omega$-consistent for $\phi(\eee^*,\xxx)$ 
over $T$, 
if the following condition holds.
\begin{eqnarray}  
& & 
\forall e \in \N \ \ \exists e^* \in \N 
\ \ \exists \ell  \in \N \ \ \forall n \geq \ell 
\ \ \forall c \in \N \ \ \
\PTM_{e}(n) \ 
\not\vdash_T \ 
\exists \xxx \ (\nnn \leq \xxx < \nnn+|\nnn|^{\ccc}) \ \ 
\phi(\eee^*,\xxx)
\nonumber
\\
\Ra \ \ \ 
& & 
\forall e \in \N \ \ \exists e^* \in \N \ \ \exists \ell \in \N 
\ \ \forall n \geq \ell \ \ \ 
\PTM_{e}(n) \ 
\not\vdash_S \ 
\exists \xxx \geq \nnn \ \ \phi(\eee^*,\xxx).
\label{eq:PTM-omega-con-2a} 
\end{eqnarray}

In the remarks below, we consider only the 
PTM-$\omega$-consistency
of theory $T$, not the PTM-$\omega$-consistency
of theory $S$ over $T$, since the PTM-$\omega$-consistency
of $S$ over $T$ follows similarly in each remark.

\vspace{7pt}

\begin{rem}\label{rem:1} 
\ 
\rm{
(Restriction of the related formulas of \PTM-$\omega$-consistency) 
\ \
\PTM-$\omega$-consistency is defined only for
$\Sigma^P_i$, $\Pi^P_i$ or $\Delta^P_i$-formulas.
This restriction is introduced from the fact that
if $\phi(\eee^*,\xxx)$ has a bounded quantifier 
$Q w < a$ with $|a| = 2^{|x|^c}$ for a constant $c$,
then
no PTM can even read ${\#}\aaa$ numeralwise. 
Since the notion of PTM-$\omega$-consistency 
is introduced to characterize a property of 
the provability of a PTM in theory $T$, 
such a restriction seems reasonable.

Actually, the proof of P $\not=$ EXP 
may imply that PA or a PT-extension of PA
is PTM-$\omega$-inconsistent for 
formula $\phi(\eee^*,\xxx)$
corresponding to the {\it formulation} of P $\not=$ EXP,  
which has a bounded quantifier with 
$\exists w < a$ with $|a| = 2^{|x|^c}$ for constant $c$.
(In other words,
the asymptotic polynomial-time unprovability
of P $\not=$ EXP does not imply
the formal unprovability of P $\not=$ EXP.)
}
\end{rem}

\vspace{7pt}

\begin{rem}\label{rem:4} 
\ 
\rm{
(Inequivalence of \PTM-$\omega$-consistency
and $\omega$-consistency) \ 
\PTM-$\omega$-consistency and $\omega$-consistency
do not imply each other.

First, we show that 
\PTM-$\omega$-consistency does not imply
$\omega$-consistency.
If we assume that \PTM-$\omega$-consistency of $T$ for $\phi(\eee^*,\xxx)$ implies
$\omega$-consistency of $T$ for $\phi(\eee^*,\xxx)$,
\PTM-$\omega$-consistency of $T$ for $\phi(\eee^*,\xxx)$ implies
consistency of $T$, since if $T$ is inconsistent,
$T$ is $\omega$-inconsistent for $\phi(\eee^*,\xxx)$.
That is,
the inconsistency of $T$ implies 
\PTM-$\omega$-{\it inconsistency} of $T$ for $\phi(\eee^*,\xxx)$. 
However, the inconsistency of $T$ implies
\PTM-$\omega$-{\it consistency} of $T$ for any formula,
since if $T$ is inconsistent, $T$ can prove any sentence and
Eq.(\ref{eq:PTM-omega-con-2}) does not hold,
which implies that $T$ cannot be \PTM-$\omega$-inconsistent. 
This is contradiction.
Therefore, \PTM-$\omega$-consistency does not imply
$\omega$-consistency.

%

Next, we show that $\omega$-consistency does not imply
\PTM-$\omega$-consistency.
Here, we assume that  
$$\NNN \models \ {\ov{\mbox{P$\not=$NP}}}.$$
It follows that theory $T =$ PA + ${\ov{\mbox{P$\not=$NP}}}$
is $\omega$-consistent since PA is $\omega$-consistent, and clearly 
$$ T \vdash \ {\ov{\mbox{P$\not=$NP}}}.$$
We now assume that $T$ is PTM-$\omega$-consistent for $\Delta^P_2$.
Then,  
$$ T \not\vdash \ {\ov{\mbox{P$\not=$NP}}},$$  
by Theorem \ref{mainthm5-dec}.
This is a contradiction.
Therefore, $T$ is PTM-$\omega$-inconsistent for $\Delta^P_2$,
while $T$ is $\omega$-consistent, if $\NNN \models \ {\ov{\mbox{P$\not=$NP}}}$.
That is, $\omega$-consistency does not imply
\PTM-$\omega$-consistency, assuming that 
$\NNN \models \ {\ov{\mbox{P$\not=$NP}}}$.
}
\end{rem}

\vspace{7pt}  
\begin{rem}\label{rem:5} 
\ 
\rm{
(Relationship between \PTM-$\omega$-consistency
and $\omega$-consistency) \ 
Although \PTM-$\omega$-consistency 
and $\omega$-consistency do not imply each other,
as described above,
the computational resource unbounded version of 
\PTM-$\omega$-consistency for $\Delta_1$-formulas
is equivalent to $\omega$-consistency for $\Delta_1$-formulas.

Now we define a computational resource unbounded version of 
\PTM-$\omega$-consistency, TM-$\omega$-consistency, as follows:
Theory $T$ is \TM-$\omega$-inconsistent for $\phi(\eee^*,\xxx)$, 
if the following two conditions hold simultaneously. 
\begin{eqnarray}
& & 
\forall e  \in \N \ \exists e^*  \in \N \  
\ \exists \ell  \in \N \ \forall n \geq \ell 
\ \forall f \in {\cal R} \ \ 
\TM_{e}(n) \ 
\not\vdash_T \ 
\exists \xxx \ (\nnn \leq \xxx < \nnn+f(|\nnn|)) \ \ 
\phi(\eee^*,\xxx),
\label{eq:PTM-omega-con-4}
\\
& & 
\exists e  \in \N \ \ \forall e^*  \in \N \ \ 
\forall \ell \in \N \ \ \exists n \geq \ell \ \ \ 
\TM_{e}(n) \ 
\vdash_T \ 
\exists \xxx \geq \nnn \ \ \phi(\eee^*,\xxx),
\label{eq:PTM-omega-con-3}
\end{eqnarray}
where ${\cal R}$ is a set of primitive recursive 
functions.
Here $T$ is a consistent primitive recursive extension 
of PA. 

See Section \ref{PT-proofs}
for the definition of $\TM_{e}(n) \vdash_T \ \ldots$, 
and see Definition \ref{C-omega-con} for
a generalized version of \PTM-$\omega$-consistency.


Eq. (\ref{eq:PTM-omega-con-4}) is equivalent to
\begin{equation}\label{eq:remark5-3}
\exists e^* \in \N \ \ 
\exists \ell  \in \N \ \forall n \geq \ell 
\ \ \forall f \in {\cal R} \ \ \
T \ 
\not\vdash \ 
\exists \xxx \ (\nnn \leq \xxx < \nnn+f(|\nnn|)) \ \ 
\phi(\eee^*,\xxx),
\end{equation}
since 
\begin{eqnarray}
& & 
\exists e  \in \N \ \ \forall e^*  \in \N \ \ 
\ \forall \ell  \in \N \ \exists n \geq \ell 
\ \exists f \in {\cal R} \ \ \
\TM_{e}(n) \ 
\vdash_T \ 
\exists \xxx \ (\nnn \leq \xxx < \nnn+f(|\nnn|) ) \ \ 
\phi(\eee^*,\xxx)
\nonumber
\\
\LRa \ \ 
& & 
\forall e^* \in \N \ \ 
\forall \ell  \in \N \ \exists n \geq \ell 
\ \exists f \in {\cal R} \ \ \
T \ 
\vdash \ 
\exists \xxx \ (\nnn \leq \xxx < \nnn+f(|\nnn|) ) \ \ 
\phi(\eee^*,\xxx).
\label{eq:remark5-4}
\end{eqnarray}
($\Ra$ is trivial, and $\La$ can be shown 
by constructing a TM that searches all proof trees, $\pi$, 
of    
$
\exists \xxx \ (\nnn \leq \xxx < \nnn+f(|\nnn|)) \ \ 
\phi(\eee^*,\xxx)
$
for all $(e^*,\ell,n,f) \in \N^3 \times {\cal R}$
in the order of the value of 
$e^*+\ell+n+|{\#}f|+|{\#}\pi|$ from 0 to greater.
)  

Eq. (\ref{eq:PTM-omega-con-3}) is equivalent to 
\begin{equation}\label{eq:remark5-2}
\forall e^*  \in \N \ \ 
\forall \ell \in \N \ \ \exists n \geq \ell \ \ \ 
T \ \vdash \ \
\exists \xxx  \geq \nnn \ \phi(\eee^*,\xxx).
\end{equation}

Since $\phi(\eee^*,\xxx)$ is a $\Delta_1$-formula
and $T$ is a consistent extension of PA,
Eq. (\ref{eq:remark5-3}) implies
\begin{equation}\label{eq:remark5-1}
\exists e^* \in \N \ \ 
\exists \ell  \in \N \ \forall x \geq \ell 
\ \ \ 
T \ 
\vdash \ \  \neg\phi(\eee^*,\xxx).
\end{equation}

Hence, 
if $T$ is \TM-$\omega$-inconsistent for $\Delta_1$-formula
$\phi(\eee^*,\xxx)$, 
$T$ is $\omega$-inconsistent for $\phi(\eee^*,\xxx)$, since
there exists $(e^*,n) \in \N^2$ such that
\begin{eqnarray*} 
& & 
\forall x \geq n  
\ \ \ 
T \ \vdash \ \  \neg\phi(\eee^*,\xxx)
\ \ \ \land \ \ \\
& &  
T \ \vdash \ \
\exists \xxx \geq \nnn \ \phi(\eee^*,\xxx) 
\end{eqnarray*} 
from Eqs. (\ref{eq:remark5-1})
and (\ref{eq:remark5-2}).  

On the other hand,
if $T$ is $\omega$-inconsistent for $\Delta_1$-formula
$\psi(\xxx)$, 
$T$ is \TM-$\omega$-inconsistent for $\phi(\eee^*,\xxx)$
($\equiv \psi(\xxx)$ for all $e^* \in \N$),
since
\begin{eqnarray*} 
& &  
\forall x \in \N \ \ 
T \ \vdash \ \  \neg\psi(\xxx)
\ \ \ \ \land \ \ \ \ 
T \ \vdash \ \
\exists \xxx \ \psi(\xxx) 
\\
\Ra \ \ 
& & 
\exists e^* \in \N \ \ 
\forall n \geq 0 \ \ \forall f \in {\cal R} \ \ \
T \ 
\not\vdash \ 
\exists \xxx \ (\nnn \leq \xxx < \nnn+f(|\nnn|) ) \ \ 
\phi(\eee^*,\xxx)
\\
& & 
\ \ \land \ \ \ \ 
\forall e^*  \in \N \ \ 
\forall n \in \N \ \ 
(\exists n \geq n) \ \ \ 
T \ \vdash \ \
\exists \xxx  \geq \nnn \ \ \phi(\eee^*,\xxx)
\end{eqnarray*} 

Thus,
TM-$\omega$-consistency for $\Delta_1$-formulas is equivalent 
to $\omega$-consistency for $\Delta_1$-formulas.
}
\end{rem}

\vspace{7pt}
\begin{rem}\label{rem:6} 
\ 
\rm{
(Provability of \PTM-$\omega$-consistency) \ 
Is PA (or another reasonable theory $T$) 
PTM-$\omega$-consistent for the related formula?
Unfortunately, we have not proven the PTM-$\omega$-consistency of PA for
$\Delta^P_2$.
Moreover, as shown in Theorem \ref{thm:ptm-omega-con-formal},
no PTM-$\omega$-consistent theory $T$, which is a consistent 
PT-extension of PA, can prove the PTM-$\omega$-consistency of PA,
although PTM-$\omega$-consistency of PA for $\Delta^P_2$ seems to  
be as natural as the $\omega$-consistency of PA.
}
\end{rem}

\vspace{7pt}

\begin{rem}\label{rem:7} 
\ 
\rm{
(Characterization of \PTM-$\omega$-consistency through axioms and deduction) \  
Assume that PA is PTM-$\omega$-consistent, and that
$T$ is a theory constructed by adding an axiom $X$ to PA
and is PTM-$\omega$-{\it inconsistent}.
Then, 
\begin{eqnarray}
& & 
\exists e \in \N \ \forall e^* \in \N \ 
\forall \ell \in \N \ \exists n \geq \ell \ \ \
\PTM_{e}(n) \ 
\vdash_{\PA} \ 
X \ \ra \ 
\exists \xxx \geq \nnn \ \ \phi(\eee^*,\xxx),
\label{eq:remark7-1}
\\
& & 
\forall e \in \N \ \exists e^* \in \N \ 
\exists \ell  \in \N \ \forall n \geq \ell \ 
\forall c \in \N \ \ \
\PTM_{e}(n) \ 
\not\vdash_{\PA} \ 
\exists \xxx \ (\nnn \leq \xxx < \nnn+|\nnn|^{\ccc}) \ \ 
\phi(\eee^*,\xxx).
\label{eq:remark7-2} 
\end{eqnarray} 

We then assume that the deduction of Eq.(\ref{eq:remark7-1}) 
is asymptotically polynomial-time, i.e.,
\begin{eqnarray} 
& & \exists e' \in \N \ \ \forall e^* \in \N \ \
\forall \ell \in \N \ \exists n \geq \ell  \ \ 
\neg Q_1 x_1 \in \N \cdots \neg Q_k x_k \in \N \ \  \exists x \geq n 
\nonumber 
\\ 
& &
\PTM_{e'}(x_1, \cdots, x_k, x) \ 
\vdash_{\PA} \ 
Y(\xxx_1, \cdots, \xxx_k) \ \ra \ 
 \ \phi(\eee^*,\xxx),
\label{eq:remark7-3}
\end{eqnarray} 
where $X \equiv Q_1 \xxx_1 \cdots Q_k \xxx_k \ Y(\xxx_1, \cdots, \xxx_k)$, 
$Q_i$ ($i=1,\ldots,k)$ are quantifiers 
and $Y(x_1, \cdots, x_k)$ is a $\Delta_1$ formula.
Here note that a polynomial (in the size of input) number of
application of logical axioms, Modus Ponens and Generalization rules 
is an asymptotically polynomial-time
deduction.

We now assume that $X$ can be asymptotically proven by a PTM over PA.
Then, 
\begin{equation}\label{eq:remark7-4}   
\exists e'' \in \N \ Q_1 x_1 \in \N \cdots Q_k x_k \in \N \ \ \ 
\PTM_{e''}(x_1, \cdots, x_k) \ 
\vdash_{\PA} \ Y(\xxx_1, \cdots, \xxx_k).
\end{equation}   
From Eqs. (\ref{eq:remark7-3}) and (\ref{eq:remark7-4}),
we obtain
$$    
\exists e \in \N \ \ \forall e^* \in \N \ \ 
\forall \ell \in \N \ \exists n \geq \ell \ \ \exists x \geq n 
\ \ \PTM_{e}(x) \ 
\vdash_{\PA} \ \phi(\eee^*,\xxx),
$$   
Hence, 
$$
\exists e \in \N \ \ \forall e^* \in \N \ \ 
\forall \ell \in \N \ \exists n \geq \ell \ \exists c \in \N \ \ 
\PTM_{e}(n) \ 
\vdash_{\PA} \ \
\exists \xxx \ (\nnn \leq \xxx < \nnn+|\nnn|^{\ccc}) \ \ 
\phi(\eee^*,\xxx),
$$
where 
$\exists \xxx \ (\nnn \leq \xxx < \nnn+|\nnn|^{\ccc}) \ 
\phi(\eee^*,\xxx) \ 
= \
\phi(\eee^*,\nnn)$,
when $c =0$ (i.e., $\exists c \in \N$). 
This contradicts
Eq.(\ref{eq:remark7-2}). 
Therefore,
if a theory $T$, which is PA $+ X$, is PTM-$\omega$-{\it inconsistent},
$X$ cannot be asymptotically proven by any PTM,
assuming that PA is PTM-$\omega$-consistent and
the deduction of Eq.(\ref{eq:remark7-1})
can be done asymptotically by a PTM.
 
Here it is worth noting that any (true) axiom $X$ {\it can} be asymptotically proven by a {\it resource unbounded}
TM over PA. The point in this remark is that
$X$ {\it cannot} be asymptotically proven by any {\it polynomial-time bounded} TM (i.e., PTM)
over PA.

}
\end{rem}

\vspace{7pt}
\begin{rem}\label{rem:8} 
\ 
\rm{
(Generalization of \PTM-$\omega$-consistency:
${\cal C}$-$\omega$-consistency)

We now generalize the concept of 
\PTM-$\omega$-consistency to
${\cal C}$-$\omega$-consistency,
where ${\cal C}$ is a (uniform)
computational class. 

Here, we introduce some concepts regarding
${\cal C}$.
Let $\U_{\cal C}$ be a universal Turing machine
specified to ${\cal C}$ in a manner similar to
$\U_{\PTM}$.
Here we omit the precise definition of $\U_{\cal C}$,
by which ${\cal C}$ is specified.
Each Turing machine in ${\cal C}$ is 
specified by $e \in \N$ as $\U_{\cal C}(e, \cdot)$.
We now introduce the following notation:
\begin{eqnarray*}
& & {\cal C}_{e}(a) \vdash_T \phi(\aaa) 
\\
& & \mbox{ \ \ }  \  
\LRa   \ \ \ 
\U_{\cal C}(e,(p,{\#}\Phi,a))={\#}\pi  \  \land  \  
\U_{\PTM}(v_{T},({\#}\phi(\aaa), {\#}\pi)) \ \mbox{accepts}.
\end{eqnarray*}

If the truth of {\bf Axiom}$_{T}(\nnn)$ 
(see Section \ref{pt-ext})
can be correctly decided 
by an algorithm of class ${\cal C}$ in $|n|$, 
on input $n$, 
we say that $T$ is ${\cal C}$-axiomizable.    
If $T$ is an extension of $T_0$ and ${\cal C}$-axiomizable,
then we say that $T$ is a ${\cal C}$-extension of
$T_0$.

\begin{definition}\label{C-omega-con}
(${\cal C}$-$\omega$-consistency) \ 

Let theory $S$ be a ${\cal C}$-extension of theory $T$.
$S$ is ${\cal C}$-$\omega$-{\it inconsistent} for $\Delta_1$-formula 
$\phi(\eee^*,\xxx)$ over $T$, 
if the following two conditions hold simultaneously. 
\begin{eqnarray} 
& & 
\forall e \in \N \ \exists e^* \in \N 
\ \exists \ell  \in \N \ \forall n \geq \ell 
\ \forall f \in F_{\cal C} \ \ \
{\cal C}_{e}(n) \ 
\not\vdash_T  
\exists \xxx \ (\nnn \leq \xxx < \nnn+f(|\nnn|)) \ \ 
\phi(\eee^*,\xxx),
\label{eq:PTM-omega-con-remark8-1} 
\\
& & 
\exists e \in \N \ \forall e^* \in \N \ 
\forall \ell \in \N 
\ \exists n \geq \ell \ \ \ 
{\cal C}_{e}(n) \ 
\vdash_S  
\exists \xxx \geq \nnn \ \ \phi(\eee^*,\xxx),
\label{eq:PTM-omega-con-remark8-2} 
\end{eqnarray} 
where
$F_{\cal C}$ is a set of primitive recursive functions, $f$,
such that $\U_{\cal C}(e,x)$
can do an existential search with
$f(|x|)$ steps (e.g., 
decide 
$\exists y (x \leq y < x+f(|x|)) \ g(y)=0$
by search of $y$ for $x, x+1, \ldots, x+f(|x|)$).

Theory $S$ is ${\cal C}$-$\omega$-{\it consistent} for $\phi(\eee^*,\xxx)$ 
over $T$,
if theory $S$ is not ${\cal C}$-$\omega$-inconsistent for 
$\phi(\eee^*,\xxx)$ over $T$.

Theory $S$ is ${\cal C}$-$\omega$-consistent for $\Sigma^P_i$
($\Pi^P_i$, $\Delta^P_i$, resp.) over $T$,
if $S$ is ${\cal C}$-$\omega$-consistent for any $\Sigma^P_i$ 
($\Pi^P_i$, $\Delta^P_i$, resp.) formula 
$\phi(\eee^*,\xxx)$ over $T$.

Theory $T$ is ${\cal C}$-$\omega$-consistent for $\phi(\eee^*,\xxx)$
($\Sigma^P_i$, $\Pi^P_i$, $\Delta^P_i$, resp.),
if $T$ is PTM-$\omega$-consistent for $\phi(\eee^*,\xxx)$
($\Sigma^P_i$, $\Pi^P_i$, $\Delta^P_i$, resp.) over $T$.

\end{definition}

The following definition is equivalent to the above:
Theory $S$ is ${\cal C}$-$\omega$-consistent for $\phi(\eee^*,\xxx)$ 
over $T$, 
if the following condition holds.
\begin{eqnarray}  
& & 
\forall e \in \N \ \ \exists e^* \in \N 
\ \ \exists \ell  \in \N \ \ \forall n \geq \ell 
\ \ \forall f \in F_{\cal C} \ \ \
{\cal C}_{e}(n) \ 
\not\vdash_T \ 
\exists \xxx \ (\nnn \leq \xxx < \nnn+f(|\nnn|) ) \ \ 
\phi(\eee^*,\xxx)
\nonumber
\\
\Ra \ \ \ 
& & 
\forall e \in \N \ \ \exists e^* \in \N \ \ 
\exists \ell  \in \N \ \ \forall n \geq \ell \ \ \ 
{\cal C}_{e}(n) \ 
\not\vdash_S \ 
\exists \xxx \geq \nnn \ \ \phi(\eee^*,\xxx).
\label{eq:PTM-omega-con-remark8-3} 
\end{eqnarray} 

}
\end{rem}

\subsection{Unprovability of 
${\ov{\mbox{P$\not=$NP}}}$ under PTM-$\omega$-Consistency}
\label{unprov-P-noteq-NP}

\begin{lemma}\label{mainthm1-dec}

Let theory $T$ be a consistent PT-extension of PA.  Then,
$$
\forall e \in \N \ \ \exists e^* \in \N \ \ \exists n \in \N \ \ 
\forall x \geq n \ \ \
\PTM_{e}(x) \ 
\not\vdash_T \  \neg\DecSAT(\eee^*,\xxx).
$$


\end{lemma}

\begin{proof}

Since 
$$
\DecSAT(\eee,\xxx) \ \equiv \
\CD[\SAT(\xxx)](\eee, \lc \CSAT \rc, \xxx)
$$ 
(see Section \ref{formaldef-pnp})
we obtain this theorem immediately from 
Theorem \ref{rbic-2nd-dec}.

\begin{flushright}
$\dashv$
\end{flushright}
\end{proof}

\begin{lemma}\label{lemma-sigma}

Let theory $T$ be a consistent PT-extension of PA.  

Let 
$
\Sigma[e^*,c] \equiv 
\{ \exists \xxx \ (\aaa \leq \xxx < \aaa+|\aaa|^{\ccc}) \ \ 
\neg\mbox{\DecSAT}(\eee^*, \xxx)
\mid  a \in \N \}
$, 
and
$\Size_{\Sigma[e^*,c]}(a) = |a|^{c+1}$. 
$$ 
\forall e \in \N \ \ \exists e^* \in \N \ \ \forall n \in \N \ \ 
\forall c \in \N \ \ \
\PTM_{e}(n) \ 
\not\vdash_T \ 
\exists \xxx \ (\nnn \leq \xxx < \nnn + |\nnn|^{\ccc}) \ \  
\neg\mbox{\DecSAT}(\eee^*, \xxx).
$$

\end{lemma}

\begin{proof}

Let ${\cal E}$ be a subset of $\N$ such that
$e \in {\cal E}$ if and only if  
$\U_{\PTM}(e,\cdot)$ is a PTM 
as follows:
\begin{itemize}
\item
Let $\phi(\xxx) \equiv \SAT(\xxx)$.
Let $\Phi \equiv \{ \phi(\aaa) \mid a \in \N  \}$,
and $\Phi' \equiv \{ \neg\phi(\aaa) \mid a \in \N  \}$. 
Let 
$
\Psi[c] \equiv \{ \psi(\ccc, \aaa, \sss) 
\mid a \in \N  \ \land \ s < 2^{m^c-1} \ \},
$ 
where
\begin{eqnarray*}
& & 
\psi(\ccc, \aaa, \sss) \ \equiv \  
\psi(\aaa, \Bit(\sss,\0) ) 
\land \psi(\aaa+\1, \Bit(\sss,\1) ) \land
\cdots \land \psi(\aaa+|\aaa|^{\ccc} \hik \1, 
\Bit(\sss,|\aaa|^{\ccc} \hik \1) ),  
\\
& & 
\psi(\xxx,\yyy) \equiv 
( \phi(\xxx) \land \yyy=\0 ) \ \lor \ ( \neg\phi(\xxx) \land \neg(\yyy=\0) ),
\\
& & 
\Size_{\Phi}(a) = \Size_{\Phi'}(a) = |a|,  \ \ \mbox{and} \ \
\Size_{\Psi[c]}(a) = |a|^{c+1}. 
\end{eqnarray*}
\item
Syntactically check whether the input has the form of 
$(d,{\#}{\Psi}[c],(a,s))$,
then follow the specification below.
Otherwise, there is no particular specification on the input.
\item
For all $i = 0,1,\ldots,|a|^c-1$,
simulate either one of 
$\U_{\PTM}(e,(d,{\#}{\Phi}, a+i))$ 
and $\U_{\PTM}(e,(d,{\#}{\Phi}',$ $a+i))$
by some rule (e.g.,
$\U_{\PTM}(e,(d,{\#}{\Phi},a+i))$ 
is simulated if and only if 
$a+i$ is even.)
\item
$\U_{\PTM}(e,(d,{\#}{\Phi},a+i))$ accepts
(and $\U_{\PTM}(e,(d,{\#}{\Phi},a+i))$ rejects)
if and only if 
$\U_{\PTM}(e,(d,{\#}{\Phi}',$ $a+i))$ rejects
(and $\U_{\PTM}(e,(d,{\#}{\Phi},a+i))$ accepts).
\item
Accept $(d,{\#}{\Psi}[c],(a,s))$
if and only if 
$$
\Bit(s,i) = 0
\ \ \ \LRa \ \ \
\U_{\PTM}(e,(d,{\#}{\Phi},a+i)) \ \mbox{ accepts},  
$$
for all $i = 0,1,\ldots,|a|^c-1$.
\end{itemize}

Note that ${\cal E}$ can be primitive recursive
by adopting a syntactically checkable canonical coding of 
the above-mentioned specification on $\U_{\PTM}$.
In other words, only $e$, for which PTM $\U_{\PTM}(e, \cdot)$
is specified in the canonical coding,
is in ${\cal E}$.
(Even if $\U_{\PTM}(e', \cdot)$ has the same functionality
as $\U_{\PTM}(e, \cdot)$ with $e \in {\cal E}$, 
unless $e'$ adopts the canonical coding,
$e' \not\in {\cal E}$.

\begin{claim}
For any $c \in \N$ and   
for any $e \in {\cal E}$, 
\begin{eqnarray}
& &
\PA \vdash \ \ 
\forall \nnn \ \ \forall \sss < \2^{|\nnn|^{\ccc} \hik \1} 
\nonumber
\\ 
& & \mbox{ \ } \ \ \ 
( \ \mbox{\CA}[\psi(\ccc,\nnn,\sss)](\eee, \lc {\Psi}[c] \rc, \nnn, \sss)
\ \ \ \ra \ \ \
\forall \xxx \ (\nnn \leq \xxx < \nnn + |\nnn|^{\ccc}) \ \ 
\mbox{\CD}[\phi(\xxx)](\eee, \lc {\Phi} \rc, \xxx) \ ).
\label{eq:lemma-sigma-claim} 
\end{eqnarray}

\end{claim}

\begin{proof}

From the construction of $\U_{\PTM}(e,\cdot)$ with $e \in {\cal E}$,
for any constant $c \in \N$,
\begin{eqnarray}
\PA \vdash & & \ 
\forall \nnn \ \ \forall \sss < \2^{|\nnn|^{\ccc} \hik \1} 
\nonumber
\\ 
& & 
( \ \mbox{\PTM-\Acc}[\psi(\ccc,\nnn,\sss)](\eee, \lc {\Psi}[c] \rc, \nnn, \sss)
\nonumber
\\
& & \mbox{ \ } 
\ \ \ \lra \ \ \
\forall \iii < |\nnn|^{\ccc} \ \ 
( \ \Bit(\sss,\iii) = \0  \ \ \lra \ \ 
\mbox{\PTM-\Acc}[\phi(\nnn+\iii)](\eee, \lc {\Phi} \rc, \nnn+\iii) \ )
\ ).
\label{eq:lemma-sigma-claim-1} 
\end{eqnarray}

In addition,
from the construction of $\U_{\PTM}(e,\cdot)$ with $e \in {\cal E}$,
\begin{equation}\label{eq:lemma-sigma-claim-2} 
\PA \vdash \ \ 
\forall \xxx \ 
( \ 
\mbox{\PTM-\Acc}[\phi(\xxx)](\eee, \lc {\Phi} \rc, \xxx) \ )
\ \ \lra \ \ 
\neg\mbox{\PTM-\Acc}[\phi(\xxx)](\eee, \lc {\Phi}' \rc, \xxx) \ )
\ ).
\end{equation}

On the other hand,
\begin{eqnarray}
\PA \vdash & & \ 
\forall \nnn \ \ \forall \sss < \2^{|\nnn|^{\ccc} \hik \1} 
\nonumber
\\ 
& & 
( \ \psi(\ccc, \nnn, \sss) 
\ \ \ \lra \ \ \
\forall \iii < |\nnn|^{\ccc} \ \ 
( \ \Bit(\sss,\iii) = \0  \ \ \lra \ \ \phi(\nnn+\iii) \ )
\ ).
\label{eq:lemma-sigma-claim-3} 
\end{eqnarray}

By Eqs. (\ref{eq:lemma-sigma-claim-1}), 
(\ref{eq:lemma-sigma-claim-2}) and (\ref{eq:lemma-sigma-claim-3}),
we obtain 
\begin{eqnarray*}
\PA \vdash & & \ 
\forall \nnn \ \ \forall \sss < \2^{|\nnn|^{\ccc} \hik \1} 
\\ 
& & 
( \ 
\mbox{\CA}[\psi(\ccc,\nnn,\sss)](\eee, \lc {\Psi}[c] \rc, \nnn, \sss)
\\ 
& & 
\ \ \lra \ \ 
( \ 
\mbox{\PTM-\Acc}[\psi(\ccc,\nnn,\sss)](\eee, \lc {\Psi}[c] \rc, \nnn, \sss)
\ \ \land \ \
\psi(\ccc, \nnn, \sss) 
\ ) 
\\
& & 
\ \ \ra \ \ 
\forall \iii < |\nnn|^{\ccc} \ \ 
( \ 
( \ \Bit(\sss,\iii) = \0  \ \ \lra \ \ 
\mbox{\PTM-\Acc}[\phi(\nnn+\iii)](\eee, \lc {\Phi} \rc, \nnn+\iii) \ )
\\
& & \mbox{ \  \  \ \  \ \ \ \  \ \ \ \ \ \ \ } 
\ \ \land \ \ 
( \ \Bit(\sss,\iii) = \0  \ \ \lra \ \ \phi(\nnn+\iii) \ )
\ )
\\
& & 
\ \ \ra \ \
\forall \iii < |\nnn|^{\ccc} \ \ 
( \ \phi(\nnn+\iii) \ \ \lra \ \ 
\mbox{\PTM-\Acc}[\phi(\nnn+\iii)](\eee, \lc {\Phi} \rc, \nnn+\iii) \ )
\\
& & 
\ \ \lra \ \
\forall \xxx \ (\nnn \leq \xxx < \nnn + |\nnn|^{\ccc}) \ \ 
\mbox{\CD}[\phi(\xxx)](\eee, \lc {\Phi} \rc, \xxx) \ )
\ ).
\end{eqnarray*}

\begin{flushright}
$\dashv$
\end{flushright}
\end{proof}

Therefore,
for any $c \in \N$ and   
for any $e \in {\cal E}$, 
\begin{eqnarray}
\PA \vdash & & \ 
\forall \nnn \ \ \forall \sss < \2^{|\nnn|^{\ccc} \hik \1} 
\nonumber
\\ 
( \ 
( \ 
& & 
\exists \xxx \ (\nnn \leq \xxx < \nnn + |\nnn|^{\ccc}) \ 
\neg\mbox{\CD}[\phi(\xxx)](\eee, \lc {\Phi} \rc, \xxx)
\ \ \ra \ \ 
\neg\mbox{\CA}[\psi(\ccc,\nnn,\sss]
(\eee, \lc {\Psi}[c] \rc, \nnn, \sss)
 \ ).
\label{eq:lemma-sigma-2}     
\end{eqnarray}

We now assume that
there exists $e \in \N$ such that 
\begin{equation}\label{eq:lemma-sigma-1}     
\forall e^* \in \N \ \exists n \in \N \ \exists c \in \N \ \ \
\PTM_{e}(n) \ 
\vdash_T \ 
\exists \xxx \ (\nnn \leq \xxx < \nnn + |\nnn|^{\ccc}) \ \  
\neg\mbox{\DecSAT}(\eee^*, \xxx).
\end{equation}
In other words, we assume there exists $e \in \N$ such that 
\begin{equation}\label{eq:lemma-sigma-3}     
\forall e^* \in \N \ \exists n \in \N \ \exists c \in \N \ \ \
\PTM_{e}(n) \ 
\vdash_T \ 
\exists \xxx \ (\nnn \leq \xxx < \nnn + |\nnn|^{\ccc}) \ \ 
\neg\mbox{\CD}[\phi(\xxx)](\eee^*, \lc {\Phi} \rc, \xxx)
\end{equation}

Then we can construct 
$\U_{\PTM}(e',\cdot)$ with $e' \in {\cal E}$ 
by using $\U_{\PTM}(e,\cdot)$
as follows:
\begin{itemize}
\item
(Input :) \ 
$(p, {\#}\Gamma[e^*,c], (n,s))$,
where $\Gamma[e^*,c] \equiv \{ 
\neg\mbox{\CA}[\psi(\ccc,\aaa,\ttt)](\eee, \lc {\Psi}[c] \rc, \aaa, \ttt) 
\mid (a,t) \in \N^2  \ \land \ t < 2^{|a|^c-1} \}$.
\item
(Output: )
G{\"o}del number of a proof tree of 
$\neg\mbox{\CA}[\psi(\ccc,\nnn,\sss)](\eee, \lc {\Psi}[c] \rc, \nnn, \sss)$
or 0.
\item
Run the following computation 
$$
\U_{\PTM}(e, (p,{\#}\Sigma[e^*,c],n))=z, \ \  
\U_{\PTM}(v_{T},({\#}\eta, z)),
$$
where 
$\eta \ \ \equiv \ \  
\exists \xxx \ (\nnn \leq \xxx < \nnn + |\nnn|^{\ccc}) \ \ 
\neg\mbox{\CD}[\phi(\xxx)](\eee^*, \lc {\Phi} \rc, \xxx)$.
\item
Compute the proof (say $\pi_2$) of 
$$
\forall \yyy \ \ \forall \ttt < \2^{|\nnn|^{\ccc} \hik \1} \ 
( \ \ 
\exists \xxx \ (\yyy \leq \xxx < \yyy + |\nnn|^{\ccc}) \ \
\neg\mbox{\CD}[\phi(\xxx)](\eee^*, \lc {\Phi} \rc, \xxx)
\ \ \ \ra \ \ \
\neg\mbox{\CA}[\psi(\ccc,\yyy,\ttt)](\eee^*, \lc {\Psi}[c] \rc, \yyy, \ttt)
 \ ),
$$
since there exists a proof of this formula by  
Eq. (\ref{eq:lemma-sigma-2}) if $e^* \in {\cal E}$.
(The computation time is finite.)
\item
Check whether $\U_{\PTM}(v_{T},({\#}\eta, z))$
accepts or rejects.
If it rejects, output 0 and halt.
If it accepts, then combine $\pi_1$ ($z = {\#}\pi_1$) and
$\pi_2$ and make a new proof tree, $\pi_3$, of 
$\neg\mbox{\CA}[\psi(\ccc,\nnn,\sss)](\eee, \lc {\Psi}[c] \rc, \nnn, \sss)$,
as follows: 
\begin{eqnarray*}
\pi_3 \ \ \equiv & & 
< \neg\mbox{\CA}[\psi(\ccc,\yyy,\ttt)](\eee^*, \lc {\Psi}[c] \rc, \yyy, \ttt),
\ \mbox{Modus Ponens} > 
\\
& & \mbox{ \ \ \ }
[ \ \pi_1, \ <\mbox{Formula A}, \ \mbox{Modus Ponens}> \ 
[ \ \pi_2, \ \mbox{Axiom X} \ ] \ ], 
\end{eqnarray*}
where 
Formula A is 
$$
\exists \xxx \ (\nnn \leq \xxx < \nnn + |\nnn|^{\ccc}) \ \ 
\neg\mbox{\CD}[\phi(\xxx)](\eee, \lc {\Phi} \rc, \xxx)
\ \ \ \ra \ \ \
\neg\mbox{\CA}[\psi(\ccc,\nnn,\sss)](\eee, \lc {\Psi}[c] \rc, \nnn, \sss)
$$
and 
Axiom X is a logical axiom, 
$$
\forall \yyy \ \ \forall \ttt < \2^{|\nnn|^{\ccc} \hik \1} 
( \ 
\exists \xxx \ (\yyy \leq \xxx < \yyy + |\nnn|^{\ccc}) \ \ 
\neg\mbox{\CD}[\phi(\xxx)](\eee^*, \lc {\Phi} \rc, \xxx)
\ \ \ \ra \ \ \
\neg\mbox{\CA}[\psi(\ccc,\nnn,\sss)](\eee^*, \lc {\Psi}[c] \rc, \nnn, \sss)
 \ )
$$
$$
\ra \ \ \
( \ \exists \xxx \ (\nnn \leq \xxx < \nnn + |\nnn|^{\ccc}) \ \ 
\neg\mbox{\CD}[\phi(\xxx)](\eee^*, \lc {\Phi} \rc, \xxx)
\ \ \ \ra \ \ \
\neg\mbox{\CA}[\psi(\ccc,\nnn,\sss)](\eee^*, \lc {\Psi}[c] \rc, \nnn, \sss)
 \ ).
$$
\item
Output $\pi_3$ for the proof tree of formula 
$\neg\mbox{\CA}[\psi(\ccc,\nnn,\sss)](\eee^*, \lc {\Psi}[c] \rc, \nnn, \sss)$.
\end{itemize}

Therefore, if we assume Eq. (\ref{eq:lemma-sigma-3}), 
there exists $e' \in {\cal E}$
such that 
$$
\exists n \in \N \ \exists c \in \N \ 
\forall s < 2^{|n|^c-1} \ \forall e^* \in {\cal E}
\ \
\PTM_{e'}(n,s) \ 
\vdash_T \ 
\neg\mbox{\CA}[\psi(\ccc,\nnn,\sss)](\eee^*), \lc {\Psi}[c] \rc, \nnn, \sss).
$$

Then, since $e' \in {\cal E}$ implies $g(e') \in {\cal E}$, 
there exists $e' \in {\cal E}$
such that 
$$
\exists n \in \N \ \exists c \in \N \ 
\forall s < 2^{|n|^c-1}
\ \
\PTM_{e'}(n,s) \ 
\vdash_T \ 
\neg\mbox{\CA}[\psi(\ccc,\nnn,\sss)](g(\eee'), \lc {\Psi}[c] \rc, \nnn, \sss).
$$

Here, for any $n \in \N$, there exists $s < 2^{|n|^c-1}$
such that 
$$
\NNN \models \ \psi(\ccc,\nnn,\sss).
$$
Hence, 
there exist $e' \in \N, \ n \in \N, \ c \in \N, \ s \in \N$ 
such that 
\begin{eqnarray*}
& & 
\NNN \models \ \psi(\ccc,\nnn,\sss) \ \ \ \land 
\\
& & 
\PTM_{e'}(n,s) \ 
\vdash_T \ 
\neg\mbox{\CA}[\psi(\ccc,\nnn,\sss)](g(\eee'), \lc {\Psi}[c] \rc, \nnn, \sss).
\end{eqnarray*}
This contradicts Corollary \ref{cor:rbic-2nd-dec},
so, Eq. (\ref{eq:lemma-sigma-1}) does not hold.

Since the contradiction occurs when $e^* =g(e')$ with
$e' \in {\cal E}$, 
we now define $g^*$ as follows:
$$ 
g^*(e) \equiv \left\{
\begin{array}{ll} 
 g(e)  & \mbox{\quad if $e \in {\cal E}$, } \\
 g(e') & \mbox{\quad if $e \not\in {\cal E}$.}          
\end{array}\right. 
$$ 
Since deciding whether $e \in {\cal E}$ or not
is primitive recursive and
the transformation of $e$ to $e'$ is also primitive recursive,
function $g^*$ is primitive recursive.
Then,
$$ 
\forall e \in \N \ \ \forall n \in \N \ \ \forall c \in \N \ \ \
\PTM_{e}(n) \ 
\not\vdash_T \ 
\exists \xxx \ (\nnn \leq \xxx < \nnn + |\nnn|^{\ccc}) \ \  
\neg\mbox{\DecSAT}(g^*(\eee), \xxx).
$$

\begin{flushright}
$\dashv$
\end{flushright}
\end{proof}

\begin{lemma}\label{lemma-PTM-omega} 
Let theory $T$ be a consistent PT-extension of PA
and \PTM-$\omega$-consistent for $\Delta^P_2$.

Then
$$
\forall e \in \N \ \ \exists e^* \in \N \ \ 
\exists \ell \in \N \ \ \forall n \geq \ell \ \ 
\PTM_{e}(n) \ 
\not\vdash_T \ \  
\exists \xxx \geq \nnn \ \ \neg\DecSAT(\eee^*,\xxx).
$$


\end{lemma}

\begin{proof}
From Lemma \ref{lemma-sigma} 
$$
\forall e \in \N \ \ \exists e^* \in \N \ \ \forall n \geq 0 \ \  
\forall c \in \N \ \ \
\PTM_{e}(n) \ 
\not\vdash_T \ 
\exists \xxx \ (\nnn \leq \xxx < \nnn + |\nnn|^{\ccc}) \ \  
\neg\mbox{\DecSAT}(\eee^*, \xxx).
$$

Since $\SAT(\xxx) \in \Sigma_1^P$ and
$\neg\SAT(\xxx) \in \Pi_1^P$,
then
$\neg\DecSAT(\eee^*,\xxx) \in \Sigma_2^P$
and 
$\neg\DecSAT(\eee^*,\xxx) \in \Pi_2^P$.
That is, $\neg\DecSAT(\eee^*,\xxx) \in \Delta_2^P$.

Therefore, if $T$ is \PTM-$\omega$-consistent for $\Delta^P_2$,
$T$ is  
\PTM-$\omega$-consistent for $\neg\DecSAT(\eee^*,\xxx) \in \Delta_2^P$.

We then obtain,  
from the definition of 
\PTM-$\omega$-consistency,
$$
\forall e \in \N \ \ \exists e^* \in \N \ \ 
\exists \ell \in \N \ \ \forall n \geq \ell \ \ 
\PTM_{e}(n) \ 
\not\vdash_T \ \  
\exists \xxx \geq \nnn \ \ \neg\DecSAT(\eee^*,\xxx).
$$

\begin{flushright}
$\dashv$
\end{flushright}
\end{proof}


\begin{theorem}\label{mainthm5-dec}

Let theory $T$ be a consistent PT-extension of PA
and \PTM-$\omega$-consistent for $\Delta^P_2$.

$$
T \not\vdash \
{\ov{\mbox{\rm P}\not=\mbox{\NP}}}.
$$

Namely, there exists no proof of 
${\ov{\mbox{\rm P}\not=\mbox{\NP}}}$ in $T$. 

\end{theorem}

\begin{proof}


Assume that 
$$
T \vdash \
{\ov{\mbox{\rm P}\not=\mbox{\NP}}},
$$
i.e., 
\begin{equation}\label{eq:mainthm5-dec}  
T \vdash \ \   
\forall \eee^* \ \ \forall \nnn \ \ \exists \xxx \geq \nnn \ \ 
\neg\DecSAT(\eee^*,\xxx).
\end{equation}

We can then construct 
PTM $\U_{\PTM}(e,\cdot)$ as follows:
\begin{itemize}
\item
(Input: ) \ $(p, {\#}\Sigma[e^*], n)  \in \N^3$,
where 
$\Sigma[e^*] \equiv 
\{ \exists \xxx \geq \aaa \ \ 
\neg\DecSAT(\eee^*,\xxx) \mid a \in \N \}$.
\item
(Output: ) \ G{\"o}del number of a proof tree of 
$\exists \xxx \geq \nnn \ \ \neg\DecSAT(\eee^*,\xxx)$.
\item
Find a proof, $\pi$, of formula 
$
\forall \yyy \ \ \forall \zzz \ \ \exists \xxx \geq \zzz \ \ 
\neg\DecSAT(\yyy,\xxx),
$
where $\pi$ exists according to the assumption,
Eq. (\ref{eq:mainthm5-dec}).
Here, the size of $\pi$ is constant in $|n|$.  
\item
Construct the following proof tree of
$\exists \xxx \geq \nnn \ \ \neg\DecSAT(\eee^*,\xxx)$:
$$
< \exists \xxx \geq \nnn \ \ \neg\DecSAT(\eee^*,\xxx), 
\mbox{Modus Ponens}> \ 
[ \ \pi, \ \mbox{Axiom X} \ ],
$$
where 
Axiom X is a logical axiom, 
$$
\forall \yyy \ \ \forall \zzz \ \ 
( \ \exists \xxx \geq \zzz \ \ 
\neg\DecSAT(\yyy,\xxx) \ ) \ \ \ 
\ra \ \ \
( \ \exists \xxx \geq \nnn \ \ \neg\DecSAT(\eee^*,\xxx) \ )
$$
\item
Output the G{\"o}del number of the proof tree.
\end{itemize}

Clearly, PTM $\U_{\PTM}(e,\cdot)$ 
outputs a correct value for all $(e^*,n) \in \N^*$.
Therefore, we obtain  
$$
\exists e \in \N \ \ \forall e^* \in \N \ \ \forall n \in \N \ \ \ 
\PTM_{e}(n) \ 
\vdash_T \ \  
\exists \xxx \geq \nnn \ \ \neg\DecSAT(\eee^*,\xxx).
$$

This contradicts Lemma \ref{lemma-PTM-omega}.
Thus,
$$
T \not\vdash \
{\ov{\mbox{\rm P}\not=\mbox{\NP}}}.
$$

\begin{flushright}
$\dashv$
\end{flushright}
\end{proof}

\noindent
{\bf Remark:} \
Theorem \ref{mainthm5-dec} and its generalization
imply
the results by Baker, Gill and Solovay \cite{BakGilSol75}
and by Hartmanis and Hopcroft \cite{HarHop76,Hartmanis78}.

First, let assume the following proposition,
which is a generalization of Theorem \ref{mainthm5-dec} and
will be formally given in Part 2 of this paper. 
\begin{proposition}\label{prop:super-C}
Let ${\cal C}$ be a (uniform)
computational class (see Remark \ref{rem:8} of Definition \ref{PTM-omega-con}),
and 
theory $T$ be a consistent ${\cal C}$-extension of PA
and ${\cal C}$-$\omega$-consistent for QBF.

Then, $T$ cannot prove 
any super-${\cal C}$-computational-lower bound.
\end{proposition}

We now assume that a relativizable proof of 
${\ov{\mbox{\rm P}\not=\mbox{\NP}}}$ 
exists for any oracle $A$ and 
that it is formalized in PA 
(or more generally, a $\omega$-consistent theory $T$). 

Then, for any oracle $A$ 
$$
\PA \vdash \
{\ov{\mbox{\rm P}^A\not=\mbox{\NP}^A}}.
$$
From Proposition \ref{prop:super-C},
PA should be PTM$^A$-$\omega$-{\it inconsistent}
for any oracle $A$.
Hence, PA should be TM-$\omega$-inconsistent,
which is equivalent to $\omega$-inconsistent 
(see Remark \ref{rem:5} of Definition \ref{PTM-omega-con}).
That is,
PA is not $\omega$-consistent.
This is a contradiction.
Thus, there exists no relativizable proof of 
${\ov{\mbox{\rm P}\not=\mbox{\NP}}}$
in PA, which corresponds to 
the result by Baker, Gill and Solovay \cite{BakGilSol75}.

Similarly, we can also obtain a result corresponding to
that by Hartmanis and Hopcroft \cite{HarHop76,Hartmanis78}
as follows:

First we assume that $T$ is a $\omega$-consistent theory.
Then, we can construct TM $M$ such that
\begin{eqnarray}
& & 
\exists e  \in \N \ \forall e^*  \in \N \  
\forall \ell  \in \N \ \exists n \geq \ell 
\ \exists f \in {\cal R} \ \ \
\PTM^{L(M)}_{e}(n) \ 
\vdash_T \ 
\exists \xxx \ (\nnn \leq \xxx < \nnn+f(|\nnn|) ) \ \ 
\phi(\eee^*,\xxx)
\nonumber
\\
\LRa \ \ 
& & 
\forall e^* \in \N \ \ 
\forall \ell  \in \N \ \ \exists n \geq \ell 
\ \ \exists f \in {\cal R} \ \ \
T \ 
\vdash \ 
\exists \xxx \ (\nnn \leq \xxx < \nnn+f(|\nnn|) ) \ \ 
\phi(\eee^*,\xxx),
\label{eq:mainthm5-dec-remark-1}
\end{eqnarray}
and
\begin{eqnarray}
& & 
\exists e  \in \N \ \ \forall e^*  \in \N \ \ 
\forall \ell \in \N \ \ \exists n \geq \ell \ \ \ 
\PTM^{L(M)}_{e}(n) \ 
\vdash_T \ 
\exists \xxx \geq \nnn \ \ \phi(\eee^*,\xxx)
\nonumber
\\
\LRa \ \ 
& & 
\forall e^*  \in \N \ \ 
\forall \ell \in \N \ \ \exists n \geq \ell \ \ \ 
T \ \vdash \ \
\exists \xxx  \geq \nnn \ \phi(\eee^*,\xxx).
\label{eq:mainthm5-dec-remark-2}
\end{eqnarray}
(For a method of constructing $M$,
see the description just after Eq. (\ref{eq:remark5-4})
in Remark \ref{rem:5} of Definition \ref{PTM-omega-con}.)

We now assume that
$$
T \vdash \
{\ov{\mbox{\rm P}^{L(M)}\not=\mbox{\NP}^{L(M)}}}.
$$
From Proposition \ref{prop:super-C},
$T$ should be \PTM$^{L(M)}$-$\omega$-{\it inconsistent}.
However, from the construction of TM $M$,
\PTM$^{L(M)}$-$\omega$-{\it inconsistent}
is equivalent to $\omega$-{\it inconsistent},
since Eqs. (\ref{eq:mainthm5-dec-remark-1})
and (\ref{eq:mainthm5-dec-remark-2}) hold.
That is,
$T$ should be $\omega$-{\it inconsistent}.
This is a contradiction.
Thus, for any $\omega$-consistent theory $T$,
there exists a TM $M$ such that 
$$
T \not\vdash \
{\ov{\mbox{\rm P}^{L(M)}\not=\mbox{\NP}^{L(M)}}},
$$
which corresponds to 
the result by Hartmanis and Hopcroft \cite{HarHop76,Hartmanis78}.

\subsection{Unprovability of 
Super-Polynomial-Time Lower Bounds in PSPACE 
under PTM-$\omega$-Consistency}
\label{unprov-super-poly-lower}

We can obtain the following theorem in a manner 
similar to that used in Section \ref{unprov-P-noteq-NP}. 

\begin{theorem}\label{thm:unprov-superpoly-lower-bounds}
Let language $L$ be in PSPACE.
Let theory $T$ be a consistent PT-extension of PA
and \PTM-$\omega$-consistent for QBF.

$$
T \not\vdash \
\forall \eee \ \ \forall \nnn \ \ \exists \xxx \geq \nnn \ \ \
\neg\CD[\phi_L(\xxx)](\eee, \lc \Phi_L \rc, \xxx),
$$

Namely, there exists no proof of 
any super-polynomial-time 
computational lower bound of $L$ in $T$.

\end{theorem}

\section{Unprovability of PTM-$\omega$-Consistency}\label{unprove-ptm-omega-dec}

This section shows that
the independence 
of P vs NP from $T$ by proving PTM-$\omega$-consistency
of $T$ for a $\Delta_2^P$-formula 
(i.e., through Theorem \ref{mainthm5-dec})
cannot be proven in theory $S$,
where $S$ is a consistent PT-extension of $T$
and is PTM-$\omega$-consistent for $\Delta_2^P$.
This result is based on the second incompleteness theorem 
of polynomial-time proofs, Theorem \ref{rbic-2nd}.

Let $T$ be a consistent PT-extension of PA, and assume that
P$\not=$NP is true.
Then, if $T$ is proven to be PTM-$\omega$-consistent
for $\neg\mbox{\DecSAT}(\eee^*,\xxx)$,
Theorem \ref{mainthm5-dec} will imply  
that $T$ is independent from P$\not=$NP. 
To prove the PTM-$\omega$-consistency of $T$ for 
$\neg\mbox{\DecSAT}(\eee^*,\xxx)$,
it is sufficient to prove that
\begin{equation}\label{eq:unprove-ptm-omega-dec-1}
\forall e \in \N \ \ \exists e^* \in \N \ \ 
\exists \ell \in \N \ \ \forall n \geq \ell \ \ \ 
\PTM_{e}(n) \ 
\not\vdash_T \ 
\exists \xxx \geq \nnn \ \ \neg\mbox{\DecSAT}(\eee^*,\xxx),
\end{equation}
since  
it has been already proven that
$$
\forall e \in \N \ \ \exists e^* \in \N 
\ \ \exists \ell  \in \N \ \ \forall n \geq \ell 
\ \ \forall c \in \N \ \ \
\PTM_{e}(n) \ 
\not\vdash_T \ 
\exists \xxx \ (\nnn \leq \xxx < \nnn+|\nnn|^{\ccc}) \ \ 
\neg\mbox{\DecSAT}(\eee^*,\xxx)
$$
by Lemma \ref{lemma-sigma}.

This section shows that
theory $S$ cannot prove Eq.(\ref{eq:unprove-ptm-omega-dec-1})
formally, if $S$ is a consistent PT-extension of $T$
and \PTM-$\omega$-consistent for $\Delta^P_2$ over $T$.
That is,  
the \PTM-$\omega$-consistency of $T$ for $\neg\mbox{\DecSAT}(\eee^*,\xxx)$
cannot be proven in $S$.
In other words,
the independence of P vs NP
from $T$ by proving the PTM-$\omega$-consistency of $T$
cannot be proven in $S$.
Here, the formal sentence of Eq.(\ref{eq:unprove-ptm-omega-dec-1})
in PA is
\begin{equation}\label{eq:unprove-ptm-omega-dec-1x}
\forall \eee \ \exists \lll  \ \forall \nnn \geq \lll \ \ \ 
\neg\mbox{\Prv}_{T}[\exists \xxx \geq \nnn \ \ \neg\mbox{\DecSAT}(h(\eee),\xxx)]
(\eee, \lc \Phi(e) \rc, \nnn),
\end{equation}
where $h$ is a primitive recursive function\footnote{
In Eq.(\ref{eq:unprove-ptm-omega-dec-1}),
there exists a primitive recursive 
function $h$ such that $e^* = h(e)$
for all $e \in \N$: i.e., 
$$
\forall e \in \N \ \ \forall \ell \in \N 
\ \ \exists n \geq \ell \ \ \ 
\PTM_{e}(n) \ 
\not\vdash_T \ 
\exists \xxx \geq \nnn \ \ \neg\mbox{\DecSAT}(h(\eee),\xxx).
$$
}, and 
$\Phi(e) \equiv \ 
\{\exists \xxx \geq \aaa \ \neg\mbox{\DecSAT}(h(\eee),\xxx) \mid 
a \in \N \}$.

This result is based on the incompleteness theorem
of polynomial-time proofs, Theorem \ref{rbic-2nd}.
To obtain this result, however, 
a slight modification is required for Theorem \ref{rbic-2nd}
as follows:
\begin{lemma}\label{lem:ptm-omega-con-asymp}
Let theory $T$ be a consistent PT-extension of PA, 
and $\Psi(e') \equiv \{ \psi(\eee',\aaa) \mid  a \in \N \}$.
$$
\forall e \in \N \ \ \exists e' \in \N \ \ \forall x \in \N
\ \ \ 
{\PTM}_{e}(x) \not\vdash_{T} \
\neg\mbox{\Prv}_{T}[\psi(\eee',\xxx)](\eee', \lc \Psi(e') \rc, \xxx).
$$

%

\end{lemma}

\begin{proof}

First, Eq. (\ref{eq:rbic-2nd-1}) is obtained in the same manner as
that of Theorem \ref{rbic-2nd}.

We then obtain  
\begin{equation}\label{eq:lem:ptm-omega-con-asymp-1}
\PA \vdash \
\forall \xxx \ \forall \yyy \ \
( \
\rho_{e,T}(\xxx) \land \neg\rho_{e,T}(\xxx) \ 
\ra \   \psi(\yyy,\xxx) \
),
\end{equation}
in place of Eq. (\ref{eq:rbic-2nd-1x}).

We then obtain the following claim
(in a manner similar to Corollary \ref{D-col1}):
\begin{claim}
Let 
$\Phi \equiv 
\{ \phi(\aaa) \mid  a \in \N \}$ and 
$\Psi(e) \equiv 
\{ \psi(\eee,\aaa) \mid  a \in \N \}$,
where $e \in \N$.
Suppose that $T$ is a consistent PT-extension of PA.
We assume
$$ 
T \vdash \ 
\forall \xxx \ \forall \yyy \ \ (\phi(\xxx) \ra \psi(\yyy,\xxx)).
$$
Then, for all $e_1 \in \N$ there exists $e_2 \in \N$ such that  
\begin{equation}\label{eq:lem:ptm-omega-con-asymp-claim} 
\forall e \in \N \ \ \
\PA \vdash \ \ \
\forall \xxx  \
( \
 \mbox{\Prv}_{T}[\phi(\xxx)](\eee_1, \lc \Phi \rc, \xxx) \ 
\ra \
 \mbox{\Prv}_{T}[\psi(\eee,\xxx)](\eee_2, \lc \Psi(e) \rc, \xxx) \ 
).
\end{equation}

\end{claim}

\begin{proof}
From the first derivability condition (D.1)
of a traditional proof theory \cite{Barwise77}
and the assumption of this lemma,
we obtain
$$ 
\PA \vdash \ 
\mbox{\Prv}_{T}(\lc 
\forall \xxx \ \forall \yyy \ \ (\phi(\xxx) \ra \psi(\yyy,\xxx))
\rc).
$$

Then,
PTM $\U_{\PTM}(e_2, \cdot )$ is constructed by using
PTM $\U_{\PTM}(e_1, (p, {\#}\Phi,\cdot))$
as follows:

\begin{enumerate}
\item
(Input :) \ 
$(p, {\#}\Psi(e), x)$ 
\item
(Output: )
G{\"o}del number of a proof tree of $\psi(\eee,\xxx)$
or 0.
\item
Run the following computation 
$$\U_{\PTM}(e_1, (p,{\#}\Phi,x))=z, \ \  
\U_{\PTM}(v_{T},({\#}\phi(\xxx),z)).$$
\item
Compute the proof (say $\pi_2$) of 
$\forall \www \ \forall \yyy \ \ (\phi(\www) \ra \psi(\yyy,\www))$,
since there exists a proof for the predicate from the assumption.
\item
Check whether $\U_{\PTM}(v_{T},({\#}\phi(\xxx),z))$
accepts or rejects.
If it rejects, output 0 and halt.
If it accepts, then combine $\pi_1$ ($z = {\#}\pi_1$) and
$\pi_2$ and make a new proof tree, $\pi_3$, for 
$\psi(\eee,\xxx)$,
as follows: 
$$
\pi_3 \ \ \equiv \ \ 
<\psi(\eee,\xxx),\mbox{Modus Ponens}> 
[\pi_1, <\phi(\xxx) \ra \psi(\eee,\xxx), \mbox{Modus Ponens}> 
[\pi_2, \mbox{Axiom X}] ], 
$$
where Axiom X is a logical axiom, 
``
$\forall \www \ \forall \yyy \ \ (\phi(\www) \ra \psi(\yyy,\www))
\ \ra \ (\phi(\xxx) \ra \psi(\eee,\xxx))$''.
\item
Output $\pi_3$ for the proof tree of formula 
$\psi(\eee,\xxx)$.
\end{enumerate}

The other part of the proof can be completed in an analogous 
manner to that in Lemma \ref{D2} except for the constructions of 
functions $h$ and $g$ to meet the above-mentioned 
construction of $\U_{\PTM}(e_2, \cdot)$) in this proof.

\begin{flushright}
$\dashv$
\end{flushright}

\end{proof}

Therefore, by setting $e \la e_2$ in 
Eq. (\ref{eq:lem:ptm-omega-con-asymp-2}),
for all $e_1 \in \N$ there exists $e_2 \in \N$ such that  
\begin{equation}\label{eq:lem:ptm-omega-con-asymp-claim-2}   
\PA \vdash \ \ \
\forall \xxx  \
( \
 \mbox{\Prv}_{T}[\phi(\xxx)](\eee_1, \lc \Phi \rc, \xxx) \ 
\ra \
 \mbox{\Prv}_{T}[\psi(\eee_2,\xxx)](\eee_2, \lc \Psi(e_2) \rc, \xxx) \ 
).
\end{equation}
Then, applying Eq. (\ref{eq:lem:ptm-omega-con-asymp-1})
to Eq. (\ref{eq:lem:ptm-omega-con-asymp-claim-2}),
we obtain that
for any $e^{+++} \in \N$, there exists $e' \in \N$ 
such that
\begin{equation}\label{eq:lem:ptm-omega-con-asymp-2} 
\PA \vdash \
\forall \xxx \
( \
\mbox{\Prv}_{T}
[\rho_{e,T}(\xxx) \land \neg\rho_{e,T}(\xxx)]
(\eee^{+++}, \lc {\cal G}^{+++} \rc, \xxx) \ 
\ra \ 
\mbox{\Prv}_{T}
[\psi(\eee',\xxx)]
(\eee', \lc \Psi(e') \rc, \xxx) \
),
\end{equation}
in place of Eq. (\ref{eq:rbic-2nd-1y}).

Hence, we obtain:  
for any $e \in \N$, there exists $e' \in \N$ 
such that
\begin{equation}\label{eq:lem:ptm-omega-con-asymp-3}   
\PA \vdash \
\forall \xxx \ \ 
( \ 
\neg\mbox{\Prv}_{T}[\psi(\eee',\xxx)](\eee', \lc \Psi(e') \rc, \xxx) \
\ra \ 
\rho_{e,T}(\xxx)
\ ),
\end{equation}
in place of Eq. (\ref{eq:rbic-2nd-2}).

The remaining part of the proof of this lemma is the same as
that of Theorem \ref{rbic-2nd}

\begin{flushright}
$\dashv$
\end{flushright}
\end{proof}

\begin{lemma}\label{lemma:ptm-sigma}
Let theory $T$ be a consistent PT-extension of PA.  
Let 
$\Phi(e') \equiv \ 
\{\exists \xxx \geq \aaa \ \ \neg\mbox{\DecSAT}(h(\eee'),\xxx) \mid 
a \in \N \}$ and
$h$ be a primitive recursive function.
\begin{eqnarray} 
& & 
\forall e \in \N \ \ \exists e' \in \N \ \ 
\exists m \in \N \ \ \forall \ell \geq m \ \ 
\forall c \in \N \ \ 
\nonumber
\\
& & 
\PTM_{e}(\ell) \ 
\not\vdash_T \ 
\exists \nnn \ (\lll \leq \nnn < \lll + |\lll|^{\ccc}) \ \  
\neg\mbox{\Prv}_{T}[\exists \xxx \geq \nnn \ \ \neg\mbox{\DecSAT}(h(\eee'),\xxx)]
(\eee', \lc \Phi(e') \rc, \nnn). \ \ \ 
\label{eq:lemma:ptm-sigma}
\end{eqnarray} 

\end{lemma}

\begin{proof}

First, we show the following claim:
\begin{claim}
Let $\Psi(d,c) \equiv 
\{ \forall \nnn \ (\aaa \leq \nnn < \aaa + |\aaa|^{\ccc}) \ \
\psi(\ddd,\nnn) \mid  a \in \N \}$
and 
$\Psi(d) \equiv 
\{ \psi(\ddd,\aaa) \mid  a \in \N \}$,
where 
$\psi(\ddd,\aaa)$ is any formula, 
$c \in \N$ and $d \in \N$.
Then, 
\begin{eqnarray}
\forall e \in \N \ \exists \widetilde{e} \in \N 
\ \forall d \in \N  \ \ & & 
\nonumber
\\
\PA \vdash \ \ 
\forall \mmm \ \ \forall \ccc \ \ \ 
(  \ & &
\mbox{\Prv}_{T}
[ \forall \nnn (\mmm \leq \nnn < \mmm + |\mmm|^{\ccc}) \ \  
\psi(\ddd,\nnn) ]
(\eee, \lc \Psi(d,c) \rc, \mmm) 
\nonumber 
\\
& & \  \ra \ \ \
\forall \nnn \ (\mmm \leq \nnn < \mmm + |\mmm|^{\ccc}) \ \ 
\mbox{\Prv}_{T}[\psi(\ddd,\nnn)]
(\widetilde{\eee}, \lc \Psi(d) \rc, \nnn) \ ) \ \ \
\label{eq:lemma:ptm-sigma-claim}
\end{eqnarray}

\end{claim}

\begin{proof}

First, we construct 
TM $\U(\widetilde{e}, \cdot )$ 
using PTM $\U_{\PTM}(e, \cdot)$
as follows:

\begin{enumerate}
\item
(Input :) \ 
$(p, {\#}\Psi(d), n)$ 
\item
(Output: )
G{\"o}del number of a proof tree of $\psi(\ddd,\nnn)$
or nothing (does not halt).

\item
Let $M_0(c) = \min\{ m \in \N \mid  m \leq n < m+|m|^c \}$  
and $M_0(c) = \max\{ m \in \N \mid  m \leq n < m+|m|^c \}$  

\item
Set $c \la 0$ and $m \la M_0(c)$. 

\item\label{item-number-1}
Run the following computation 
$$\U_{\PTM}(e, (p,{\#}\Psi(d,c),m) =z, \ \  
\U_{\PTM}(v_{T},({\#}\rho, z)),$$
where
$\rho \equiv 
\forall \xxx \ (\mmm \leq \xxx < \mmm + |\mmm|^{\ccc}) \ \
\psi(\ddd,\xxx)
$
\item
Check whether $\U_{\PTM}(v_{T},({\#}\rho,z))$
accepts or rejects.
If it accepts, go to \ref{item-number-2}. 
If it rejects, set $m \la m + 1$ and check 
whether $m > M_1$.
If $m \leq M_1$, then go to \ref{item-number-1}.
Otherwise, set $c \la c+1$,
compute $M_0(c)$ and $M_0(1)$, $m \la M_0(c)$,
and go to \ref{item-number-1}.

\item\label{item-number-2}.
Make a new proof tree, $\pi_2$, for 
$\psi(\ddd,\nnn)$,
from proof tree $\pi_1$ ($z = {\#}\pi_1$) 
for $\rho$, 
as follows: 
$$
\pi_2 \ \ \equiv \ \ 
<\psi(\ddd,\nnn),\mbox{Modus Ponens}> 
[\pi_1, \mbox{Axiom Y}], 
$$
where Axiom Y is a logical axiom, 
``
$
\forall \xxx \ (\mmm \leq \xxx < \mmm + |\mmm|^{\ccc}) \ \
\psi(\ddd,\xxx)
\ \ \ra \ \ 
\psi(\ddd,\nnn))
$''.

Output $\pi_2$ for the proof tree of formula 
$\psi(\ddd,\nnn)$.
\end{enumerate}

Here, if $c$ is a constant in $|n|$,
then TM $\U(\widetilde{e}, \cdot )$ 
should be PTM in $|n|$.

Therefore,
from the construction of $\U(\widetilde{e}, \cdot )$,
we obtain  
\begin{eqnarray*}
\forall e \in \N \ \exists \widetilde{e} \in \N
\ \forall d \in \N  
\ \ & & 
\\
\PA \vdash \ \ 
\forall \mmm \ \ \forall \ccc \ \ \ 
( \ & &
\mbox{\Prv}_{T}
[ \forall \nnn (\mmm \leq \nnn < \mmm + |\mmm|^{\ccc}) \ \
\psi(\ddd,\nnn) ]
(\eee, \lc \Psi(d,c) \rc, \mmm)
\\
& & \ \ra \ \ \
\forall \nnn \ (\mmm \leq \nnn < \mmm + |\mmm|^{\ccc}) \ \ 
\mbox{\Prv}_{T}[\psi(\ddd,\nnn)]
(\widetilde{\eee}, \lc \Psi(d) \rc, \nnn) \ )
\end{eqnarray*}

\begin{flushright}
$\dashv$
\end{flushright}
\end{proof}

We now assume that
\begin{eqnarray*} 
& &
\exists e \in \N \ \ \forall e' \in \N \ \ 
\forall m \in \N \ \ \exists \ell \geq m \ \ 
\exists c \in \N \ \ \\
& &
\PTM_{e}(\ell) 
\vdash_T \ 
\exists \nnn \ (\lll \leq \nnn < \lll + |\lll|^{\ccc}) \ \  
\neg\mbox{\Prv}_{T}
[\exists \xxx \geq \nnn \ \ \neg\mbox{\DecSAT}(h(\eee'),\xxx)] (\eee',
\lc \Phi(e') \rc, \nnn).
\end{eqnarray*}

Then, from Eq.(\ref{eq:lemma:ptm-sigma-claim}),
\begin{eqnarray*} 
& &
\exists e \in \N \ \ \forall e' \in \N \ \ 
\forall m \in \N \ \ \exists \ell \geq m \ \ 
\exists c \in \N \ \ \\
& &
\PTM_{e}(\ell) \  
\vdash_T \ \
\neg\mbox{\Prv}_{T}
[\forall \nnn (\lll \leq \nnn < \lll + |\lll|^{\ccc})
\exists \xxx \geq \nnn \ \ \neg\mbox{\DecSAT}(h(\eee'),\xxx)]
(\eee', \lc \Phi(e',c) \rc, \lll).
\end{eqnarray*}

This contradicts Lemma \ref{lem:ptm-omega-con-asymp}.

\begin{flushright}
$\dashv$
\end{flushright}
\end{proof}

We obtain the following lemma immediately from Lemma \ref{lemma:ptm-sigma} 
and the PTM-$\omega$-consistency of $S$.

\begin{lemma}\label{lem:ptm-omega-con-asymp2}
Let theory $T$ be a consistent PT-extension of PA, $S$ be a consistent
PT-extension of $T$, and $S$ be PTM-$\omega$-consistent for
$\Delta_2^P$ over $T$.  Let $\Phi(e) \equiv \
\{\exists \xxx \geq \aaa \ \neg\mbox{\DecSAT}(h(\eee),\xxx) \mid 
a \in \N \}$ and $h$ be a primitive recursive function.
$$
\forall e \in \N \ \ \exists e' \in \N \ \ 
\exists m \in \N \ \ \forall \ell \geq m \ \ \ 
\PTM_{e}(\ell) \ 
\not\vdash_S \ 
\exists \nnn \geq \lll \ \ 
\neg\mbox{\Prv}_{T}[\exists \xxx \geq \nnn \ \ \neg\mbox{\DecSAT}(h(\eee'),\xxx)]
(\eee', \lc \Phi(e') \rc, \nnn).  $$

\end{lemma}

\begin{theorem}\label{thm:ptm-omega-con-formal}
Let theory $T$ be a consistent PT-extension of PA, and 
$S$ be a consistent PT-extension of $T$
and PTM-$\omega$-consistent for $\Delta_2^P$ over $T$.
$$ 
S \not\vdash \ \ \
\forall \eee \ \ \exists \lll \ \ \forall \nnn \geq \lll \ \ \ 
\neg\mbox{\Prv}_{T}[\exists \xxx \geq \nnn \ \ \neg\mbox{\DecSAT}(h(\eee),\xxx)]
(\eee, \lc \Phi(e) \rc, \nnn).  
$$

Namely,
the PTM-$\omega$-consistency of $T$ 
for $\neg\mbox{\DecSAT}(\eee^*,\xxx)$, which is sufficient to prove $T
\not\vdash {\ov{\mbox{\rm P}\not=\mbox{\NP}}}$, cannot be proven in
$S$ (see Eqs. (\ref{eq:unprove-ptm-omega-dec-1}) and
(\ref{eq:unprove-ptm-omega-dec-1x})).

\end{theorem}

\begin{proof}

Let assume that 
$$ 
S \vdash \ \ \
\forall \eee \ \ \exists \lll \ \ \forall \nnn \geq \lll \ \ \ 
\neg\mbox{\Prv}_{T}[\exists \xxx \geq \nnn \ \ \neg\mbox{\DecSAT}(h(\eee),\xxx)]
(\eee, \lc \Phi(e) \rc, \nnn).  
$$ 
Then 
$$
\forall e \in \N  \ \   
S \vdash \ \
\exists \lll \ \ \forall \nnn \geq \lll \ \ \ 
\neg\mbox{\Prv}_{T}[\exists \xxx \geq \nnn \ \ \neg\mbox{\DecSAT}(h(\eee),\xxx)]
(\eee, \lc \Phi(e) \rc, \nnn).  
$$ 
This implies
$$
\forall e \in \N  \ \   
S \vdash \ \
\forall \lll \ \ \exists \nnn \geq \lll \ \ \ 
\neg\mbox{\Prv}_{T}[\exists \xxx \geq \nnn \ \ \neg\mbox{\DecSAT}(h(\eee),\xxx)]
(\eee, \lc \Phi(e) \rc, \nnn).  
$$ 
This means that there exists $e^*
\in \N$ such that 
$$
\forall e \in \N  \ \ \forall \ell \in \N \ \ \  
\PTM_{e^*}(\ell) \ \vdash_S \ \
\exists \nnn \geq \lll \ \ \ 
\neg\mbox{\Prv}_{T}[\exists \xxx \geq \nnn \ \ \neg\mbox{\DecSAT}(h(\eee),\xxx)]
(\eee, \lc \Phi(e) \rc, \nnn).  
$$ 

This contradicts Lemma
\ref{lem:ptm-omega-con-asymp2}.

Thus, 
$$ S \not\vdash \ \
\forall \eee \ \ \exists \lll \ \ \forall \nnn \geq \lll \ \ \ 
\neg\mbox{\Prv}_{T}[\exists \xxx \geq \nnn \ \ \neg\mbox{\DecSAT}(h(\eee),\xxx)]
(\eee, \lc \Phi(e) \rc, \nnn).  
$$

\begin{flushright}
$\dashv$
\end{flushright}
\end{proof}


\section{Unprovability of the Security of 
Computational Cryptography}\label{sec:crypto}

This section will show that
the security of any computational
cryptographic scheme is unprovable
in the standard notion of the modern cryptography, 
where an adversary is modeled to be 
a polynomial-time Turing machine.

First we will introduce a very fundamental cryptographic 
problem, the intractability of totally inverting
a one-way function by a deterministic PTM
(polynomial-time Turing machine).
Modern computational cryptography
is based on the assumption of the existence of one-way functions
\cite{Goldreich01}.\footnote{Although a one-way function is
usually defined against probabilistic PTMs, one-wayness against
deterministic PTMs is more fundamental than that against probabilistic
PTMs. For example, if a function is one-way against probabilistic PTMs, 
then the function will be also one-way against deterministic 
PTMs. That is, proving the one-wayness of a function against
probabilistic PTMs always implies proving that against deterministic PTMs.  
However, the reverse is not always true.}
In other words, to prove any level of security 
of such a computational cryptosystem
implies proving the one-wayness 
(the intractability of total inversion by any deterministic PTM)
of an underlying function.
Therefore, if it is impossible to prove the one-wayness of any 
function, it will be also impossible to prove any level of security
of any computational cryptographic scheme.

This section will show that
the intractability of totally inverting
a function by a deterministic PTM
is impossible to prove formally in the standard modern cryptographic setting.

\begin{definition}\label{one-way}
Let $n \in \N$, and  
$ f_n:$ $\Zn \ \ra  \ \Znc$ 
be a function
with parameter $n$ and ${\cal F} \equiv \{ f_n \mid n \in \N \}$
be a set of functions, where $c$ is a constant.

${\cal F}$ is called {\it one-way} if
there is no (deterministic) PTM $\U_{\PTM}(e, \cdot)$
such that
for all $x=(n, y, z) \in \N \ \times \ \Znc \ \times \ \Zn$ 
$\U_{\PTM}(e,x)$ outputs $w \in \Zn$
if there exists $w$ such that $y=f_n(w)$, and outputs nothing otherwise.
Here 
$\Size(x)=\Size(n, y, z) = |n|$.
\end{definition}

\begin{definition}\label{d-one-way}
Let ${\cal F} \equiv \{ f_n \mid n \in \N \}$
be a set of functions (see Definition \ref{one-way}).
Let $\mbox{\rm Inv}$ be an inversion oracle (a deterministic algorithm or
 a table)
such that $\mbox{\rm Inv}(n,y)$ outputs one of $\{w \mid y=f_n(w) \}$ 
if there exists $w$ such that $y=f_n(w)$, and outputs nothing otherwise.  

${\cal F}$ is called {\it decisionally one-way} if, 
for any inversion oracle $Inv$,
there is no (deterministic) PTM $\U_{\PTM}(e, \cdot)$
such that,
for all $x=(n, y, z) \in \ \N \ \times \ \Znc \ \times \ \Zn$, 
$\U_{\PTM}(e,x)$ accepts 
if and only if 
$\mbox{\rm Inv}(n,y) > z$.
(Note that $w$
is uniquely determined for each Inv and $(n,y)$.)

\end{definition}

\begin{lemma}\label{lemma-one-way}
${\cal F}$ is {\it one-way} if and only if
${\cal F}$ is {\it decisionally one-way}.
\end{lemma}

\begin{proof}
It is trivial that
if a PTM can invert $f_n$, it can also solve the 
corresponding decisional problem.

On the other hand,
we will show that
if a PTM can solve the 
decisional problem of $f_n$, then 
there exists a PTM that can invert $f_n$.
In other words,
$f_n$ can be completely inverted by using the solution of 
the decisional problem as a black-box 
$|n|$ times.
Here, we use binary search. 
Given a problem $(n, y)$ to invert $f_n$, 
queries to $\U_{\PTM}(e, \cdot)$
are $(n, y, \lf n/2 \rf)$, 
$(n, y, \lf (3/4)n \rf)$ (if the answer to the previous query is
accept), $\ldots$.         
Repeating this binary search $|n|$ times yields
an integer $v \in \Zn$.
Then, check whether $y=f_n(v)$ holds or not.
If it holds, set $w=v$. 
Otherwise, set $w$ to a null string
(or decide that there exists no value of $w \in Zn$ 
such that $y=f_n(w)$).

Therefore,
${\cal F}$ is  {\it one-way} if and only if
no PTM can solve the corresponding decisional problem.

\begin{flushright}
$\dashv$
\end{flushright}
\end{proof}

As shown in Lemma \ref{lemma-one-way},
the one-wayness of function family ${\cal F}$ can be characterized by
the intractability of the decisional problem, which can be also characterized by
a formula in theory $T$ as shown below.

\begin{definition}\label{fdef-one-way}
Let $R_{\cal F}^{Inv} \subset \N^4$ be a relation with respect to
inversion oracle $Inv$
such that
$(e, n, y, z) \in R_{\cal F}^{Inv}$ if and only if 
$(n, y, z) \in \ \N \ \times \ \Znc \ \times \ \Zn$, and
$\U_{\PTM}(e,(n, y, z))$ accepts 
if and only if 
$\mbox{\rm Inv}(n,y) > z$.

%
\end{definition}

\begin{lemma}\label{lemma-one-way2}
If ${\cal F}$ is one-way,
P$\not=$NP.
\end{lemma}

\begin{proof}
 
${\cal F}$ is one-way,
if and only if,
for any inversion oracle $Inv$,
there is no (deterministic) PTM $\U_{\PTM}(e, \cdot)$
such that,
for all $(n, y, z) \in \ \N \ \times \ \Znc \ \times \ \Zn$, 
$\U_{\PTM}(e,(n, y, z))$ accepts 
if and only if 
$(e, n, y, z) \in R_{\cal F}^{Inv}$.

Language $\{(e, n, y, z) \in R_{\cal F}^{Inv} \}$ is clearly NP. 
Therefore, if P = NP,
there is no one-way function family.

\begin{flushright}
$\dashv$
\end{flushright}
\end{proof}

We then obtain the following lemma 
immediately from Lemma \ref{lemma-one-way2}
and Theorem \ref{mainthm5-dec}.

\begin{lemma}\label{lemma-one-way5}
For any theory $T$ which is a consistent PT-extension of PA
and \PTM-$\omega$-consistent for $\Delta_2^P$,
there exists no proof of 
the one-wayness of any function family in $T$.
\end{lemma}

%

To study the (im)possibility of proving the one-wayness
of a function family, ${\cal F} \equiv \{ f_n \mid n \in \N \}$,
we need to make a model of provers (and 
adversaries). We now present a reasonable model of
provers.

\begin{definition}\label{def-model}
(Model of a prover in computational cryptography)

A {\it prover} is a PTM, which, given the (finite size of) description of
a cryptographic problem, outputs a
proof of the problem in a theory $T$ that
is a constant PT-extension of PA and
PTM-$\omega$-consistent for $\Delta^P_2$.
\end{definition}

This model should be justified by the fact
that an adversary is modeled to be a PTM 
in the definition of the one-wayness of a function family
in the above definition, which is the standard setting
in modern cryptography.
In other words,
the models of a prover and adversary should be
equivalent, since both prover and adversary are 
theoretical models of our human being
who analyzes the security of a one-way function family
to prove the security or to break it.
The key part of this model is that 
theory $T$ available for a prover to prove the security 
should be PTM-$\omega$-consistent,
since a prover is assumed to be a PTM.
This is because PTM-$\omega$-{\it inconsistent} theory
may include an unreasonably strong axiom 
(e.g., ${\ov{\mbox{P$\not=$NP}}}$ itself) that no PTM can 
prove asymptotically in PA. 

We now obtain the following theorem from
Lemma \ref{lemma-one-way5}.

\begin{theorem}\label{imp-crypto}
Under the prover model of Definition \ref{def-model}, 
there exists no proof of 
the one-wayness of any function family.
\end{theorem}

Note that PTM is just a one possible model of 
the feasible computation for our human being.
Even if in the future we have to change the 
feasible computation model of our human being,
the impossibility result of 
Theorem \ref{imp-crypto}
remains unchanged, because 
the computational models of prover and adversary
should be equivalent in any feasible computation model.
We will show similar results in various
computational classes in Part 2 of this paper. 

In addition, combining the result \cite{ImpRud89} with 
Theorem \ref{mainthm5-dec}
yields the following consequence:
 
\begin{theorem}\label{imp-trapdoor}
Under the prover model of Definition \ref{def-model}, 
there exists no proof of 
the existence of a (black-box) reduction 
from a one-way permutation to a secret key agreement. 
\end{theorem}

\section{Proof Complexity}\label{sec:p-complex}

In order to characterize the computational complexity 
to recognize the feasibility (triviality) of 
a theory to prove a statement,
this section introduces a proof complexity.

\begin{definition}\label{def:complexity}
We say that the {\it proof complexity} of 
$\phi$ is 
$O({\cal C})$
if 
there exists a proof of $\phi$ in a theory
$T$ that is a consistent PT-extension of PA
and ${\cal C}'$-$\omega$-consistent
for any class ${\cal C}'$ that
includes ${\cal C}$.
We say that the {\it proof complexity} of 
$\phi$ is $\Omega({\cal C})$
if 
there exists no proof of $\phi$ in any theory
$T$ that is a consistent PT-extension of PA
and ${\cal C}$-$\omega$-consistent.

\end{definition}

We now assume that 
PA is ${\cal C}$-$\omega$-consistent
for any computational class ${\cal C}$, which includes 
a computational class with constant-time complexity, $O(1)$.

Then, we obtain the following result:

%
%
\begin{itemize}
\item
Let $\ov{\mbox{P$\not=$EXP}}$ be a sentence that
formalizes the statement of P$\not=$EXP in PA 
in a manner similar to that in Section \ref{formaldef-pnp}.

The proof complexity of 
$\ov{\mbox{P$\not=$EXP}}$ is 
$O(1)$ (i.e., a {\it constant}-time),
since 
$\ov{\mbox{P$\not=$EXP}}$ can be proven in 
PA.

\item
Let Con(PA) be a sentence that
formalizes the consistency of PA in PA.
In other words,
$$
\mbox{Con}(\PA) \ \equiv \ 
\forall \xxx \  \neg\mbox{\PTM-\Acpt}(\vvv_{\PA}, \lc \bot \rc, \xxx).
$$
(For the notation and related result,  
see Lemmas \ref{dif-prov} and \ref{con-poly-proof}.)

The proof complexity of Con($\PA$) is 
$O(\mbox{P})$ (i.e., a {\it polynomial}-time),
since 
$\mbox{Con}(\PA)$ cannot be proven in PA
by the second G{\"o}del incompleteness theorem, 
but
it has a polynomial-time proof over PA
(Lemma \ref{con-poly-proof}), and
there exists a PTM-$\omega$-consistent theory $T$
(e.g., $T \equiv \PA + \mbox{Con}(\PA)$) 
that proves Con(PA).

\item 
The proof complexity of ${\ov{\mbox{\rm P}\not=\mbox{\NP}}}$
is $\Omega(\mbox{P})$
(i.e., {\it super-polynomial}-time),
since ${\ov{\mbox{\rm P}\not=\mbox{\NP}}}$
cannot be proven in a PTM-$\omega$-consistent theory 
(Theorem \ref{mainthm5-dec}).

\item
The proof complexity of 
$\sigma \ (\equiv \ 
\forall \eee \ \exists \lll \ \forall \nnn \geq \lll \ \   
\neg\mbox{\Prv}_{T}[\exists \xxx \geq \nnn \ \ \phi(\xxx)]
(\eee, \lc \Phi \rc, \nnn) )
$
is $\Omega(\mbox{P})$
(i.e., {\it super-polynomial}-time),
from Theorem \ref{thm:ptm-omega-con-formal}.
Therefore, the proof complexity of the independence 
of P vs NP from $T$
by proving the PTM-$\omega$-consistency
is also $\Omega(\mbox{P})$.

\end{itemize}

\section{Informal Observations}\label{subsec:phil}

If we assume Hypotheses 1 and 2 in Section 
\ref{sec:our-result-implication}, 
our main theorem implies that
P vs NP is independent from PA.
As the next step, it is natural to try to 
prove Hypothesis 2 (PTM-$\omega$-consistency of PA).
Since PA cannot prove Hypothesis 2,
a theory $T$ to prove Hypothesis 2 
should include an axiom, $X$, outside PA.
What axiom is appropriate for this purpose?

Usually it is not so easy for mathematicians/logicians 
to select/determine an appropriate axiom that would be 
widely recognized as feasible.
An extreme strategy is to adopt Hypothesis 2 itself  
as the new axiom, but such an axiom would not be
accepted as feasible.
Then, what is the criterion of a feasible axiom?
Unfortunately we now have no candidate.
Here we note that consistency and 
$\omega$-consistency are too weak as such a criterion
since PA + Hypothesis 2 is $\omega$-consistent (i.e., consistent)
if Hypothesis 2 is true.  
Currently, the feasibility of an axiom is decided only by 
whether it is widely accepted by many mathematicians/logicians
to be feasible.
 
Our result may suggest a criterion for
the feasibility of an axiom/theory.

Although axiom $X$ is outside PA (i.e., PA cannot prove $X$),
there exists an {\it asymptotic} proof of $X$ over PA,
if $X$ is true.
In other words, a Turing machine can 
produce an asymptotic proof of $X$ over PA.
We then consider the computational complexity
of a Turing machine that can produce an asymptotic proof of $X$ 
over PA.
According to Theorem \ref{thm:ptm-omega-con-formal},
theory $T = \PA + X$ to prove Hypothesis 2
should be PTM-$\omega$-{\it inconsistent}, and 
Remark \ref{rem:7} of Definition \ref{PTM-omega-con} shows that 
$X$ {\it cannot} be asymptotically proven by any 
{\it polynomial-time bounded} TM (i.e., PTM) over PA,
under some assumption.

If the computational capability of human beings
(along with our available/feasible computing facilities) 
is modeled as a polynomial-time Turing machine, 
which is widely accepted as a feasible computation model, 
our result implies that 
no human being can prove
axiom $X$ asymptotically over PA.
This may imply that axiom $X$
cannot be perceived as a feasible (or trivially true) 
statement by human beings, since it is 
beyond our capability to prove (or recognize the truth of) 
it even asymptotically over PA.
If so, 
a theory $T$ in which Hypothesis 2 can be proven
should include an axiom that cannot be perceived as feasible  
by human beings.
That is,
Hypothesis 2 cannot be proven
in any feasible theory $T$, which is
widely recognized to be feasible by 
mathematicians/logicians (i.e., human beings). 
In other words, even if Hypotheses 1 and 2 are
true and P vs NP is independent from PA,
such an independency cannot be proven
(through proving Hypothesis 2)  
in any feasible theory $T$ for us.
Similarly, even if Hypothesis 1 is true, 
$\ov{\mbox{P$\not=$NP}}$ 
may not be proven
in any feasible theory 
for human beings.

Con(PA),
which is a formal sentence representing 
the consistency of PA in PA,
is also unprovable in PA.
That is, a theory $T$ to prove Con(PA)
should include an axiom, $Y$, outside PA. 
In contrast with the above-mentioned case of
proving Hypothesis 2, 
Con(PA) can be asymptotically proven by a polynomial-time 
(more precisely, linear-time) 
Turing machine over PA
(Lemma \ref{con-poly-proof}),
and can be proven in a PTM-$\omega$-consistent theory,
PA + Con(PA), if PA is PTM-$\omega$-consistent.
Although Con(PA) would not be accepted as a feasible
axiom, the fact that Con(PA) can be proven in 
a PTM-$\omega$-consistent theory may imply
the existence of a feasible axiom, $Y$, for us
such that $T$ = PA + $Y$ can prove Con(PA) and $T$ is
PTM-$\omega$-consistent.
Actually, Gentzen \cite{Gentzen38} 
proved Con(PA) in a feasible theory
for us, which is in ZF (formal theory of set theory)
and whose additional axiom, $Y$, to PA is 
regarding transfinite induction
(corresponding to the axiom of foundation in ZF).

The relationship between G{\"o}del's incompleteness theorem
and our result is similar to that between recursion theory
and computational complexity theory.
Recursion theory studies (un)computability
on Turing machines, which are widely accepted
as the most general computation model
(the Church-Turing thesis),     
while computational complexity theory studies
(un)computability on 
a much more restricted computation model, 
a {\it feasible computation model for us
(human beings)}, i.e., polynomial-time
Turing machines (PTMs).
The major difference in the computation
model of recursion theory and the 
computational complexity theory
is that the former is resource {\it unbounded},
while the latter is resource {\it bounded}
(polynomial-time bounded).

G{\"o}del's incompleteness theorem 
is a result on unprovabilty
in the most general formal theories,
consistent theories 
(or slightly restricted theories, $\omega$-consistent
theories), that include PA, 
while
our main theorem is a result on unprovabilty
in much more restricted formal theories,
{\it feasible formal theories for us
(human beings)}, i.e., PTM-$\omega$-consistent theories,
that include PA.
The major difference in the formal theory
of G{\"o}del's incompleteness theorem and
our main theorem   
is that the former considers only the feasibility of 
the theory for 
a resource {\it unbounded} machine
(i.e., consistency or $\omega$-consistency), 
while the latter considers the feasibility of the theory for 
a resource {\it bounded}
(polynomial-time bounded) machine
(i.e., PTM-$\omega$-consistency).
In fact, as shown in Remark \ref{rem:5} 
of Definition \ref{PTM-omega-con},
the resource {\it unbounded} version of   
PTM-$\omega$-consistency is 
$\omega$-consistency.

Here, it is worth noting that
it should be controversial to decide 
the feasibility of a theory by PTM-$\omega$-consistency,
where all axioms and deductions in a theory should be
asymptotically proven by a PTM,
but that it might be similar to the situation
in computational complexity theory where 
it should have been controversial to characterize 
a feasible computation by class P,
since class P clearly includes many infeasible
computations for us such as $n^{10000}$ computational 
complexity in input size $n$.

Therefore, it may be reasonable to consider that
class P is introduced to characterize 
an {\it infeasible} computation,
rather than to characterize a {\it feasible} computation. 
That is, we consider that 
a computation outside P is {\it infeasible},
or an infeasible computation is characterized as
a super-polynomial-time computation class (super-P),
since almost all computations in super-P
are actually infeasible except a very small fraction of
super-P such as a computation with $O(n^{\log\log\log{n}})$
complexity
(In contrast, almost all computations in P are 
infeasible such that a computation with $n^c$ complexity
is infeasible for $c > 20$,
and only a small fraction of P is feasible).

Similarly, it may be reasonable to consider that
PTM-$\omega$-{\it inconsistency}, 
rather than PTM-$\omega$-{\it consistency}, 
is introduced to
characterize {\it infeasible} theories.  
In fact, as we mentioned above,
it is considered to be difficult for us (or PTMs)
to perceive the feasibility (triviality) of an axiom
of a PTM-$\omega$-{\it inconsistent} theory,  
since an axiom of a PTM-$\omega$-{\it inconsistent} theory 
{\it cannot} be proven even asymptotically by any 
PTM over PA, under some assumption
(Remark \ref{rem:7} of Definition \ref{PTM-omega-con}).
Our main theorems imply that 
P$\not=$NP (or any super-polynomial-time lower bound 
in PSPACE) is provable only in such an {\it infeasible} theory.

Note that our results do not 
deny the possibility of proving P=NP 
in a feasible theory for us,
if P=NP is true. 

G{\"o}del's second  incompleteness theorem
has a positive significance in that
it helps us to separate two distinct theories, $T$ and $S$,
because $T \vdash {\rm Con}(S)$ implies that
$T \not= S$  (and $T \supset S$) since $S \not\vdash {\rm Con}(S)$
by G{\"o}del's second incompleteness theorem.
Using this idea, the results of this paper may
provide some hint of the computational
capability of human beings.

Let $M$ be a machine whose computational capability 
is unknown.
If $C$ is a computational class, 
our result helps us to characterize the computational power of $M$ 
relative to $C$,
because $M \vdash_T {\rm SuperLowerBound}(C)$ 
where theory $T$ is feasible for $M$ implies that
the computational power of $M$ should be beyond $C$.
Here SuperLowerBound($C$) denotes a formula
to represent the super-$C$ computational lower bound in PA.
If we assume $M$ to be a computational model of
human beings,
then 
our obtained computational lower bound result
of $M \vdash_T {\rm SuperLowerBound}(C)$
in a feasible theory $T$ for us
implies the upper bound of our computational power. 
For example, 
we have already obtained a proof of a super-AC$^0$ lower bound
\cite{FurSaxSip84,Smolensky87}.
This fact means that the computational
power of human beings may exceed AC$^0$.

This result may also give us some hint as to why
all known results of computational lower bounds inside PSPACE are 
limited to very weak or restricted computational classes.
If the computational capability of human beings is considered to 
far exceed the target computational class for lower bound proof
(e.g., the target class is AC$^0$),
then it is likely that we may produce a proof of the lower bound
statement in a feasible theory for us.
However, if our computational capability 
is comparable to (or is not much beyond) the target computational 
class for lower bound proof,
then it may be very unlikely that we can provide its proof
in a feasible theory for us.
In other words, the best result of computational lower bounds 
may suggest the computational capability of human beings.


\section{Concluding Remarks}\label{sec:concl}

This paper introduced 
a new direction for studying 
computational complexity lower bounds;
resource bounded unprovability
(Sections \ref{sec:formalization} and \ref{sec:rbicthm})
and 
resource bounded undecidability 
(Sections \ref{sec:pt-dec} and \ref{sec:rbicthm-dec}).
This approach can be generalized to various systems by generalizing 
verification machines, $\U_{\PTM}(v_{T}, \cdot)$ in 
proof systems (Section \ref{sec:formalization}) and  
$\U(v, \cdot)$ in decision systems (Section \ref{sec:pt-dec}).

As mentioned in Section \ref{subsec:phil},
the relationship between G{\"o}del's incompleteness theorem
and our result is similar to that between recursion theory
and computational complexity theory.
Recursion theory studies (un)computability
on the most general computation model,
Turing machines (TMs),     
while computational complexity theory studies
(un)computability on 
a much more restricted computation model, 
a {\it feasible computation model for us}, i.e., polynomial-time
Turing machines (PTMs),
where PTMs are a resource (polynomial-time) 
bounded version of TMs.
G{\"o}del's incompleteness theorem 
is a result on unprovabilty
in the most general formal theories,
consistent theories 
(or slightly restricted theories, $\omega$-consistent
theories), that includes PA,
while
our main theorem is a result on unprovabilty
in much more restricted formal theories,
{\it feasible theories for us}, i.e., 
PTM-$\omega$-consistent theories,
that includes PA,
where PTM-$\omega$-consistent theories
are a resource (polynomial-time) bounded version of
$\omega$-consistent theories.
Note that a statement 
(e.g., $\ov{\mbox{P$\not=$NP}}$ ) unprovable  
in a PTM-$\omega$-consistent theory $T$
is independent from $T$ in our result, 
while
a statement (e.g., Con($T$) ) unprovable in 
an ($\omega$-)consistent theory $T$
is dependent on $T$ in G{\"o}del's incompleteness theorem.


In Part 2,
we will extend these results to other computational 
classes and show that:
for all $i \geq 1$, 
a super-$\Pi^P_i$ lower bound and 
a super-$\Sigma^P_i$ lower bound 
cannot be proven in  
a $\Sigma^P_i$-$\omega$-consistent theory
and a $\Pi^P_i$-$\omega$-consistent theory,
respectively.
For all $i \geq 1$, 
a super-$\mbox{\rm AC}^{i-1}$ lower bound
and a super-$\mbox{\rm NC}^i$ lower bound
cannot be proven in  
an $\mbox{\rm AC}^{i-1}$-$\omega$-consistent theory and
an $\mbox{\rm NC}^i$-$\omega$-consistent theory,
respectively.
In addition, Part 2 will present similar results on
probabilistic and 
quantum computational classes, since 
a probabilistic TM and quantum TM can be simulated 
by a classical deterministic TM; they can be formulated in PA
in a manner similar to that in Part 1. Thus,
for example, we will show that 
a super-BPP lower bound cannot be 
proven in  
a BPP-$\omega$-consistent theory and that
a super-BQP lower bound cannot be proven in 
a BQP-$\omega$-consistent theory.

\section*{Acknowledgments}

The authors would like to thank Noriko Arai, Toshiyasu Arai, 
Amit Sahai, Mike Sipser, Jun Tarui and Osamu Watanabe
for their invaluable comments and suggestions.
We would also like to thank anonymous reviewers of 
ECCC and FOCS'04 for valuable comments on previous 
versions of our manuscript.


\end{document}